\documentclass[aps,prb,twocolumn,showpacs,reprint,superscriptaddress,floatfix]{revtex4}

\usepackage{graphicx}
\usepackage{epsfig}
\usepackage{array}
\usepackage{verbatim}

\usepackage{times}
\usepackage{amsmath}
\usepackage{amssymb}
\usepackage{amsfonts}
\usepackage{latexsym}

\usepackage{multirow}
\usepackage{booktabs}

\usepackage{color}

\def\upa{\uparrow}
\def\dna{\downarrow}

\usepackage[colorlinks,bookmarks=false,citecolor=blue,linkcolor=red,urlcolor=blue]{hyperref}

\def\be{\begin{equation}}
\def\ee{\end{equation}}
\def\bea{\begin{eqnarray}}
\def\eea{\end{eqnarray}}

\def\vec{\mathbf}
\def\mc{\mathcal}

\begin{document}

\date{\today}
\title{Quantum magnetism on the Cairo pentagonal lattice}
\author{I. Rousochatzakis}
\affiliation{Max Planck Institut f$\ddot{u}r$ Physik komplexer Systeme, N\"othnitzer Str. 38, 01187 Dresden, Germany}
\affiliation{Institute for Theoretical Solid State Physics, IFW Dresden, 01171 Dresden, Germany}

\author{A. M. L\"auchli}
\affiliation{Institut f\"ur Theoretische Physik, Universit\"at Innsbruck, Technikerstr. 25, A-6020 Innsbruck, Austria}
\affiliation{Max Planck Institut f$\ddot{u}r$ Physik komplexer Systeme, N\"othnitzer Str. 38, 01187 Dresden, Germany}

\author{R. Moessner}\affiliation{Max Planck Institut f$\ddot{u}r$ Physik komplexer Systeme, N\"othnitzer Str. 38, 01187 Dresden, Germany}

\begin{abstract}
We present an extensive analytical and numerical study of the antiferromagnetic Heisenberg model
on the Cairo pentagonal lattice, the dual of the Shastry-Sutherland lattice with a close realization in the $S=5/2$ compound Bi$_2$Fe$_4$O$_9$. 
We consider a model with two exchange couplings suggested by the symmetry of the lattice,  
and investigate the nature of the ground state as a function of their ratio $x$ and the spin $S$.  
After establishing the classical phase diagram we switch on quantum mechanics in a gradual way 
that highlights the different role of quantum fluctuations on the two inequivalent sites of the lattice. 
The most important findings for $S=1/2$ include:  (i) a surprising interplay between a collinear and a four-sublattice orthogonal phase due 
to an underlying order-by-disorder mechanism at small $x$ (related to an emergent $J_1$-$J_2$ effective model with $J_2\gg J_1$),  
and (ii) a non-magnetic and possibly spin-nematic phase with d-wave symmetry at intermediate $x$. 
\end{abstract}


\maketitle

\section{Introduction}
Geometric frustration is at the heart of strong correlations in many models of quantum magnetism.\cite{HFM,Diep,Richter}  
Given that the most elementary building block of frustration is the triangle, lattice spin models with triangular units have been the minimal candidate models for realizing 
novel phases of matter in frustrated magnetism and have thus been explored widely over the years.  
Perhaps the most celebrated example in two dimensions is the kagome lattice antiferromagnet (AFM), an array of corner-sharing triangles, where frustration leads to an 
extensive number of classical ground states (GS).\cite{HFM} As a result, quantum fluctuations play a non-trivial role and may even favor a spin-liquid state, as it is currently believed for the 
extreme quantum-mechanical S=1/2 case.\cite{HFM}

Another elementary unit with built-in frustration is the pentagon. 
As in the case of the triangle, an AFM exchange interaction on a single pentagon 
favors a coplanar classical state\cite{SchmidtLuban} which is frustrated, in the sense that not all sides of the pentagon are fully satisfied. 
This state is a spiral with pitch angle $4\pi/5$ which, compared to the 120$^\circ$ angle in the single triangle case, might suggest a lower degree of frustration.  

However, there is a generic aspect of pentagonal lattice models which underlies a deeper degree of frustration and complexity.  
This is related to the fact that there is no Bravais lattice of pentagons and so, unlike the 120$^\circ$-state which can be easily ``tiled'' over the triangular lattice, 
it is not a priori evident that the single pentagon minimum is a good starting point for the description of the global low-energy physics.  
In fact, even in the finite-size case of the dodecahedron, which is a uniform tiling of pentagons on the geometry of a sphere, the GS is not related 
to the single pentagon minimum.\cite{SchmidtLuban}

A related aspect of 2D pentagonal lattices is that they often consist of two or more inequivalent sites and bonds (see below). 
Together with the above, one then expects non-trivial classical and quantum-mechanical phases in pentagon-based lattice models, 
and this is a direction in frustrated magnetism which is largely unexplored.

\begin{figure}[!b]
\includegraphics[clip=true,width=0.65\linewidth]{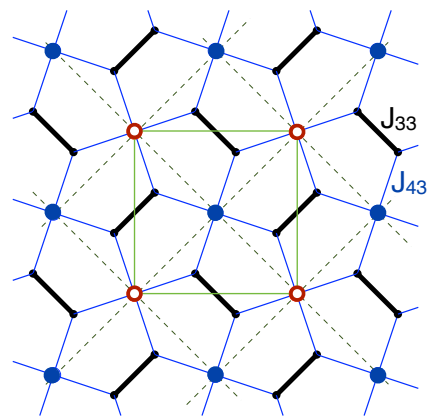}
\caption{(Color online) The Heisenberg model on the Cairo pentagonal lattice considered in this article. 
Solid thick and thin bonds stand for the $J_{33}$ and $J_{43}$ exchange couplings. 
The (green) rectangle in the middle denotes the 6-site unit cell of the Cairo lattice. 
The four-fold coordinated sites form a tilted square lattice (denoted by thin dashed lines) which is further divided into two square sublattices 
denoted by large open (red) and filled (blue) circles.}
\label{Fig:lattice}
\end{figure}

As mentioned above, the 2D plane has the generic property that cannot be tiled using regular pentagons. 
One alternative, which is realized in the pentagonal Penrose lattice, is to combine pentagons with other motifs in order to fill the void spaces.\cite{Kepler} 
Another is to use irregular pentagons which leads to the 14 pentagonal tesselations known so far.\cite{Tilings1,Tilings2} 
Among them, there are two tilings that have attracted interest in frustrated magnetism.  
The first is what is called ``the pentagonal lattice'' in some literature (see e.g. Fig.~2 in Ref.~[\onlinecite{Moessner}]), and the second is the 
``Cairo pentagonal lattice''\cite{Raman,Urumov,Ressouche,Ralko,Rojas,Iliev} which is the dual of the Shastry-Sutherland lattice (see Fig.~9 of Ref.~[\onlinecite{Raman}]), 
and is shown in Fig.~\ref{Fig:lattice}. 
 
The Cairo lattice is the main subject of this study. Its main features can be seen in Fig.~\ref{Fig:lattice}. 
First, there are two inequivalent sites with coordination numbers 3 and 4, and likewise there are two inequivalent bonds 
connecting 3-fold with 3-fold sites (thin blue) and 3-fold with 4-fold sites (thick black). 
In particular, the 4-fold sites form a square lattice which is represented by dashed lines in Fig.~\ref{Fig:lattice}. 
The Cairo lattice has a square Bravais lattice with a unit cell (green rectangle in the middle of Fig.~\ref{Fig:lattice}) of six sites, 
four of which are 4-fold and two are 3-fold coordinated. 

Among the previous studies on the Cairo lattice, we point out the Ising model study by Urumov\cite{Urumov} and Rojas {\it et al.}\cite{Rojas},
and that by Ralko\cite{Ralko} who studied a Hubbard model with hard-core bosons (equivalent to an XXZ model under a staggered magnetic field).
Our own study focuses on the Heisenberg model described by the Hamiltonian 
\be\label{eqn:Ham}
\mc{H} = \sum_{\langle ij \rangle} J_{ij}~\vec{S}_i \cdot \vec{S}_j ~,
\ee
where the sum runs over the nearest-neighbors $\langle ij\rangle$ of the Cairo lattice. 
As shown in Fig.~\ref{Fig:lattice}, we consider two different exchange interactions (which is the minimal number imposed by the symmetry of the Cairo lattice), 
$J_{33}$ and $J_{43}$, and focus on the regime where both are antiferromagnetic.  
By tuning the ratio $x\equiv J_{43}/J_{33}$ and the spin $S$, we shall be able to drive the system through a number of phases some 
of which have strong quantum-mechanical origin. As we are going to show below, the rich physics of this model is intimately connected 
to the presence of two inequivalent sites and bonds in the lattice. 

Besides the general motivation outlined above, there is another theoretical motivation for looking at this specific lattice model. 
This goes back to the work by Raman {\it et al.}\cite{Raman} who proposed a connection between Quantum Dimer models\cite{QDM} (QDM) and 
the so-called Klein models using a well controlled decoration procedure on a number of frustrated lattices. 
The Cairo lattice has most ingredients for resonating valence bond physics,\cite{QDM} since it is non-bipartite and has large resonance loops with even-length.\cite{Raman} 
And indeed, one of the main predictions of that study is that the QDM on the Cairo pentagonal lattice has an extended spin liquid GS. 
This leads to the question whether such a phase would survive if one includes dynamics out of the singlet manifold, and so a study of 
the $S=1/2$ Heisenberg model is a natural extension.

In parallel, it turns out that the Cairo pentagonal lattice is not only of purely theoretical interest, 
since the magnetism of Bi$_2$Fe$_4$O$_9$ offers a somewhat close realization of this model with $S=5/2$.\cite{Ressouche,Iliev} 
This compound was originally synthesized in the seventies by Shamir {\it et al.},\cite{Shamir} but it has attracted recent interest 
since it is a common by-product in the synthesis of the well known multiferroic compound BiFeO$_9$. 
In fact, Singh {\it et al.},\cite{Singh} have shown that Bi$_2$Fe$_4$O$_9$ also shares some magnetoelectric properties. 

A magnetic characterization in single crystals of Bi$_2$Fe$_4$O$_9$ has been given by Ressouche {\it et al.}\cite{Ressouche}
Despite the large Curie-Weiss temperature $\theta\simeq 1670$ K, this material orders magnetically at much lower temperatures $T_N\simeq 238$ K, 
which is the standard signature of magnetic frustration. 
The most nontrivial finding is the nature of the low-T phase: It is a coplanar configuration, whereby the 
4-fold coordinated Fe$^{3+}$ spins form four orthogonal sublattices, while the 3-fold coordinated spins bind antiferromagnetically with each other 
and in the direction of the local exchange field exerted by their neighboring 4-fold coordinated sites. 

As explained by Ressouche {\it et al.},\cite{Ressouche} Bi$_2$Fe$_4$O$_9$ is not a perfect realization of the Cairo lattice model. 
The first reason is that each pentagonal unit of the $ab$-plane comprises seven physical spins, 
since there are two ferromagnetically (FM) coupled Fe$^{3+}$ spins residing at each 4-fold site. 
Secondly, the minimal microscopic model description of this compound comprises three in-plane and two out-of-plane exchange pathways.\cite{Ressouche} 
Despite this, the classical configuration of Bi$_2$Fe$_4$O$_9$ seems to survive in a much larger parameter space,  
since the same state appears also in our more symmetric version of the model.

\section{Main results and organization of the article}\label{Sec:MainResults} 
Our target is the GS phase diagram of the model in the $x$-$S$ plane. 
To accomplish this goal and establish a fairly good understanding of the various phases of the model 
we begin with the purely classical limit (Sec.~\ref{Sec:classical}) and then we switch on quantum-mechanics in a gradual way 
by using different levels of approximations and complexity. In doing so we shall also be able to highlight 
the different role of quantum fluctuations at the two inequivalent sites of the model. 

\begin{figure*}[!t]
\includegraphics[clip=true,width=\linewidth]{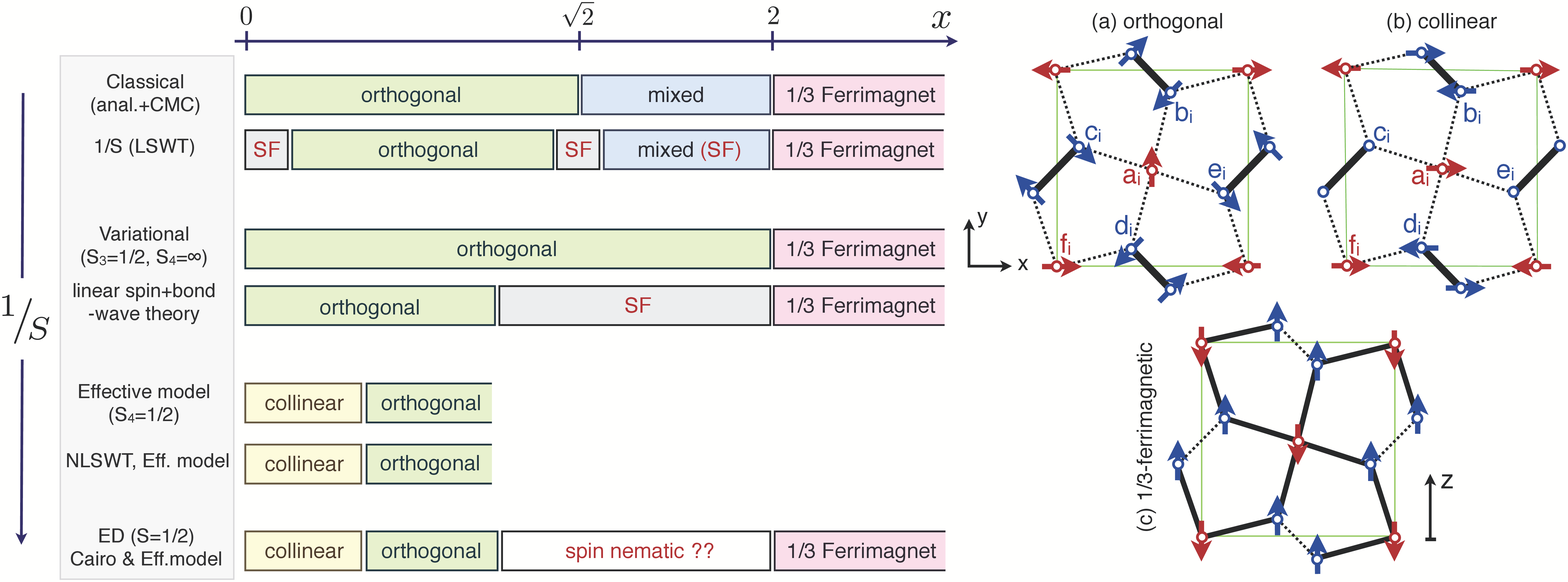}
\caption{The Heisenberg model on the Cairo pentagonal lattice. Left panel: Evolution of the phase diagram as we go from the classical (top) to the quantum limit (bottom). 
Quantum fluctuations are included gradually, in a way that highlights the different role of the two inequivalent sites iof the lattice. 
The regions indicated by ``SF'' in the second and fourth lines are the ones with strong semiclassical fluctuations. 
Right panel: The main magnetic phases appearing in the phase diagram. The mixed phase (not shown) interpolates between the orthogonal (a) and the 1/3-ferrimagnetic phase (c) 
as described in the text. In (a) and (b) we also show our labeling scheme for the corresponding bosonic operators (for each of the 6 sites of the unit cell enclosed by the green solid line)  
that appear in LSW theory.  Note that in the collinear phase (b), the spins in half of the $J_{33}$ bonds (e.g. the ones labeled by $c_i$ and $e_i$) 
do not feel any exchange field from the neighboring 4-fold sites.}
\label{Fig:methods}
\end{figure*}

Our theoretical predictions for the phase diagram in the $x$-$S$ plane are presented in Fig.~\ref{Fig:methods}.  
The classical (large-$S$) phase-diagram is shown at the top line of Fig.~\ref{Fig:methods}.  
It consists of three magnetic phases: 
(i) the orthogonal phase ($x<\sqrt{2}$) found for Bi$_2$Fe$_4$O$_9$, 
(ii) a collinear 1/3-ferrimagnetic phase ($x>2$),   
and (iii) an intermediate ($\sqrt{2}<x<2$) mixed phase which combines both (i) and (ii). 
We should note here that the 1/3-ferrimagnetic phase appears also for Ising spins\cite{Rojas} and for hard-core bosons.\cite{Ralko}

Our first step to include quantum fluctuations in the problem is by a standard linear spin wave theory (Sec.~\ref{Sec:LSWT}). 
Our results for the renormalized spin lengths indicate three regions (see second line of Fig.~\ref{Fig:methods}) 
with strong quantum fluctuations: 
(i) at the 3-fold sites for small $x$, 
(ii) at the 4-fold sites as we approach the classical transition between the orthogonal and the mixed phase ($x=\sqrt{2}$), and 
(iii) at the 4-fold sites throughout the mixed phase. 
These are the regions where new competing phases might appear as we approach the extreme $S=1/2$ quantum limit. 
By contrast the 1/3-ferrimagnetic phase seems to survive quantum fluctuations. 

One may immediately realize that the strong fluctuations at the 3-fold sites at small $x$ result from the following generic feature.  
At small $x$, the 3-fold sites prefer to bind into quantum-mechanical singlets which are very different from the product up-down configuration of the classical ansatz.   
This strong tendency to form singlets on the $J_{33}$ dimers can be captured by a modified variational ansatz  (Sec.~\ref{Sec:Variational})
whereby the 3-fold dimers are treated fully quantum-mechanically and the 4-fold sites still classically.  

A numerical minimization of this ansatz for $S=1/2$ gives the phase diagram shown in the third line of Fig.~\ref{Fig:methods}. 
The first important result is that the orthogonal state is now stabilized up to $x=2$, i.e. the mixed state does not survive in this ansatz for low enough spins $S$.  
In addition, the present ansatz predicts a finite staggered polarization on the $J_{33}$-dimers, apart from the strong tendency to form singlets. 
This can be explained by the fact that the local exchange field exerted from the 4-fold sites in their orthogonal configuration 
is staggered and can thus admix a finite triplet $|t_0\rangle$ amplitude with the singlet wavefunction as soon as $x$ is finite. 

We may go one step further and include stronger quantum fluctuations by performing an expansion around the orthogonal variational GS found for $x<2$. 
The natural way to do this is to perform a standard Holstein-Primakoff expansion for the 4-fold spins and a bond-wave expansion\cite{Bhatt,Judit} for the 3-fold dimers. 
The corresponding quadratic theory is presented in Sec.~\ref{Sec:LSBWT} and its predictions for $S=1/2$ are shown in the fourth line of Fig.~\ref{Fig:methods}.  
The main result is the presence of very strong fluctuations above $x\sim 1$, which provides strong evidence that the variational treatment 
is not a good starting point in the window $1\!<\!x\!<\!2$, and that a new phase will be stabilized in this regime for low enough spins $S$.

The last place to look for quantum fluctuations is at the 4-fold sites (Sec.~\ref{Sec:EffModel}), which have been treated classically or semi-classically up to now. 
It turns out that these fluctuations are responsible for some very rich physics at low energies.  
The main reason for this is that the $x=0$ limit has a highly degenerate GS manifold since the 4-fold sites are free to point up or down in this limit. 
As a result, the low-energy physics at small $x$ is governed by effective interactions between the 4-fold sites 
which are mediated by the virtual fluctuations of the $J_{33}$-dimers out of their singlet GS. 
By integrating out these high-energy fluctuations up to fourth order in $x$, we have derived an effective low-energy theory for spins $S=1/2$, 
which governs the interactions between the 4-fold sites up to $x\sim 0.4-0.6$. 

The first key result of the effective theory is that the nearest-neighbor coupling $J_1$ is much smaller than the 
next-nearest-neighbor coupling $J_2$ due to destructive quantum interferences.  
Thus, at small enough $x$, the effective model reduces to the well studied\cite{Chandra,Schultz,Bishop,Caprioti,Sirker,Darradi} 
$J_1$-$J_2$ AFM model on the square lattice with $J_2\gg J_1$. 
An immediate consequence is that there must exist a critical value of $x$  (see fifth line of Fig.~\ref{Fig:methods}) 
below which the orthogonal phase becomes unstable in favor of the collinear phase through an order-by-disorder mechanism. 

The second important result from the effective theory is the appearance, in fourth-order in $x$, of a 4-spin coupling term $K$ 
which involves the 4 spins in every plaquette of the square lattice, and is similar to the well-known ring exchange term.\cite{Roger} 
Our exact diagonalizations provide evidence that this plaquette term is the one that actually drives the system 
into the intermediate ($1<x<2$) quantum-mechanical phase mentioned above.

It is worth noting that the collinear phase cannot not be detected using the above linear semiclassical theories. 
The reason is that the collinear phase is not the GS of either the fully classical or the variational ansatz, and therefore one must include 
interactions between the spin-waves or spin+bond waves respectively in order to stabilize this phase. 
To demonstrate this, we have employed (Sec.~\ref{Sec:NLSWT}) a spin-wave expansion around the collinear phase using the effective model Hamiltonian 
and keeping up to quartic terms in the interactions. 
The results from a self-consistent mean-field decoupling show that the collinear phase can indeed be stabilized at low enough $x$ (fifth line of Fig.~\ref{Fig:methods}).

It is by now quite clear that the phase diagram becomes richer as we approach the extreme $S=1/2$ limit.  
We have seen for instance that out of the three classical phases, the mixed phase does not survive quantum fluctuations for low enough spins $S$.   
We have also uncovered an interesting order-by-disorder mechanism which is at play in the small-$x$ regime and gives rise to the collinear phase for $S=1/2$.   
We have also learned from the spin+bond-wave expansion that a new phase is to be expected at intermediate $x$ for low enough $S$.    
However the nature of this phase is not known yet.  
In particular, one important question is whether the small $x$ limit is a good perturbative limit for the description of this phase or not. 
If the answer is yes we must next identify the effective term that drives the transition, 
which is a difficult task given that our fourth-order effective model should be valid only up to $x\sim 0.4-0.6$. 

To solve these open issues we look at the full quantum-mechanical $S=1/2$ problem using exact diagonalizations (ED) on finite-size clusters (Sec.~\ref{Sec:ED}). 
Our numerical results on the original Cairo lattice model confirm the presence of two nearly decoupled AFM sublattices for small $x$ which is the physics we expect 
from the effective model at large $J_2/J_1$. Both the symmetries of the low-energy spectra and the GS correlations show 
signatures of the orthogonal and the collinear phase, but we are not able to pinpoint the exact transition between the two phases 
given that the locking between the two AFM sublattices present at large $J_2/J_1$ takes place at very large length scales.\cite{CCL,Weber} 

The results also establish the presence of a new phase at intermediate values of $x$ before we reach the 1/3-ferrimagnetic phase. 
Apart from a GS level crossing we also find a new spectral structure at low energies. 
In particular, the magnetization process in a field now shows steps of $\Delta S_z=2$ which is typical for collinear spin-nematic phases.\cite{Shannon,Penc} 
Moreover, this phase seems to be adiabatically connected to the GS manifold of the $x=0$ limit,  
which suggests that it is driven by one of the couplings in the effective model. 

To clarify this issue we have also performed ED in the effective model on the square lattice but with unconstrained $J_1/J_2$ and $K/J_2$. 
We have found that the low-energy spectral structure in the regime $K \gg J_{1,2}$ is very similar to the one in the Cairo lattice model at intermediate $x$. 
This suggests that the intermediate phase is driven by one of the three topologically different 4-spin exchange terms of the model, which we identify as the one 
which has an enlarged SU(2)$\times$SU(2) symmetry. 
We show that this term favors a spin-nematic phase with d-wave symmetry similar to the one found by Shannon {\it et al}.\cite{Shannon} 
The main difference is that here the symmetry breaking seems to take place in one of the two sublattices of the square lattice only, while the symmetry of the other sublattice 
remains intact. 

The remaining part of the article is organized along the main lines described above. 
We shall mainly focus on the central aspects and predictions of each separate approach 
and relegate technical details to the Appendices. 


\section{Classical limit}\label{Sec:classical} 
In the classical limit we find three different GS's as a function of $x$. 
At large $x$ the lattice becomes effectively bipartite and one can minimize the energy by a collinear arrangement 
of up and down spins on the 3-fold and the 4-fold sites respectively (see Fig.~\ref{Fig:methods}(c)). 
Since the number of 3-fold sites is twice the number of 4-fold sites, $N_3=2N_4$, 
this is a ferrimagnetic configuration with a total magnetization of 1/3.  This phase remains stable down to $x=2$. 

In the opposite limit of small $x$, the classical GS is the orthogonal configuration shown in Fig.~\ref{Fig:methods}(a), which is the one found for Bi$_2$Fe$_4$O$_9$.\cite{Ressouche}  
Here the 4-fold sites form an orthogonal 4-sublattice configuration, while the nearest neighbor 
3-fold sites align antiferromagnetically to each other and at the same time point opposite to the total exchange field exerted by the neighboring 4-fold sites.  
This phase remains stable up to $x=\sqrt{2}$. 
We note here that the orthogonal configuration of the 4-fold sites has been found previously on some ring-exchange\cite{Roger} models on the square lattice, see e.g. Refs.~[\onlinecite{Chubukov,Lauchli}]. 

This leaves a window between $x=\sqrt{2}$ and $2$ where the spins find a compromise between the
two phases by combining both into a mixed phase. Namely, at $x=\sqrt{2}$ the 3-fold and 4-fold sites begin to tilt out of the plane but in opposite directions to each other. 
In particular, the projection of this non-coplanar configuration onto the xy-plane gives the orthogonal phase while the projection along the z-axis gives the 1/3 ferrimagnetic state.  
At $x=2$ the spins are completely aligned along the z-axis.

\begin{figure}[!t]
\includegraphics[clip=true,width=0.85\linewidth]{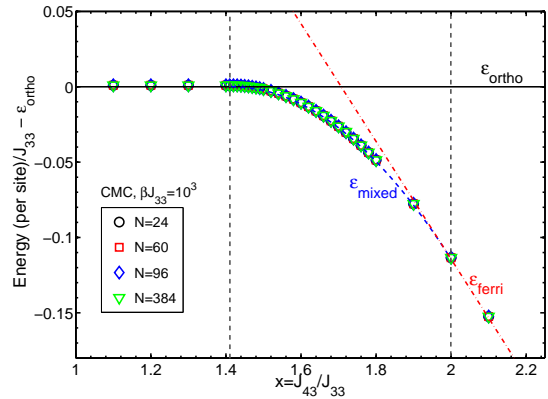}
\caption{ Classical GS energy per site (measured from the energy of the orthogonal phase) 
as a function of $x=J_{43}/J_{33}$.  The lines stand for the analytical expressions given in Eqs.~(\ref{eqn:Emixed}-\ref{eqn:Eferri}), 
and the symbols are numerical data from CMC simulations at $\beta J_{33}=10^3$.}
\label{fig:EnergyCMC}
\end{figure}

All three classical phases are special cases of the ansatz (see right panel of Fig.~\ref{Fig:methods}(a)) 
\bea
\vec{S}_{a_i} &=& S \left( p_i \cos\theta ~\vec{e}_y - \sin\theta ~\vec{e}_z \right)\\
\vec{S}_{f_i} &=&  S \left( p_i \cos\theta ~\vec{e}_x - \sin\theta ~\vec{e}_z \right) \\
\vec{S}_{b_i} = \vec{S}_{d_i} &=&  S \left( -p_i \cos\theta' ~\vec{e}_+ + \sin\theta' ~\vec{e}_z \right) \\
\vec{S}_{c_i} = \vec{S}_{e_i} &=&  S \left( p_i \cos\theta' ~\vec{e}_- + \sin\theta' ~\vec{e}_z \right)
\eea
where $p_i = e^{i \vec{Q} \cdot \vec{R}_i}$, $\vec{Q}=(\pi,\pi)$, and $\vec{e}_\pm = \frac{\vec{e}_x \pm \vec{e}_y}{\sqrt{2}}$. 
The two angles $\theta$ and $\theta'$ account for the tilting out of the xy-plane of the 4-fold and the 3-fold sites respectively. 
For the orthogonal and the 1/3-ferrimagnetic phases, $\theta=\theta'=0$ and $\theta=\theta'=\pi/2$ respectively, while 
for the mixed phase $\sin\theta=\sqrt{2-4/x^2}$, and $\sin\theta' = \frac{x}{2} \sin\theta$. 
The corresponding energies per site are 
\bea
\label{eqn:Eortho} \varepsilon_{\text{ortho}}/S^2 &=& -(1+2\sqrt{2}x)/3 \\
\label{eqn:Emixed} \varepsilon_{\text{mixed}}/S^2 &=& -1-x^2/3~,\\
\label{eqn:Eferri} \varepsilon_{\text{ferri}}/S^2 &=& (1-4 x)/3 ~.
\eea

To confirm that these phases correspond to the global minima we have also performed Classical Monte Carlo (CMC) calculations at low temperatures using the Metropolis algorithm.  
The average energies per site in some representative low-T equilibrium ensembles are shown in Fig.~\ref{fig:EnergyCMC} for a number of cluster sizes.
The results are in excellent agreement with the above picture. In particular they confirm the presence of the intermediate mixed phase. 

We note here that the energy of the mixed phase is only slightly below the energies of the neighboring orthogonal and ferrimagnetic phases, 
and this suggests that this phase may be quite fragile against quantum fluctuations. 
Indeed, the variational treatment presented below in Sec.~\ref{Sec:Variational} will demonstrate that the mixed phase is unstable for $S=1/2$ as soon as 
we include the quantum fluctuations on the 3-fold sites.

\section{Semiclassical expansion}\label{Sec:LSWT}
\subsection{Linear Spin-Wave theory}
Our next step is to assess the strength of quantum fluctuations by performing a separate semiclassical expansion around each of the three classical GS's. 
The Cairo pentagonal lattice has 6 sites (two 4-fold and four 3-fold coordinated) per unit cell.   
So we introduce six bosonic operators denoted by $a_i, b_i, c_i, d_i, e_i, f_i$ to describe the transverse fluctuations on each site of the $i$-th unit cell. 
For each site we define a local quantization $z$-axis (see Fig.~\ref{Fig:lattice}(a)-(c)) and perform a standard Holstein-Primakoff expansion, namely 
\be
S_{a_i}^z = S - a_i^+ a_i ,~  S_{a_i}^+ \simeq \sqrt{2S}~ a_i ~,
\ee
and similarly for the remaining sites of the unit cell.  
The standard procedure\cite{White,Blaizot} for the diagonalization of the resulting quadratic Hamiltonian  
is outlined in App.~\ref{App:LSWT}. 
Here we shall focus on the two main quantities of interest: the renormalization of the GS energy 
and the renormalization of the local spin lengths by harmonic fluctuations. 
 
\begin{figure}[!b]
\includegraphics[clip=true,width=0.9\linewidth]{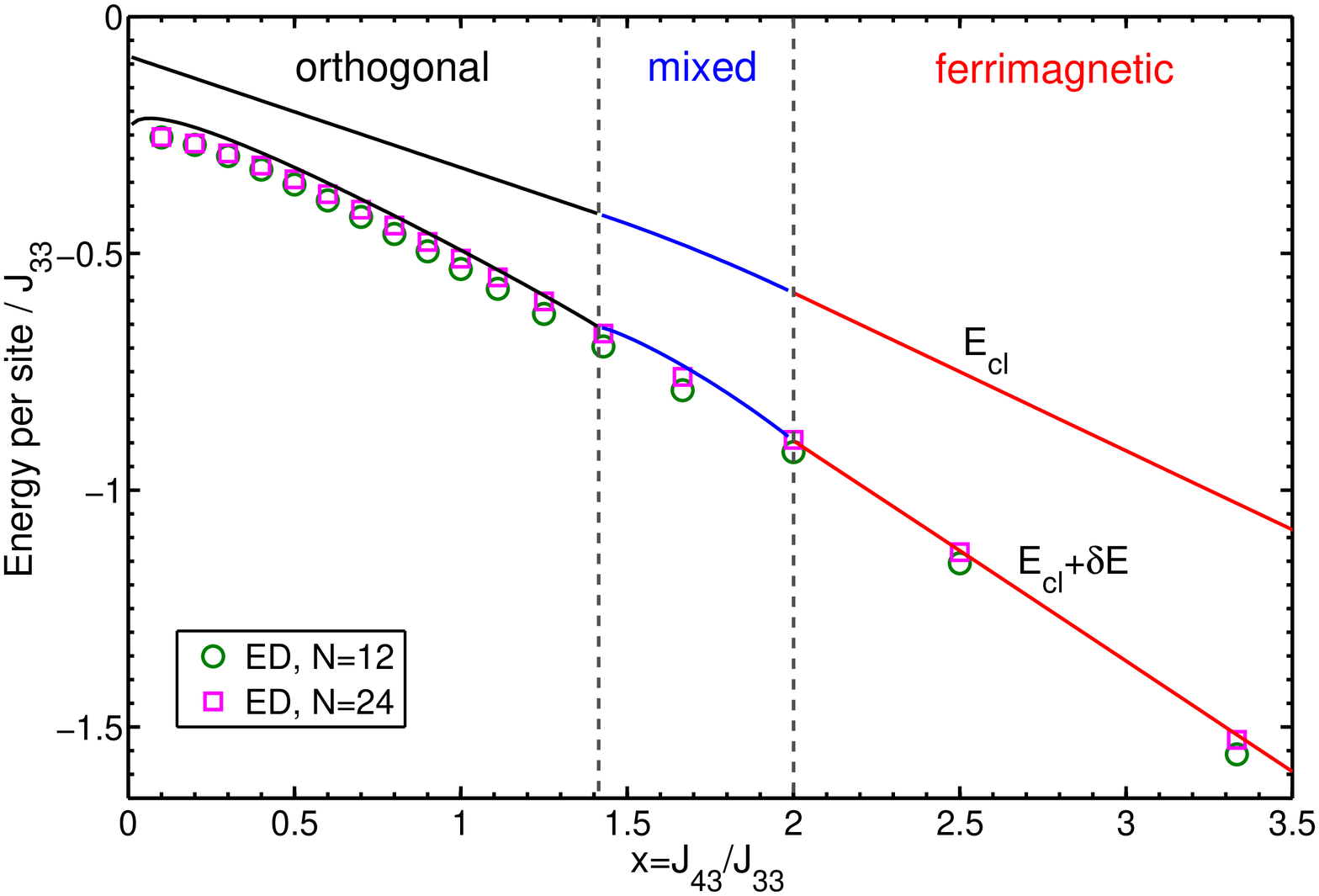}
\includegraphics[clip=true,width=0.9\linewidth]{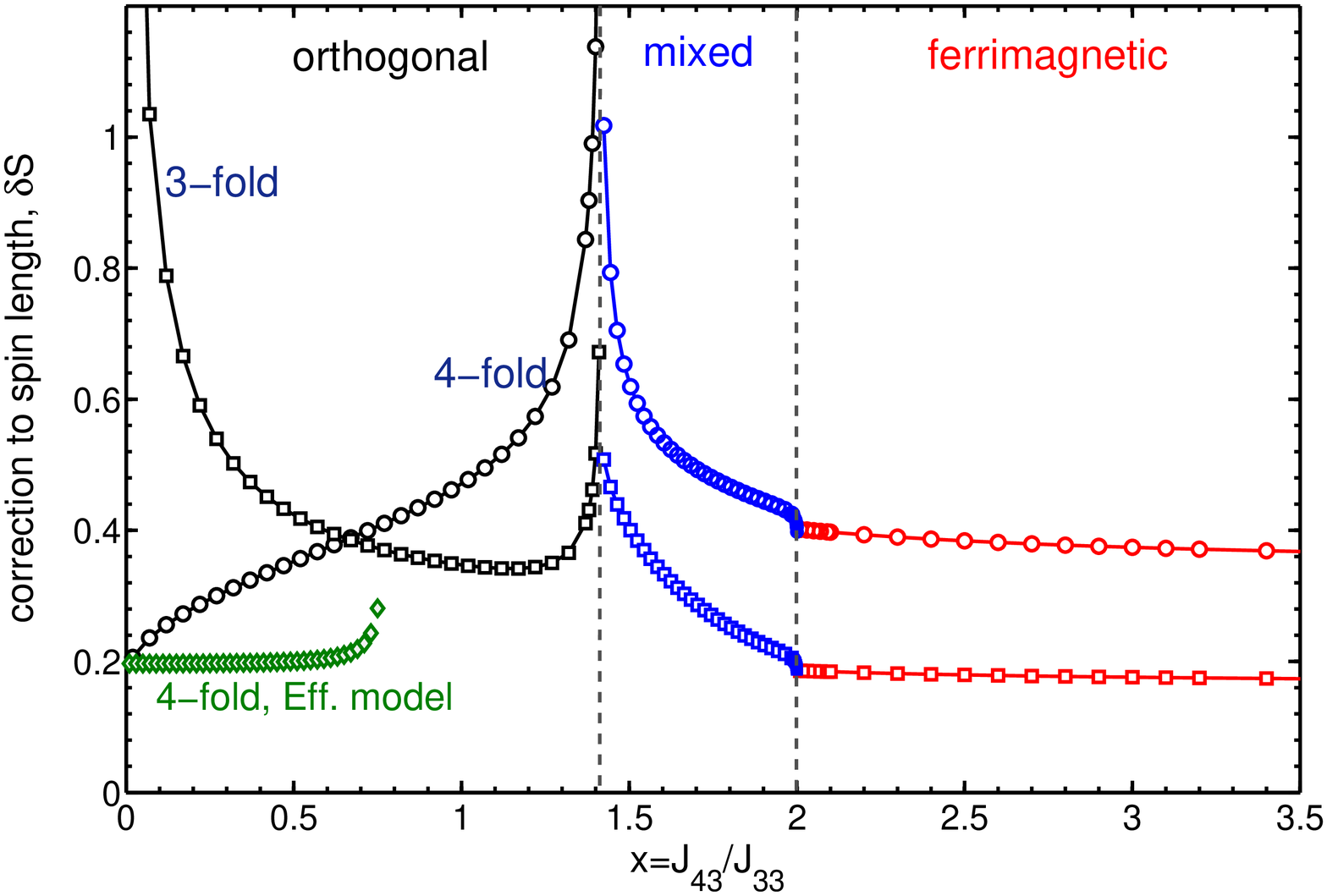}
\caption{Upper panel: Classical ($E_{cl}$) and semiclassical ($E_{cl}+\delta E$) GS energy per site from LSW theory   
around the orthogonal (black), the mixed (blue), and the ferrimagnetic state (red), as a function of $x=J_{43}/J_{33}$.  
For comparison, we also show (with symbols) the GS energy per site obtained from a fully quantum-mechanical numerical calculation (ED) on two clusters with 12 and 24 sites 
with periodic boundary conditions. Lower panel: Correction to the local spin length of the 4-fold (circles) and the 3-fold (squares) coordinated sites. 
The (green) diamonds show the results from LSW theory around the orthogonal phase in the effective model (see Sec.~\ref{Sec:EffModel}).}\label{fig:LSWT}
\end{figure}

\subsection{Results}
In the following we present the predictions of LSW theory for spins $S=1/2$. 
We first discuss the harmonic correction $\delta E$ to the GS energy. 
This is shown in the upper panel of Fig.~\ref{fig:LSWT}, where we also make a comparison to 
exact diagonalization results on clusters with 12 and 24 sites and S=1/2 (open symbols).  
We see that by including harmonic corrections to the classical energy one obtains a remarkably good agreement with the full quantum-mechanical results on finite-size clusters 
(the agreement becomes better with increasing number of sites especially at larger $x$).  
However as we are going to show below this agreement is quite deceptive for $S=1/2$.  
Although the semiclassical theory accounts for most of the energy, there is quite a lot of rich physics taking place at much smaller energies compared to the bare scale $J_{33}$, 
and this happens because the $x=0$ limit is singular, in the sense that it has a highly-degenerate GS manifold (see Sec.~\ref{Sec:EffModel} below).

We next examine the quadratic correction $\delta S$ to the local spin lengths of the two inequivalent sites of the lattice, 
which are shown in the lower panel of Fig.~\ref{fig:LSWT}. 
The first important feature to note is the upturn of $\delta S$ for the 3-fold sites as we approach the limit $x=0$. 
This is a manifestation of the singular nature of the $x=0$ limit mentioned above.  
In the semiclassical treatment we are expanding around the product up-down state on the $J_{33}$-dimers, 
and thus we cannot capture the actual tendency to form quantum-mechanical singlet wavefunctions at small $x$. 
Below we shall correct for this effect for the case of $S=1/2$ by integrating out the large $J_{33}$ energy scale of the $J_{33}$-singlets  
and by deriving an effective Hamiltonian model for the 4-fold sites only. 
In doing this it will become immediately apparent that the orthogonal phase is actually unstable at small $x$ and $S=1/2$ 
against a collinear magnetic phase which is stabilized by an underlying order-by-disorder effect.   

A second feature in our results for $\delta S$ is its upturn for both types of sites around the transition between the orthogonal and the mixed phase. 
This is clearly a sign of strong quantum fluctuations, and suggests that another possibly non-magnetic phase might be stabilized for $S=1/2$ in this regime. 
Our ED results for $S=1/2$ will indeed provide strong evidence for a non-magnetic state in this regime. 
In particular, as we show in the following section, the mixed phase becomes unstable altogether by including quantum fluctuations on the $J_{33}$ dimers only.  

Third, the harmonic corrections in the ferrimagnetic phase seem to be almost independent of $x$ and suggest that this phase 
probably survives quantum fluctuations. Indeed our ED results confirm this.   

Finally we should note that, except for the special region around $x=0$, the correction for the 4-fold sites is generally larger than that of the 3-fold sites. 
For $x>\sqrt{2}$ they differ by about a factor of two.  This is a manifestation of the very different role of quantum fluctuations on the two inequivalent sites of the lattice. 
It is also at odds with the simple intuition that higher coordination sites tend to behave more classically, 
but seems to be a consistent feature in a number of 2D lattices (see e.g. Ref.~[\onlinecite{Anu}]).

\section{Switching on quantum mechanics on the 3-fold dimers: Variational ansatz}\label{Sec:Variational}
In the purely classical description one treats each spin as a classical vector pointing in some fixed direction in spin space.
However, as we discussed above, in the limit of small $x$, the nearest neighbor 3-fold sites prefer to bind into singlets, which are 
locally entangled quantum-mechanical states that are very different from the up-down (or down-up) product state of the classical ansatz.  
In particular, the latter contributes an energy of $-S^2$, while the singlet wavefunction 
contributes $-S(S+1)$.\footnote{We should note here that by including the correction $\delta E_1$ (see above) 
from harmonic fluctuations around the classical up-down state we recover the correct energy $-S(S+1)J_{33}$ in the limit of $x\to 0$). 
Still however, the up-down state is very different from a singlet wavefunction.} 
In order to capture this strong-singlet physics at small $x$ we introduce a variational wavefunction $|\Psi_{\text{var}}\rangle$ 
which still treats the 4-fold sites as classical vectors but leaves complete quantum-mechanical freedom for the 3-fold sites.
Furthermore, to include both the orthogonal and the mixed phase as special cases of this ansatz 
we assume that the unit cell of $|\Psi_{\text{var}}\rangle$ is twice the unit cell of the Hamiltonian with ordering wavevector at $(\pi,\pi)$. 
In addition, we do not put any restriction on the directions of the 4-fold classical vectors.  
Thus the variational parameters in this ansatz are the 6 polar angles corresponding to three 4-fold classical vectors 
(by global SO(3) symmetry, the fourth 4-fold vector is forced to point in a fixed direction). 
In effect, this ansatz amounts to solving quantum-mechanically, for each set of the above parameters, 
the problem of an AFM S=1/2 dimer in the presence of two local fields of arbitrary directions.

\begin{figure}[!t]
\includegraphics[clip=true,width=0.7\linewidth]{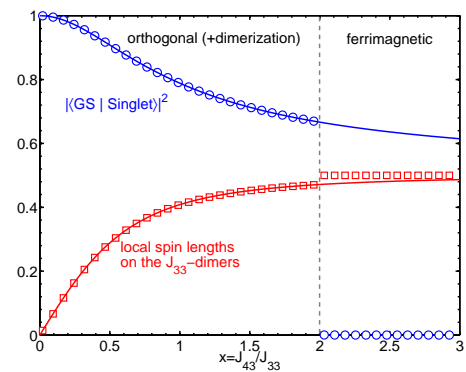}
\caption{Comparison between the numerical minimization of the variational ansatz described in the text  (symbols) 
and the analytical prediction (solid lines) from the problem of an AFM dimer in the presence of a staggered field 
$h_s=\sqrt{2} S x$ (see App.~\ref{App:Dimer}). 
Circles show the squared overlap between the optimal quantum-mechanical state $|\Phi\rangle$ of a dimer and the full singlet state, 
and squares show the local spin lengths on the $J_{33}$-bonds.}
\label{fig:variational}
\end{figure}

The first major result from the variational minimization is that the optimal GS corresponds to the orthogonal coplanar configuration up to $x=2$ and 
to the ferrimagnetic state (where the $J_{33}$ bonds form $|\!\!\uparrow\uparrow\rangle$ triplets) for $x>2$. 
Hence including quantum-mechanics on the 3-fold sites makes the mixed phase unstable for low enough spin $S$, which confirms our expectation that this 
compromise between the orthogonal and the ferrimagnetic phase is fragile. 

We have double-checked this important result by searching for the minimum energy of the same variational problem but in the more restricted parameter 
space whereby the 4-fold classical spins are tilted away from the $xy$-plane by an angle $\theta$ (i.e. as in the mixed phase configuration). This angle is the only variational parameter which makes the problem of finding the minimum much more tractable numerically. In fact, this problem is equivalent to that of an AFM dimer 
in the presence of a staggered field $\vec{h}_s = \sqrt{2} x S \cos\theta~\vec{e}_x$, plus a uniform field $\vec{h}_u = 2 x S \sin\theta~\vec{e}_z$, i.e. it is an extension of the staggered field-only case treated in App.~\ref{App:Dimer}. 
And indeed, the GS of this problem has $\theta=0$ (coplanar phase) for $0\le x < 2$, 
but $\theta=\pi/2$ (ferrimagnetic phase) for $x>2$. 

We now look at the main quantities of interest in the above optimal variational state for $S=1/2$. 
The first is the overlap of the optimal GS on the dimer with the exact singlet wave-function.  
The second is the polarization of the two 3-fold sites forming a dimer. 
Both quantities signify the amount of triplet admixture in the GS, and they are shown in Fig.~\ref{fig:variational}, where they are also compared to 
the corresponding analytical predictions given in App.~\ref{App:Dimer}. 
Our first comment is the overlap remains quite large up to $x=2$ which signifies that the strong $x=0$ coupling limit is a good perturbative limit 
for discussing the physics of the full $S=1/2$ quantum problem even at intermediate $x$. 
Another feature is that the polarization of the 3-fold sites becomes immediately finite as soon as we switch on a finite $x$. 
This happens because we are dealing with a quantum-mechanical dimer in the presence of a staggered field (i.e., the local exchange fields at the two sites 
of the dimer are opposite to each other in the coplanar phase) which admixes the triplet $|t_0\rangle$ into the singlet GS as soon as $x$ is finite (see details in App.~\ref{App:Dimer}). 
This is in contrast to the case of a AFM dimer in a uniform field which can polarize the dimer only above a critical value which is set by the singlet-triplet gap.


\begin{figure}[!t]
\includegraphics[clip=true,width=0.8\linewidth]{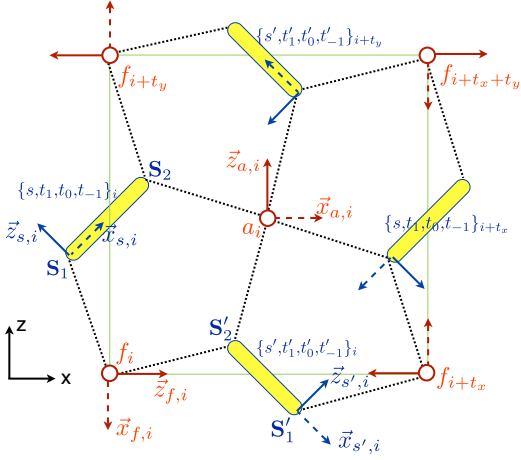}
\caption{Bosonic operators and local quantization axes (for the four-fold sites and the first site of each dimer) in the orthogonal state, where all spins lie on the xz-plane.}
\label{fig:SpinBondWaves}
\end{figure}

\section{Quadratic fluctuations around the variational ansatz}\label{Sec:LSBWT}
\subsection{Linear Spin+Bond-Wave Theory}
In the previous variational wavefunction the $J_{33}$-dimers were treated fully quantum-mechanically 
but the 4-fold sites were treated as classical vectors. So we refine our treatment to include the quadratic fluctuations around 
the variational state, by performing a semi-classical spin wave expansion for the 4-fold sites 
and a bond-wave expansion\cite{Bhatt,Judit} for the $J_{33}$-dimers. 

Figure \ref{fig:SpinBondWaves} shows the unit cell of the lattice and the orthogonal variational state around which we expand. 
For the 4-fold sites $a_i$ and $f_i$,  we perform a Holstein-Primakoff expansion 
$S_{a_i}^z = S - a_i^+ a_i$,  $S_{a_i}^+ \simeq \sqrt{2S}~ a_i$ (and similarly for $\vec{S}_{f_i}$) 
where we use the local quantization axes shown in Fig.~\ref{fig:SpinBondWaves}. 

Using the above representation for the 4-fold spins, we may regroup the various terms in the Hamiltonian as follows
\be
\mc{H}=\mc{H}^{(33)}+\mc{H}^{(43)}_{1}+\mc{H}^{(43)}_2 \equiv \mc{H}_0 +\mc{H}^{(43)}_2 
\ee
where $\mc{H}^{(33)}$ contains the $J_{33}$ coupling terms, while $\mc{H}^{(43)}_1$   
contains the parts that come from the constant $S$ from the z-component of the 4-fold spins, 
and $\mc{H}^{(43)}_2$ contains the remaining portion from the $J_{43}$ coupling terms. 
More explicitly, 
\bea
\mc{H}_0 &=& \sum_{i} 
J_{33}~\vec{S}_{1,i} \cdot \vec{S}_{2,i} -h_s\vec{z}_{s,i} \cdot \left( \vec{S}_{1,i} -\vec{S}_{2,i} \right) \nonumber \\
&+& J_{33}~\vec{S}'_{1,i} \cdot \vec{S}'_{2,i} -h_s\vec{z}_{s',i} \cdot \left( \vec{S}'_{1,i} -\vec{S}'_{2,i} \right) 
\eea
where $h_s=\sqrt{2} x S $. Therefore $\mc{H}_0$ describes two independent dimers in the presence of staggered fields, a problem that is solved in  App.~\ref{App:Dimer}. 

\begin{figure*}[!t]
\includegraphics[clip=true,width=0.99\linewidth]{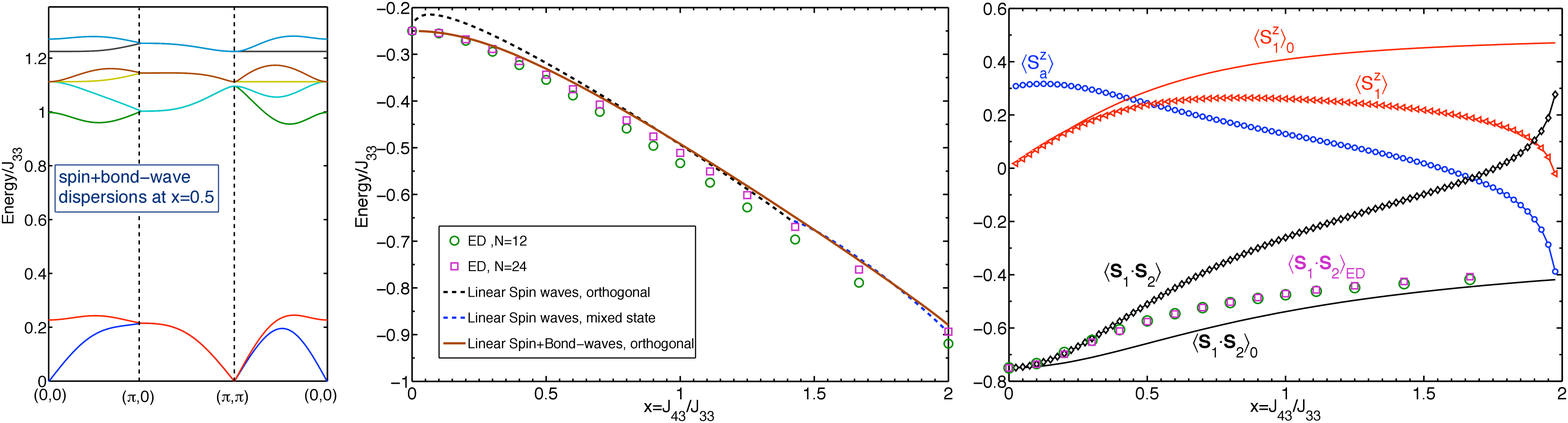}
\caption{Main results from the spin+bond-wave calculation which captures the quadratic fluctuations around the variational GS. 
Left panel: The 8 branches of hybridized spin+bond wave excitations at $x=0.5$ (see text). 
Middle panel:  GS energy per site as a function of $x$. Comparison between linear spin wave theories around the orthogonal and the mixed state (dashed lines), 
the spin+bond-wave theory, and ED data from the 12 and 24 site clusters. 
Right panel: Various GS expectation values as a function of $x$.
$\langle S_a^z \rangle$ stands for the local spin length of the 4-fold sites. 
$\langle S_1^z \rangle_0$ and $\langle S_1^z \rangle$ stand for the staggered polarization on the $J_{33}$-dimers in the variational GS and when we include quadratic fluctuations, respectively, and similarly for $\langle\vec{S}_1\!\cdot\!\vec{S}_2\rangle_0$ and $\langle\vec{S}_1\!\cdot\!\vec{S}_2\rangle$. 
The symbols are ED results from the 12 (triangles) and 24 (circles) site clusters.}
\label{Fig:Mixed}
\end{figure*}

To account for the fluctuations driven by the remaining portion $\mc{H}^{(43)}_2$ of the Hamiltonian we shall make a bond-wave expansion.  
We first introduce eight bosonic operators which create the singlet $|s\rangle=\frac{1}{\sqrt{2}}\left( |\!\uparrow\downarrow\rangle-|\!\downarrow\uparrow\rangle \right)$ 
and the three triplet states $|t_1\rangle = |\!\uparrow\uparrow\rangle$, $|t_{-1}\rangle = |\!\downarrow\downarrow\rangle$ and 
$|t_0\rangle = \frac{1}{\sqrt{2}}\left( |\!\uparrow\downarrow\rangle+|\!\downarrow\uparrow\rangle \right)$,  
for each of the two $J_{33}$-dimers per unit cell. 
These bosons will be denoted by $\{s,t_1,t_0,t_{-1}\}_i$ and $\{s',t_1',t_0',t_{-1}'\}_i$ (see Fig. \ref{fig:SpinBondWaves}).  
The spin operators of the two sites of each dimer have the following bosonic representation\cite{Bhatt,Judit}
\bea
S_{1,2}^z &=& \pm \frac{1}{2} \left( t_0^+ s + s^+ t_0 \right)+\frac{1}{2} \left( t_1^+ t_1 - t_{-1}^+ t_{-1} \right) \\
S_{1,2}^+ &=& \mp \frac{1}{\sqrt{2}} \left( t_1^+ s -s^+ t_{-1} \right) + \frac{1}{\sqrt{2}} \left( t_1^+ t_0 + t_0^+ t_{-1} \right)
\eea
We also introduce the bosons which create the two lowest eigenstates of $\mc{H}_0$ (see Eqs.~(\ref{Eq:e1}) and (\ref{Eq:e2})) 
for the two separate dimers in the unit cell, namely 
\bea
\psi_1^+ &=& u~s^+ + v~t_0^+, ~~~~~ \psi_2^+ = v~s^+ -u~t_0^+~,\\
\psi_1'^+ &=& u ~s'^+ + v ~t'^+_0, ~~~ \psi_2'^+ = v~s'^+ -u~t'^+_0 ~,
\eea
where the constants $u, v$ are defined in App.~\ref{App:Dimer}. 
To expand around the variational GS we take the constraint $\psi_1^+ \psi_1 + \psi_2^+ \psi_2 + t_1^+ t_1 + t_{-1}^+ t_{-1} =1$, and replace the right hand side 
with a large number $M$. We then assume that the bosons $\psi_1$ and $\psi_1'$ are condensed and perform a large-$M$ expansion\cite{Judit} 
\bea
\psi_1 \simeq \sqrt{M} - \frac{1}{2\sqrt{M}} (\psi_2^+ \psi_2 + t_1^+ t_1 + t_{-1}^+ t_{-1})~.
\eea
Replacing in the Hamiltonian and keeping only quadratic terms in the expansion we arrive at
\be\label{eq:Hmixed}
\mc{H}\simeq  E_0 + \frac{1}{2} \sum_k \vec{A}_k^+ \cdot
\left(\begin{array}{cc}
\vec{C}_k & \vec{D}_k \\
\vec{D}_k^+ & \vec{C}_{-k}^T
\end{array}\right)
\cdot \vec{A}_k
\ee
where $E_0/N_{uc} =(2M+3) \epsilon_1  - \epsilon_2  -1/2 -4x v u M/\sqrt{2}$, $N_{uc}=6 N$ is the number of unit cells, 
$\epsilon_{1,2}$ are the single-dimer energies given in Eqs.~(\ref{Eq:e1}) and (\ref{Eq:e2}),  
\begin{widetext}
\begin{equation}
\begin{array}{cc}
\vec{A}_k^+ = \left(
\begin{array}{cccccccccccccccc}
a_k^+, & f_k^+, & \psi_{2,k}^+, & t_{1,k}^+, & t_{-1,k}^+, & \psi'^+_{2,k}, & t'^+_{1,k}, & t'^+_{-1,k}, & a_{-k}, & f_{-k}, & \psi_{2,-k}, & t_{1,-k} ,& t_{-1,-k},& \psi'_{2,-k}, & t'_{1,-k} ,& t'_{-1,-k} 
\end{array}
\right)~,
\end{array} 
\end{equation}
\end{widetext}
and the 8$\times$8 matrices $\vec{C}_k$ and $\vec{D}_k$ are given explicitly in App.~\ref{App:SpinBondWaves}.  
To diagonalize this quadratic Hamiltonian we search for a Bogoliubov transformation $\vec{A}_k = \vec{V}_k \cdot \tilde{\vec{A}}_k$ as described in App.~\ref{App:LSWT},
in terms of new bosons $\tilde{a}_{k}, \tilde{f}_{k}, \tilde{\psi}_2, \ldots, \tilde{t}'_{-1,k}$ for which 
\bea
\mc{H} &\simeq& E_0+ \frac{1}{2} \sum_k \Big( \omega_{1k} + \ldots + \omega_{8k} \Big) \nonumber \\
&+&  \sum_k  \Big( \omega_{1k} ~\tilde{a}_k^+ \tilde{a}_k + \ldots + \omega_{8k} ~\tilde{t}'^+_{-1,k} \tilde{t}'_{-1,k} \Big) ~.
\eea

\subsection{Results}
In the left panel of Fig.~\ref{Fig:Mixed} we show the eight spin+bond-wave branches of excitations along some symmetry directions in the BZ
and for $x=0.5$. The spectrum consists of 2 low-lying modes and 6 high-energy modes around $E=J_{33}$. 
These modes arise respectively from the two spin-wave modes of the 4-fold sites and the 6 $J_{33}$-triplet modes per each unit cell. 
At $x=0$, both the 4-fold sites and the $J_{33}$-dimers are isolated and thus all modes are completely localized. In particular, 
the two spin-wave modes have zero energy and the 6 triplet excitations cost energy $J_{33}$. 
The hybridization caused by a finite $x$ then gives rise to the dispersion structure shown in the left panel of Fig.~\ref{Fig:Mixed}.

Of particular interest are the Goldstone modes which appear in the spectrum due to the fact that the variational state breaks the continuous SO(3) symmetry. 
From the nature of the variational state one expects two Goldstone modes, one at $\vec{k}=0$ and another at the ordering wave-vector 
$\vec{Q}=(\pi,\pi)$. However, the left panel of Fig.~\ref{Fig:Mixed} shows that we actually have 3 gapless modes, one at zero momentum and two at 
$\vec{Q}$. The extra gapless mode at $\vec{Q}$ appears also in the linear spin-wave dispersions (not shown here) around the orthogonal state 
and is spurious in both cases. Such a spurious gapless mode appears also in the $J_1$-$J_2$ model around the collinear phase in the large $J_2$ regime 
where it may be lifted by including nonlinear terms.\cite{Uhrig} 
In the present case, these non-linear terms actually select a different state than the orthogonal one.

We now turn to the quadratic GS energy given by
\be
E =  E_0+\frac{1}{2} \sum_k \left( \omega_{1k} +\ldots + \omega_{8k} \right)~.
\ee 
Our results as a function of $x$ are shown in the middle panel of Fig.~\ref{Fig:Mixed}, where we also make a comparison with the semi-classical spin-wave energies 
from Sec.~\ref{Sec:LSWT}. Comparing with ED data on finite clusters with 12 and 24 sites, we see that the present Spin+Bond-Wave theory gives a better 
agreement than the Spin-Wave theory especially at small $x$.  However at larger values of $x$, the mixed spin+bond-wave expansion does not deliver a better energy than
the pure semiclassical expansion. 
 
Let us now look at the spin length of the 4-fold sites which is given by 
\be
\langle \vec{S}_a^z \rangle = S-\frac{1}{N_{uc}} \sum_k \sum_{n=9}^{16} | V_k(1,n) |^2 ~,
\ee 
and can also be calculated numerically by integrating over the BZ. The result is shown in the right panel of Fig.~\ref{Fig:Mixed} as a function of $x$. 
We see that quadratic fluctuations destroy completely the long-range order at the 4-fold sites around $x\sim 1.5$. 
This means that there exist quite severe quantum fluctuations in this regime which might destabilize the anstaz phase.

A similar conclusion arises by looking at the behavior of the exchange energy on the $J_{33}$-dimers as well as their staggered polarization. 
These quantities can be calculated using the expressions 
\bea
\langle \vec{S}_1\!\cdot\!\vec{S}_2\rangle \!&=&\! \frac{1}{4}-u^2 + (2u^2-1) \langle \psi_2^+\psi_2 \rangle + 2 u^2 \langle t_1^+t_1\rangle \\
\langle S_{1,2}^z\rangle \!&=&\! \pm u v \left( 1-2 \langle\psi_2^+\psi_2\rangle- 2\langle t_1^+t_1\rangle \right) 
\eea
where we have used $\langle t_1^+t_1\rangle=\langle t_{-1}^+t_{-1}\rangle$  (due to time reversal symmetry), and  
\bea
\langle\psi_2^+\psi_2\rangle &=& \frac{1}{N_{uc}} \sum_k \sum_{n=9}^{16} |V_k(3,n)|^2 \\
\langle t_1^+ t_1 \rangle &=& \frac{1}{N_{uc}} \sum_k \sum_{n=9}^{16} |V_k(4,n)|^2 ~.
\eea
The results are shown in the right panel of Fig.~\ref{Fig:Mixed} where they are compared to the corresponding behavior in the variational GS, i.e., without quantum fluctuations. 
We see that the exchange energy becomes quickly reduced in magnitude with $x$, and it even crosses over to positive values above $x\sim 1.8$. 
This behavior is drastically different from the corresponding result $\langle\vec{S}_1\!\cdot\!\vec{S}_2\rangle_0$ in the variational GS and from the ED data. 
Thus the effect of quadratic fluctuations is quite strong.

\section{Full quantum $S=1/2$ limit: Effective low-energy theory for small $x=J_{43}/J_{33}$}\label{Sec:EffModel}
Up to now, the spins at the 4-fold sites were treated classically or semi-classically.  
However it turns out that there are quite strong quantum-mechanical effects on these sites which modify to a large extent the picture 
we have so far and brings rich physics which takes place at a smaller energy scale. 

At $J_{43}=0$ the system consists of isolated $J_{33}$-dimers which form singlets, and free spins (4-fold sites) which are free to point up or down. 
This highly degenerate GS manifold is lower in energy from excited states by at least $J_{33}$ which is the cost of promoting one singlet into a triplet. 
By switching on a small $x$ the 4-fold spins begin to interact with each other through the virtual fluctuations of the 3-fold dimers out of the singlet manifold. 
By integrating out these fluctuations one may derive an effective model description for the low energy sector using degenerate perturbation theory. 
Specific details of this expansion are provided in App.~\ref{App:PertTheory}. 
Here we summarize the main results. 

It turns out that an expansion up to fourth order in $x$ provides the essential low-E physics at small $x$. 
The major interactions between the 4-fold sites are depicted in Fig.~\ref{fig:LowestProcesses}. The first important insight comes already in second order perturbation theory. 
Namely that the effective n.n. coupling $J_1$ vanishes due to quantum interference since the two possible paths shown in Fig.~\ref{fig:LowestProcesses} have opposite amplitude.
The overall minus sign originates in the fact that the singlet wavefunction is antisymmetric with respect to the interchange of the two sites of the $J_{33}$-dimer. 
By contrast there is only one possible path connecting n.n.n sites and this gives a finite $J_2$ coupling in 2nd order. 
A nonzero $J_1$ appears first in third order together with a renormalization for $J_2$.  
In fourth order, in addition to a renormalization of $J_1$ and $J_2$, one also obtains a four-spin exchange term that involves the spins of each square plaquette. 
For a plaquette with a horizontal dimer (see the 7th-cluster in Table~\ref{T1}), this term reads 
\be\label{Kp1.Eq}
\hat{\mc{K}}\!=\!(\vec{S}_1\!\cdot\!\vec{S}_2)(\vec{S}_3\!\cdot\!\vec{S}_4) \!-\! \frac{1}{2}(\vec{S}_1\!\cdot\!\vec{S}_4)(\vec{S}_2\!\cdot\!\vec{S}_3)
\!+\! (\vec{S}_1\!\cdot\!\vec{S}_3)(\vec{S}_2\!\cdot\!\vec{S}_4) 
\ee
or, in pictorial form,
\begin{eqnarray}\label{Kp2.Eq}
\hat{\mc{K}}&=& \parbox{1.8in}{\epsfig{file=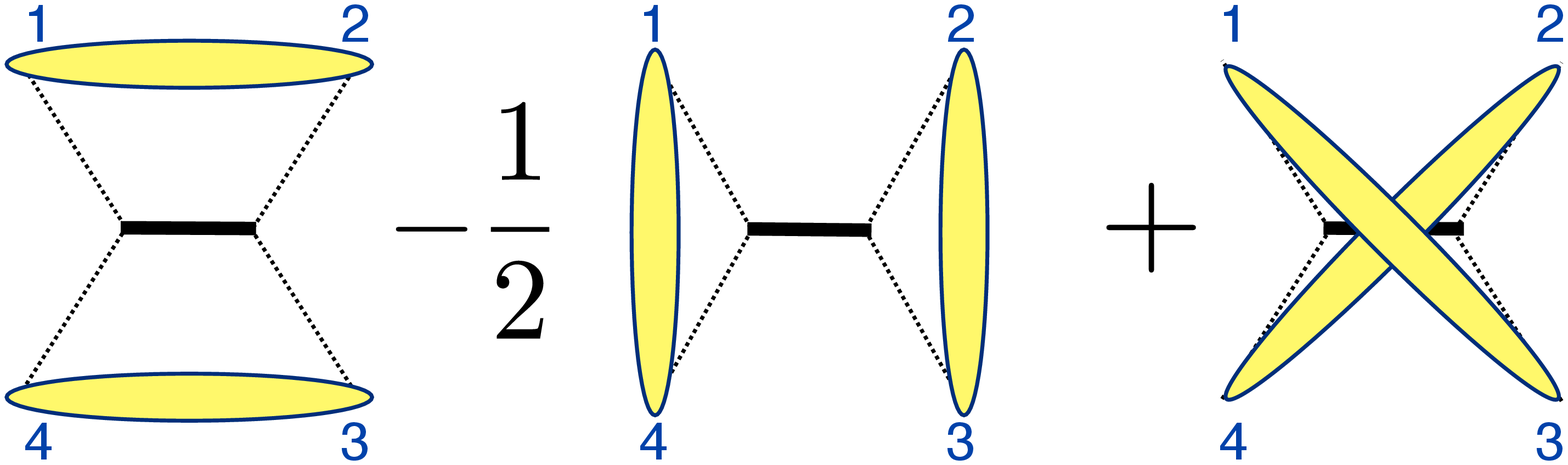,width=1.8in,clip=}} \nonumber \\
&\equiv& \hat{\mc{K}}_{h}-\frac{1}{2} \hat{\mc{K}}_{v}+\hat{\mc{K}}_{x}~.
\eea
This term is different from the usual ring exchange process\cite{Roger} on the square lattice. 
In particular, the coefficients of $\hat{\mc{K}}_{h}$ and $\hat{\mc{K}}_{v}$ are different here 
(the actual relative factor of $-1/2$ is not generic but is expected to be modified in higher orders of perturbation theory), 
and this reflects the lack of the $C_4$ symmetry around the center of each square plaquette in the underlying parent Hamiltonian on the Cairo lattice. 
In essence the full symmetry of the underlying Cairo lattice is only manifest in fourth order of perturbation theory, 
which is a good reason as to why one should push the perturbation theory at least up to 4th order. 
Finally the coefficient of the last term $\hat{\mc{K}}_x$ is the same as that of $\hat{\mc{K}}_{h}$ and this reflects the underlying symmetry of exchanging 
$\vec{S}_2$ and $\vec{S}_3$ (or $\vec{S}_1$ and $\vec{S}_4$) in all 4-th order processes. However this is not a generic feature although it might still hold up to some order higher than fourth.  

\begin{figure}[!t]
\includegraphics[clip=true,width=0.99\linewidth]{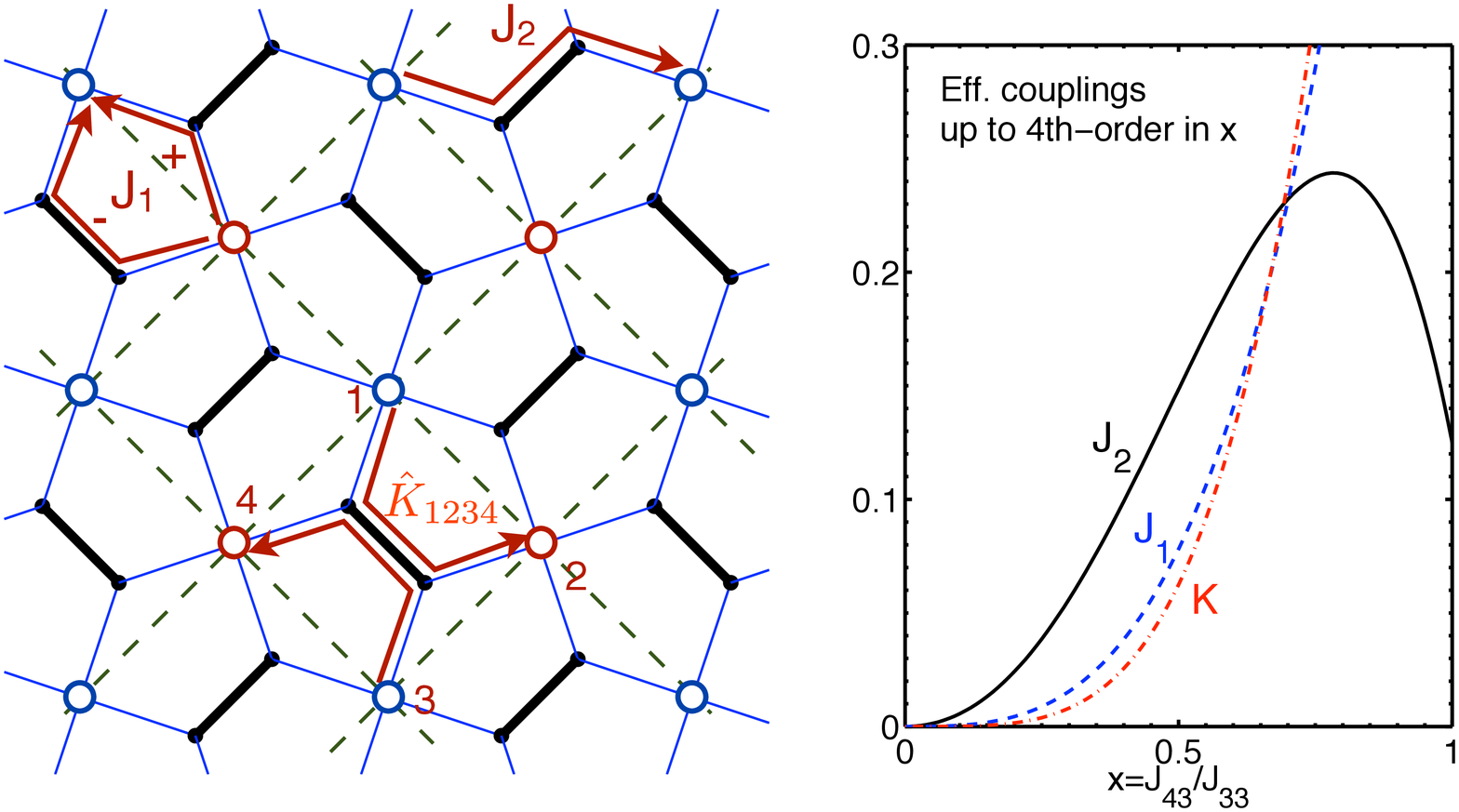}
\caption{Left: Lowest order processes in the effective model.  
The 4-fold sites form a tilted square lattice (dashed lines). In second order in $x$,  $J_1$ vanishes because of quantum interference between two different paths (shown by arrows) with opposite amplitude. By contrast, $J_2$ is finite since there is a single path. In fourth order in $x$, a 4-spin term appears that invokes the 4 spins around a square plaquette.
Right: Dependence of $J_1$, $J_2$ and $K$ on $x$, as given by Eqs.~(\ref{eqn:j1})-(\ref{eqn:k}).}
\label{fig:LowestProcesses}
\end{figure}

\begin{figure*}[!t]
\includegraphics[clip=true,width=0.99\linewidth]{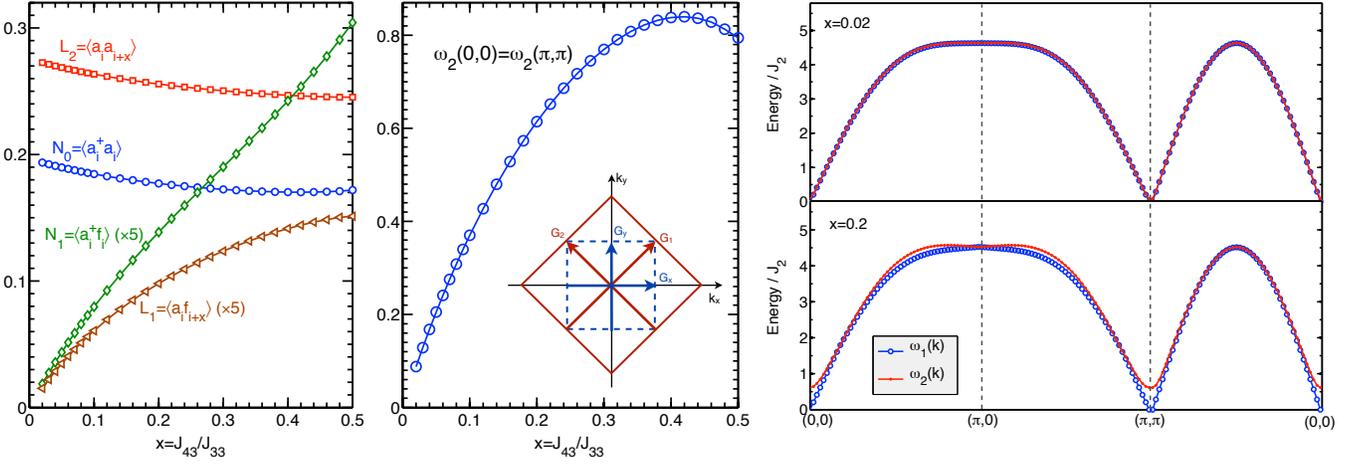}
\caption{Left panel: Self-consistent solution for the mean-field pairing fields that appear in the interacting spin wave theory around the collinear phase. 
Middle panel: The energy (in units of $J_2 (x)$) of the second excitation mode at $\vec{k}=0$ and at $\vec{k}=(\pi,\pi)$. 
Inset: The BZ of the effective model is enclosed by the dashed (blue) lines and is set by the reciprocal vectors $\vec{G}_x$ and $\vec{G}_y$. 
At small enough $x$ the effective model reduces to the $J_1$-$J_2$ model on the square lattice whose unit cell contains one site only. 
Its BZ is enclosed by the solid (red) lines and is set by the vectors $\vec{G}_1$ and $\vec{G}_2$. 
Right panel: Spin-wave dispersions from NLSWT around the collinear phase for two representative values of $x$.}
\label{fig:NLSWT}
\end{figure*}

Altogether, up to fourth order in $x$ we have a low-energy effective $J_1$-$J_2$-$K$ model on the square lattice with parameters  
\bea
J_1&=& \frac{1}{2}x^3+\frac{1}{4}x^4  \label{eqn:j1}\\
J_2&=& \frac{1}{2}x^2+\frac{3}{4}x^3-\frac{9}{8}x^4  \label{eqn:j2}\\
K&=& x^4 \label{eqn:k}
\eea 
up to fourth order in $x$. As can be seen in the right panel of Fig. \ref{fig:LowestProcesses}, these functions show a rapid increase above $x\sim 0.4$, 
so we expect that the fourth order series expansion does not converge beyond $x\sim 0.4-0.6$.

From the above effective theory we get the following insights into the low-energy physics of the problem at small $x$. 
The first insight comes from the fact that we are dealing with a dominant $J_2$ coupling in the square lattice. 
This gives two square sublattices which are decoupled from each other and order antiferromagnetically. 
Now the angle $\theta$ between the two N\'eel vectors is not fixed at the classical level even when one includes a finite $J_1$, 
since the total exchange field exerted at a given site of one sublattice from the four neighboring spins of the other sublattice adds up to zero. 
By contrast, the classical energy per site of the 4-spin exchange term goes as $+K S^4 \left( \cos^2\theta  + 2 \right)/6$, and thus the plaquette 
term favors the orthogonal state with $\theta=\pi/2$. 
However up to this point we have neglected quantum fluctuations. 
At sufficiently small $x$, $J_1$ will dominate over $K$, and will drive quantum fluctuations 
which at the harmonic level are known\cite{Chandra} to favor one of the two collinear phases with $\theta=0$ or $\pi$ (see Fig.~\ref{Fig:methods}(b)).   
Hence there must be a critical value of $x$ below which the orthogonal state becomes unstable towards the $Z_2$ collinear phase.

It is should be noted here that out of the three members of Eq.~(\ref{Kp2.Eq}), it is the combination $\hat{\mc{K}}_{h}-\frac{1}{2}\hat{\mc{K}}_{v}$ 
that is responsible for the selection of the orthogonal phase at the classical level, since the energy of the third term $\hat{\mc{K}}_x$ 
does not depend on the angle $\theta$.

\section{Non-Linear Spin-Wave (NLSW) theory in the Effective model}\label{Sec:NLSWT}
A LSW expansion around the orthogonal phase in the effective model does not capture the instability to the collinear phase at small $x$. 
This can be seen in the lower panel of Fig. \ref{fig:LSWT} , where we have also included (by green diamonds) the the results from such a calculation 
(specific details are again relegated to App.~\ref{App:LSWT})) for the correction $\delta S$ of the 4-fold sites. We see that $\delta S$ 
does not show any anomaly down to $x=0$ (where it approaches the same limiting value with that obtained by LSW
theory in the full Cairo lattice model).   

A LSW expansion around the collinear phase does not work either, since the collinear phase is not the classical minimum and the resulting Hamiltonian matrix is not positive definite.  
To stabilize the collinear state, one is then lead to include anharmonic corrections to the theory. 
This situation is analogous to the case of the triangular AFM in a field, where the up-up-down state is not the classical minimum but is stabilized by the leading 1/S corrections to the 
linear theory.\cite{Alicea} 

Here we have performed a similar Non-Linear Spin-Wave (NLSW) expansion around the collinear phase. 
The quartic terms are treated by a standard mean-field decoupling based on the following (real) pairing fields which must be finite in the collinear phase
\bea
N_0 &=& \langle a_i^+ a_i \rangle = \langle f_i^+ f_i \rangle \nonumber\\
N_1 &=& \langle a_i^+ f_i \rangle = \langle a_i^+ f_{i+x+y} \rangle \nonumber\\
L_1 &=& \langle a_i f_{i\pm x} \rangle = \langle a_i f_{i\pm y} \rangle \\
L_2 &=&  \langle a_i a_{i\pm x} \rangle = \langle a_i a_{i\pm y} \rangle 
= \langle f_i f_{i\pm x} \rangle = \langle f_i f_{i\pm y} \rangle ~.\nonumber 
\eea 
Here $N_0$ is the onsite correction to the spin length, $N_1$ is the short-range correlationalong the FM lines of the collinear phase, 
$L_1$ is the short-range correlation along the AFM lines of the collinear phase, and finally $L_2$ is the AFM correlation between next-nearest neighbors.
The mean-field decoupling leads to a quadratic theory of the same form with Eq.~(\ref{eqn:Hlswt}) and
with the matrices $\vec{C}_k$ and $\vec{D}_k$ given explicitly in App.~\ref{App:NLSWT}. 

Given the presence of the two inter-penetrating N\'eel sub-lattices, which is dictated by the dominant $J_2$ coupling, we expect that 
$L_2$ is finite even at $x=0$, while both $N_1$ and $L_1$ should vanish at $x=0$ and become finite as soon as the order-by-disorder effect takes place. 
Our results from the numerical self-consistent solution for the above pairing fields are shown in the left panel of Fig.~\ref{fig:NLSWT}. 
As expected, both $N_0$ and $L_2$ are finite (and strong) at $x=0$ with little dependence on $x$, while $N_1$ and $L_1$ approach zero at small $x$.  

The right panel of Fig.~\ref{fig:NLSWT} shows the development of a finite gap for the second modes at $\vec{k}=0$ and $\vec{k}=(\pi,\pi)$. 
As we show in the inset, these momenta map to the $\vec{k}=(0,\pi)$ and $\vec{k}=(\pi,\pi)$ mode of the BZ of the $J_1$-$J_2$ model which we get 
if we neglect the K-term at very small $x$. So the restoration of the gap at these k-points is consistent with our expectation of having two Goldstone modes in the 
collinear phase which, in the BZ of the $J_1$-$J_2$ model, sit at $\vec{k}=(0,0)$ and at $\vec{k}=(\pi,0)$ (which is the ordering vector of the state about which we performed our semiclassical expansion). 
The lower panel of Fig.~\ref{fig:NLSWT} shows the dispersion of the two spin-wave branches along certain symmetry directions of the BZ.


\begin{figure}[!t]
\includegraphics[clip=true,width=0.59\linewidth]{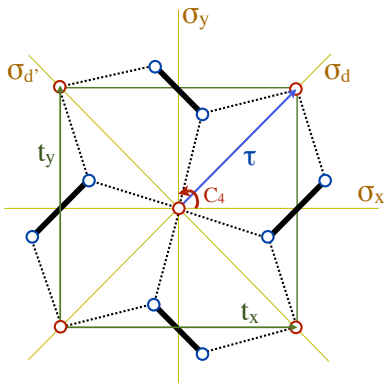}
\caption{The unit cell of the Cairo pentagonal lattice along with a clarification of the space group symmetries. 
The vectors $\vec{t}_x$ and $\vec{t}_y$ denote the primitive translations of the Bravais square lattice. 
In addition to the $C_4$ rotations around the 4-fold coordinated sites we also have the four non-symorphic operations 
$(\sigma_i | \boldsymbol{\tau})$ which stand for reflections $\sigma_i$ ($i=x$, $y$, $d$, and $d'$) followed by the non-primitive translation $\boldsymbol{\tau}$. 
In the absence of the 3-fold sites, this point group reduces to $C_{4v}$ which is the point group of the square lattice.}
\label{fig:sym}
\end{figure}

\begin{table}[!t]
\caption{Character table of the little group of $\vec{k}=0$ which is isomorphic to $C_{4v}$. 
The decomposition of the 5 IR's (first column) into modes with well 
defined angular momenta $l$ are given inside the parentheses.}\label{tab:C4v}
\begin{ruledtabular}
\begin{tabular}{l|ccccc}
 & $E$ & $C_2$ & $C_4$                     & $(\sigma_x|\boldsymbol{\tau})$  & $(\sigma_d|\boldsymbol{\tau})$ \\
&         &            & $C_4^{-1}$               & $(\sigma_y|\boldsymbol{\tau})$  & $(\sigma_{d'}|\boldsymbol{\tau})$ \\
\midrule[0.8pt]
$A_1~(l=0)$  & 1 & 1 & 1 & 1 & 1 \\
$A_2~(l=0)$ & 1 & 1 & 1 & -1 & -1 \\
$B_1~(l=\pi)$ & 1 & 1 & -1 & 1 & -1 \\
$B_2~(l=\pi)$ & 1 & 1 & -1 & -1 & 1 \\
$E~(l=\pm\frac{\pi}{2})$ & 2 & -2 & 0 & 0 & 0 \\
\end{tabular}
\end{ruledtabular}
\end{table}

\section{Exact diagonalizations in the Cairo lattice for $S=1/2$}\label{Sec:ED}
In the remaining part of the article we discuss our exact diagonalization results from finite-size clusters with periodic boundary conditions and with spin $S=1/2$.   
We have investigated both the Cairo lattice model, as well as different variations of the effective model in the square lattice. 

The main results from the Cairo clusters can be summarized as follows. 
At $x\sim 2$, we find a GS levelcrossing to a state with total spin $N/6$, which is the onset of the 1/3-ferrimagnetic phase (Sec.~\ref{SubSec:OneThird}).  
In the opposite regime of small $x$ (Sec.~\ref{SubSec:Smallx}), we find a singlet GS and a low-energy structure which proves the 
presence of two nearly decoupled sublattices, in agreement with the effective model.  
This is explicitly demonstrated by comparing the spectra of the 24-site Cairo cluster to that of the effective 8-site square cluster (Sec.~\ref{SubSubSec:8site}).  
We also examine a number of GS properties at small $x$, such as spin-spin, dimer-dimer, and vector-chiral correlations (Sec.~\ref{SubSubSec:Correlations}), as well as 
the low-energy symmetry properties of the spectrum (Sec.~\ref{SubSubSec:TOS}).
These results show a strong competition between the collinear and the orthogonal state at low $x$, 
and highlight the fact that the locking between the two sublattice N\'eel vectors (favored by the large $J_2/J_1$ in the effective model) takes place at large length scales.   

At intermediate $x$ ($x\sim 1.2$ for the 24-site cluster) we find a GS level crossing to a new singlet state, accompanied by a whole 
reorganization of the low-lying excitations and their symmetries.   
This intermediate phase is discussed separately in Sec.~\ref{Sec:IntPhase}.

\begin{figure*}[!t]
\includegraphics[clip=true,width=0.684\linewidth]{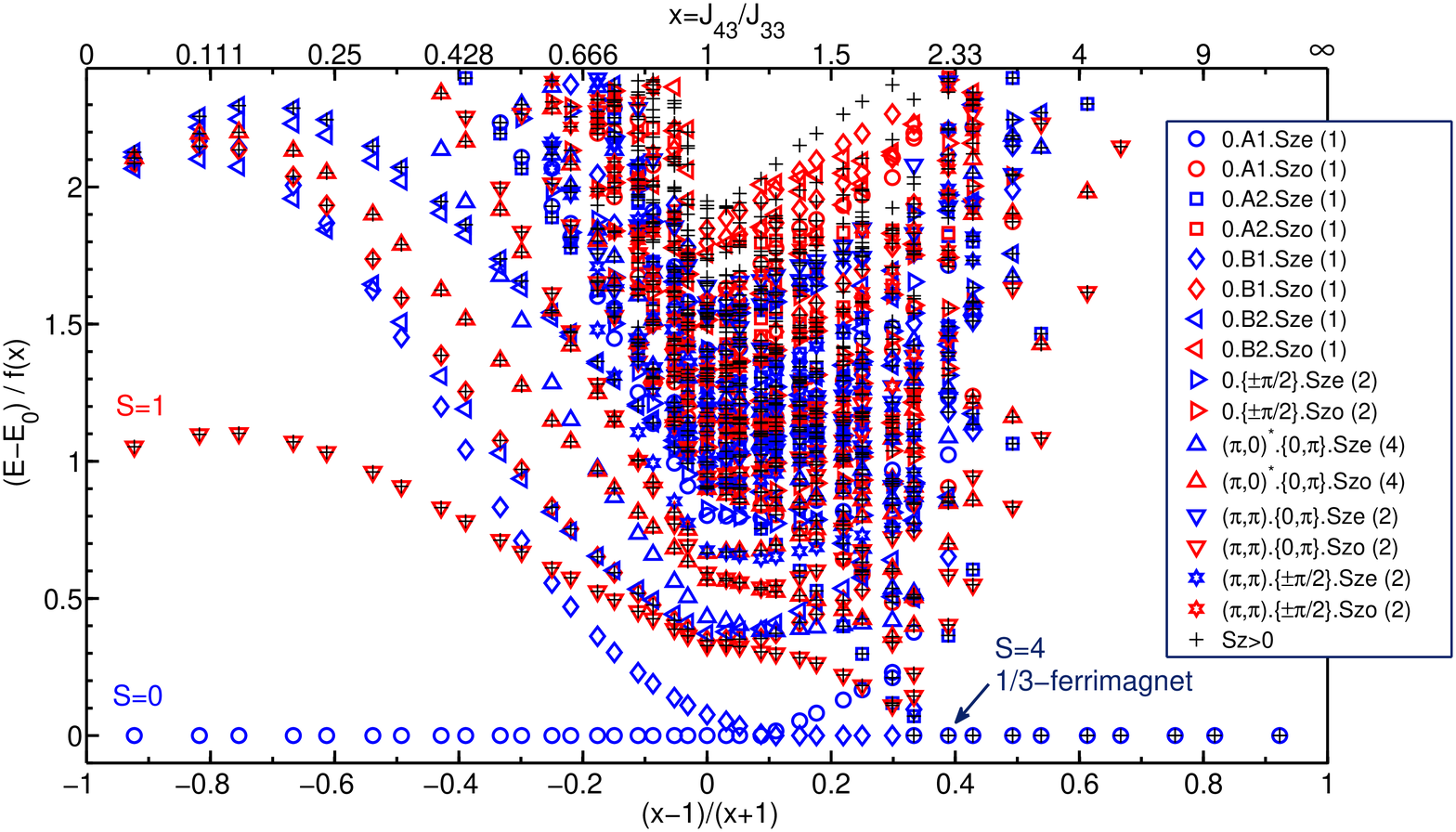}~\includegraphics[clip=true,scale=0.316]{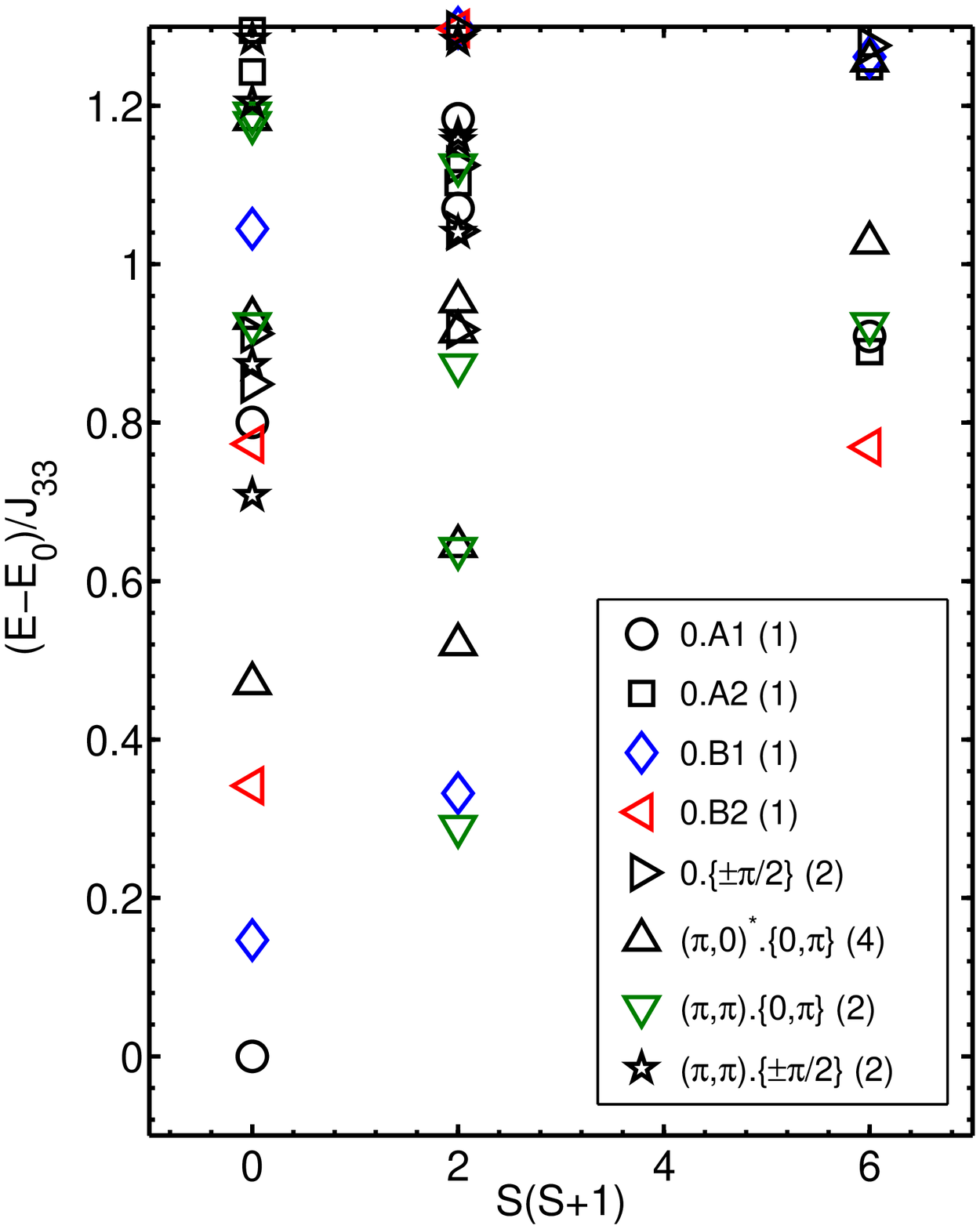}~~~~
\caption{Left panel: Low-energy spectrum in the 24-site Cairo pentagonal cluster as a function of $(x-1)/(x+1)$ (bottom x-axis), where $x=J_{43}/J_{33}$ (top x-axis).  
The energy is measured from the respective GS and is scaled with the function $f(x) = J_{33} x^2$ for $x<1$ and $f(x)=J_{43}$ for $x>1$. 
Open symbols stand for $S_z=0$ states while crosses denote $S_z>0$ states. The legends specify the different symmetry sectors for this cluster.  
The last portion of the symmetry label accounts for the parity under spin inversion, with ``Sze'' and ``Szo'' specifying respectively even and odd sectors. 
The arrow denotes the 1/3-ferrimagnetic state (here with total spin $S=4$) which becomes the GS for $x\gtrsim1.96$. 
Right panel: Low-energy spectrum of the 24-site Cairo pentagonal cluster versus the total spin $S$ for $x=0.8$. Different symbols correspond to different 
symmetry sectors as shown by the legends. The parentheses next to the legends denote the degeneracy of each sector (apart the Zeeman degeneracy). }
\label{fig:Pentag24_LESpectrumvsx}
\end{figure*}

\subsection{Finite-size clusters and symmetries of the Cairo lattice}\label{SubSec:SpaceGroup}
As we mentioned above, the Cairo pentagonal lattice has a unit cell of six sites. 
It turns out that the largest finite-size cluster (with periodic boundary conditions) 
that has the full point group symmetry of the infinite lattice and which is also accessible by our computational capabilities 
has 24-sites (the next symmetric cluster has 48 sites which is too large). 
As we are going to see, most of the valuable information presented below comes from this 24-site cluster, exactly because it has all point group symmetries. 
We have also investigated clusters with 12, 18 and 30 sites which are however either too small or lack some of the point group symmetries of the infinite lattice. 

Before we present and analyze our ED results it is also useful to understand the space group structure of the Cairo lattice. 
Its Bravais lattice is the square lattice with primitive translations $\vec{t}_x$ and $\vec{t}_y$ 
along the $x$- and $y$-axis respectively (see Fig.~\ref{fig:sym}). Apart from primitive translations, the lattice is also invariant under 
the $C_4$ rotations around the 4-fold sites as well as under four non-symorphic ``glide'' operations $(\sigma_i | \boldsymbol{\tau})$ ($i=1$-$4$), 
which are reflections followed by the non-primitive translation $\boldsymbol{\tau}$ (see Fig.~\ref{fig:sym}). 
Thus the point group $G_0$ of the Cairo lattice is isomorphic to $C_{4v}$.  

Next, we would like to discuss the Irreducible Representations (IR's) of this space group with emphasis on special points in the Brilloin Zone (BZ). 
We begin with the zero momentum sector. The little group of $\vec{k}=0$ is the full point group and its IR's are taken over from those of $C_{4v}$. 
These are shown in Table \ref{tab:C4v} where we also show their decomposition into IR's of the $C_4$ subgroup which are labeled by the angular momenta $l$. 
Specifically we have four one-dimensional IR's out of which the first two, $A_1$ and $A_2$, are s-wave states ($l=0$) 
and the remaining ones, $B_1$ and $B_2$, are d-wave states ($l=\pi$).  In addition there is a 2-dimensional sector  ``E'' which decomposes into $l=\pm\pi/2$
and will be denoted in the following by ``$0.\{\pm\pi/2\}$''.

The little group of $\vec{k}=(\pi,\pi)$ is also $C_{4v}$. However the IR's are now different from the $\vec{k}=0$ case, 
since the representation theory for non-symmorphic groups is more involved for momenta that sit on the BZ boundary.
It turns out\footnote{This follows by employing the theory of Ray representations or Herring's method (see e.g. Chapter 11 of Ref.~[\onlinecite{Inui}] and references therein).}
that there are only two 2-dim IR's for $\vec{k}=(\pi,\pi)$. The first combines $l=0$ and $l=\pi$ while the second 
combines $l=\pm \pi/2$. So in the following we shall denote these IR's by ``$(\pi,\pi).\{0,\pi\}$'' and ``$(\pi,\pi).\{\pm \pi/2\}$'' respectively.

We finally discuss the $(\pi,0)$ point. Here the little group is isomorphic to $C_{2v}$ but the IR's are not those of $C_{2v}$ since we are again dealing with 
a point on the BZ boundary. Here group theory predicts a single 2-dimensional IR\footnote{This specific case is discussed in Sec. 11.2.5 of Ref.~[\onlinecite{Batanouny}].} 
whose members have angular momenta $l=0$ and $\pi$. By including the second member $(0,\pi)$ of the star of $\vec{k}$, we get a single 4-dimensional IR which we shall denote here by ``$(\pi,0)^\ast.\{l=0,\pi\}$''.

\subsection{Large-$x$ regime: The 1/3 ferrimagnetic phase}\label{SubSec:OneThird}
Our ED results of the full model show that the 1/3 ferrimagnetic state is indeed stabilized at large $x$. 
The transition to this state can be easily identified by a GS level crossing between the lowest $S=0$ state and the lowest $S=N/6$ state. 
For all clusters investigated the transition occurs around the classical point, $x\sim 2$. 
For the 24-site cluster which is the most symmetric cluster, the transition takes place at $x\simeq 1.96$ (see left panel of Fig.~\ref{fig:Pentag24_LESpectrumvsx} below).  

\begin{figure*}[!t]
\includegraphics[clip=true,width=0.15\linewidth]{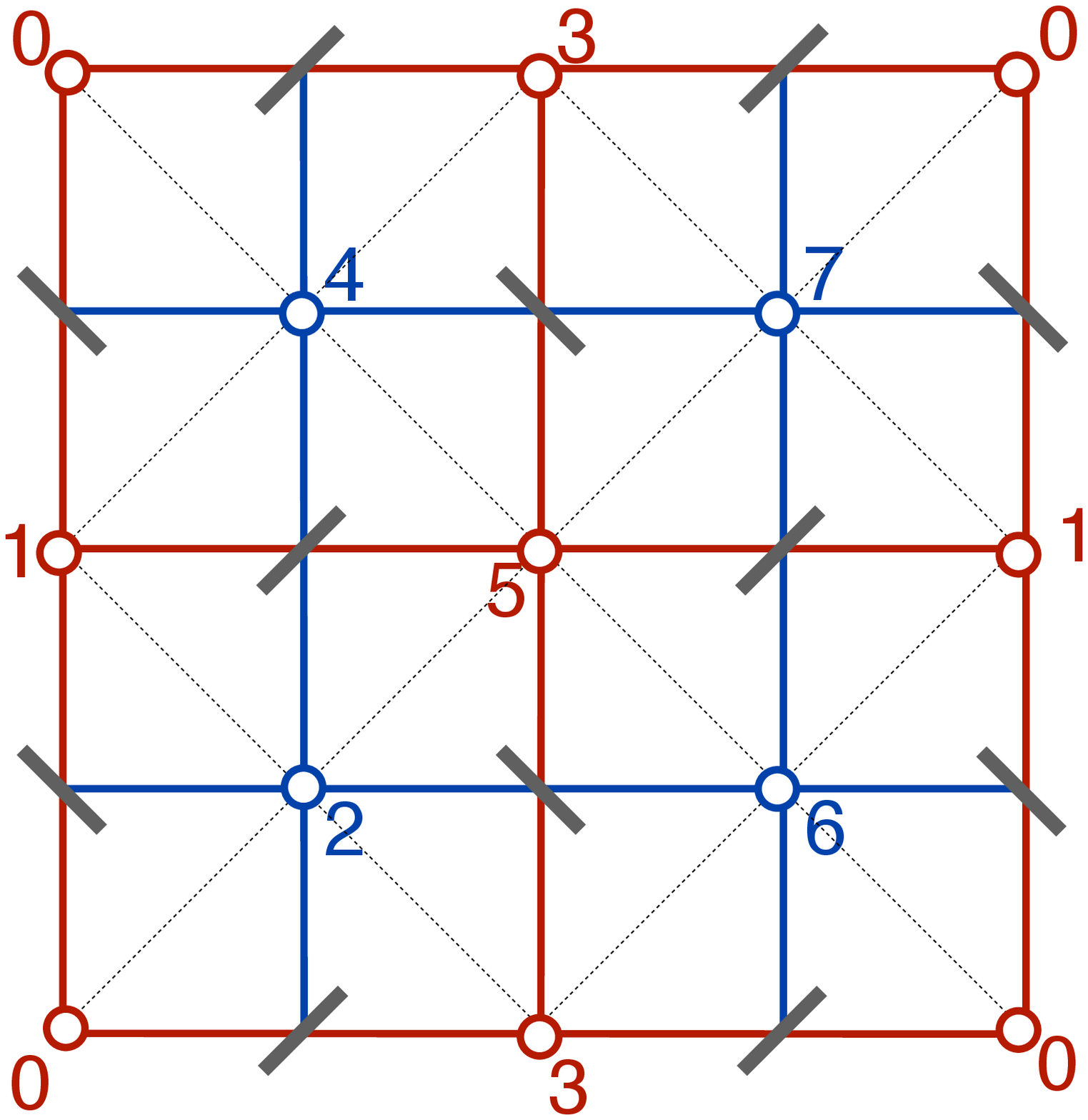}~~~\includegraphics[clip=true,width=0.85\linewidth]{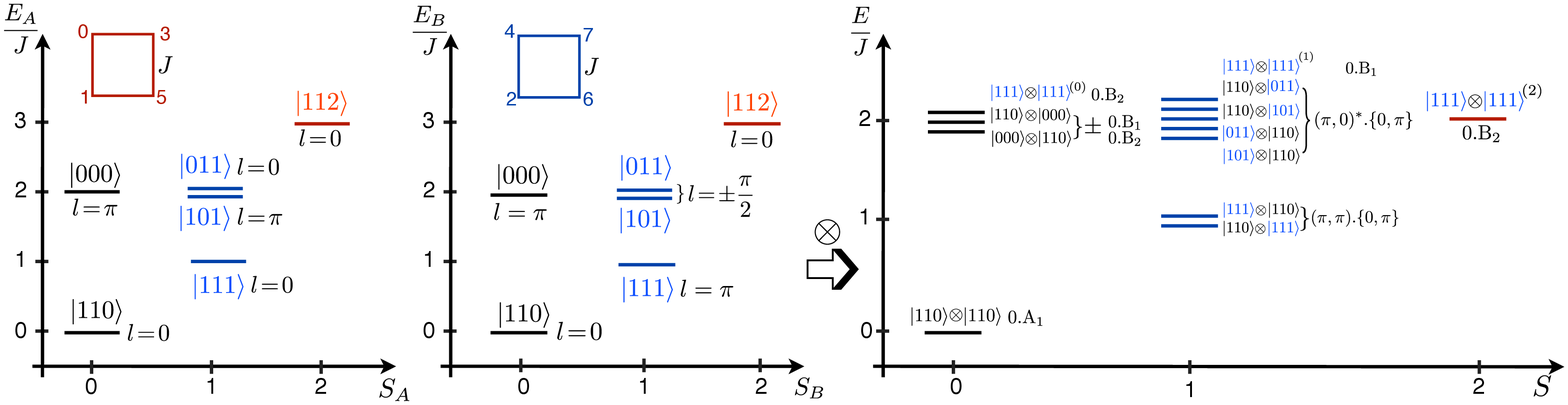}
\caption{Left panel: The effective $J_1$-$J_2$-$K$ model description of the 24-site Cairo pentagonal cluster. There are two inter-penetrating square 
plaquettes which consist of the sites (1, 0, 3, 5) and (2, 4, 7, 6). The $J_1$ and $J_2$ interactions are denoted by thin dotted and thick solid lines respectively, 
while the orientation of the $K$-term in each $J_1$-plaquette is denoted by a thick segment (which coincides with the underlying $J_{33}$-dimer of the Cairo lattice) at the middle. 
At small $x$ the two square plaquettes are decoupled from each other and only the $J_2$ coupling survives (here $J \equiv 2 J_2$). 
The total energy spectrum (last panel) results from a direct sum of the energy spectra of the two plaquettes (second and third panels). 
In addition to the spin quantum numbers we also specify the angular momentum $l$ with respect to $C_4$ rotations around the site numbered $5$. 
All degenerate levels are split by hand for clarity.}
\label{fig:plaquettespectrum}
\end{figure*}

\subsection{Small-$x$ regime: The presence of two weakly coupled AFM sublattices}\label{SubSec:Smallx}
\subsubsection{Structure of spectrum at $x\to 0$}\label{SubSubSec:8site}
Figure \ref{fig:Pentag24_LESpectrumvsx} shows the low-energy spectrum of the 24-site Cairo lattice as a function of $(x-1)/(x+1)$.
To highlight the dominance of the $J_2$ energy scale at small $x$ and to better examine the resulting splitting of the $x=0$ GS manifold, 
we have plotted the energy (always measured from the GS) in units of $x^2 J_{33}$ for $x<1$, and in units of $J_{43}$ for $x>1$.

The spectrum shows the following features in the limit of $x=0$. 
The first excitations above the singlet GS are two triplets with excitation energy $x^2 J_{33}$. At energy $2 x^2 J_{33}$ we find all together nine degenerate states, 
among which we have three singlets, five triplets and 1 quintet.    

To understand this structure we turn to the effective model description of the 24-site Cairo pentagonal cluster (see left panel of Fig.~\ref{fig:plaquettespectrum}). 
This cluster  comprises 8 four-fold coordinated sites which form two inter-penetrating square plaquettes A (sites 1, 0, 3, and 5) and B (sites 2, 4, 7, and 6).
This simplifying aspect allows to work-out explicitly the low-energy structure of the effective model for small $x$. 
Let us label the states of plaquette A by $|S_{05} S_{13} S_A\rangle$, where $\vec{S}_{05}=\vec{S}_0+\vec{S}_5$, 
$\vec{S}_{13}=\vec{S}_1+\vec{S}_3$, and $\vec{S}_A=\vec{S}_{05}+\vec{S}_{13}$. 
Similarly the states of plaquette B are labeled as $|S_{27} S_{46} S_B\rangle$.  
To lowest order in $x$ only the $J_2$ terms survive which read 
\bea
\mc{H}_2 &=& 2 J_2 \Big( \vec{S}_{46} \cdot \vec{S}_{27} + \vec{S}_{05}\cdot \vec{S}_{13} \Big) \nonumber\\
&=&  J_2 \Big( \vec{S}_A^2 + \vec{S}_B^2 -\vec{S}_{46}^2 - \vec{S}_{27}^2 - \vec{S}_{05}^2 - \vec{S}_{13}^2 \Big) ~,
\eea
where the factor of two in the first line accounts for the fact that each $J_2$ interaction appears twice in our cluster due to the periodic boundary conditions. 
Thus the two plaquettes A and B are decoupled from each other, and the full spectrum $E=E_A+E_B$ (rightmost panel of Fig.~\ref{fig:plaquettespectrum}) can be obtained by adding the two 
single-plaquette spectra (second and third panels of Fig.~\ref{fig:plaquettespectrum}) by a standard addition of angular momenta $\vec{S}=\vec{S}_A+\vec{S}_B$. 
It is then straightforward to show that the global GS is the direct product of the two $|1,1,0\rangle$ singlets of each plaquette and that it is s-wave ($l=0$). 
The lowest excited states are the triplets $|1,1,1\rangle_A\otimes|1,1,0\rangle_B$ and $|1,1,0\rangle_A\otimes|1,1,1\rangle_B$, 
which have energy $J\equiv 2 J_2$ above the GS and angular momentum $l=0$ and $\pi$ respectively. 
At energy $2J$, we get 3 singlets, 5 triplets, and 1 quintet. 

So the multiplicities and the symmetry properties of the spectrum at small $x$ 
match exactly the ones found by ED in the original 24-site Cairo cluster (Fig.~\ref{fig:Pentag24_LESpectrumvsx}). 
This confirms the main picture from the effective model of having two nearly decoupled inter-penetrating square AFM's.

\subsubsection{GS correlations}\label{SubSubSec:Correlations}
\textit{GS spin correlations}.---
Figure \ref{fig:Pentag24_LRSpin} shows the GS spin-spin correlations for the 24-site cluster. 
The left panel shows the correlation profiles at a representative value $x=0.5$, while the right panel shows the dependence as a function of $x$. 
There are two important features in these figures. First, there is a strong AFM correlation between the reference site and the 4-fold sites that belong to the same subsystem. 
This confirms the strong AFM $J_2$ coupling within each sublattice.

The second feature, which remains true in the entire range up to $x=1$, is that there are almost no correlations between the reference site and the 4-fold spins 
belonging to a different sublattice. This feature is consistent with the orthogonal phase scenario where the two N\'eel vectors are perpendicular to each other ($\theta=\pi/2$). 
It would also be consistent with the collinear phase scenario where the correlations actually vanish for finite-size clusters 
because the GS is an equal superposition of the two genuinely different collinear phases ($\theta=0$ and $\pi$).  

Hence the spin-spin correlation data cannot establish the critical value of $x$ where the transition between the orthogonal and the collinear phase takes place.  
For this, one would need much larger cluster sizes since, as we know from previous studies on the $J_1$-$J_2$ model,\cite{CCL,Weber} 
there is a very large length scale associated with the locking between the two N\'eel vectors.

\textit{GS dimer and vector-chiral correlations}.---
We now turn to some other correlations which may also be used as diagnostic tools for the collinear and the orthogonal phases.   
In the collinear phase the spins order FM in one direction and AFM in the perpendicular direction, 
so the dimer-dimer correlations along the two directions should reflect this physics.\cite{Weber}
On the other hand, in the orthogonal phase the spins lie in one plane so the orthogonal phase must show a staggered signal in the vector-chiral correlations.\cite{Chubukov,Lauchli} 
Figure \ref{fig:Pentag24_NematicityChirality} shows these two type of correlations for the same GS at a representative value of $x=0.1$. 
The results show that both types of correlations are present in the same GS and with the expected profile. 
As in the case of the spin-spin correlations, this shows that the two phases compete with each other 
but the 24-site cluster is too small to discriminate between the two.

\subsubsection{Excitations: Low-energy Towers of States}\label{SubSubSec:TOS}
We next turn to the low energy excitations and check whether we see any signatures of the collinear and the orthogonal states. 
This is done by looking at the symmetry structure of the so-called Anderson towers of states.\cite{Anderson}   
It is by now established, following the seminal work by Bernu \textit{et al.}~\cite{Bernu1,Bernu2} and Lecheminant \textit{et al.}~\cite{Lecheminant_J1J2,Lecheminant_kagome},  
that a given magnetic phase in the thermodynamic limit shows up in finite-size spectra through the clear formation of a 
tower of states which scale as $S(S+1)/N$ and is well separated from higher excitations. 
A wavepacket out of this infinite tower would be stationary in the thermodynamic limit and would correspond to the given classical state. 
Not surprisingly then, the multiplicities and symmetry properties of this set of states are intimately connected to the symmetries that are broken in the classical phase and can 
actually be derived by group theory alone.\cite{Bernu1,Bernu2,Lecheminant_J1J2,Lecheminant_kagome,polytopes,TOSMisguich} 

Now, the collinear and the orthogonal phase break the full symmetry group of the Hamiltonian in a different way, 
so the structure of the corresponding tower of states should be very different from each other. 
In App.~\ref{App:Towers} we derive the symmetry content of the two towers using group theory. 

\begin{figure}[!t]
\includegraphics[clip=true,width=0.99\linewidth]{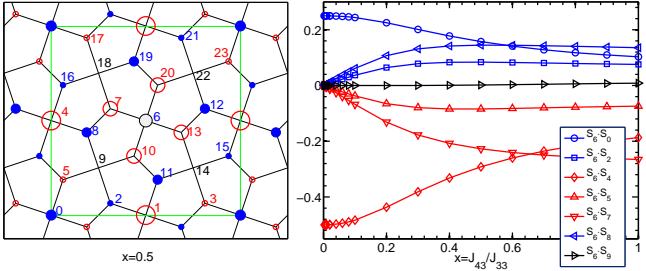}
\caption{Left: Ground state spin-spin correlation profiles $\langle\vec{S}_6\cdot \vec{S}_i\rangle$ at $x=0.5$ for the 24-site Cairo pentagonal cluster. 
Filled (blue) circles denote positive correlations while open (red) circles denote negative correlations. 
Right: Correlations between the reference site (6) and the set of all symmetry inequivalent sites of the cluster as a function of $x$, up to $x=1$. 
Note that the limiting correlation values at $x=0$ are consistent with the exactly known GS wavefunction of the 8-site effective cluster (see text).}
\label{fig:Pentag24_LRSpin}
\end{figure}

The final predictions for the lowest total spin $S$ sectors of the towers are given in Tables \ref{tab:ortho} and \ref{tab:col}.  
Specifically, the collinear phase comes with 2 states per total spin $S$, while the orthogonal phase should show $2S+1$ states (not related to the Zeeman degeneracy) at a given spin sector $S$.  
The symmetry properties of these states with respect to $C_4$ rotations as well as the four non-symorphic operations $(\sigma_i|\boldsymbol{\tau})$ are shown 
in Tables \ref{tab:ortho} and \ref{tab:col}.

These towers should be now compared to the low-E excitations of the 24-site cluster which are shown in the right panel of Fig.~\ref{fig:Pentag24_LESpectrumvsx} as a function of the total spin $S$, 
at a representative value $x=0.8$. In the singlet sector we find the ``0.A1'' ground state, a ``0.B1'' state nearby and another ``0.B2'' state slightly higher in energy. 
The first one belongs to the orthogonal tower but we may also think of the pair ``0.A1'' and ``0.B2'' as parts of the collinear tower. 
The ``intruder'' state ``0.B1'' does not belong to any of the towers and indeed at small enough $x$ this state is slightly higher than the ``0.B2'' state. 
At $x=0.8$ this state has very low energy because it is about to become the GS above $x\sim 1.2$ (see Sec.~\ref{Sec:IntPhase} below).

Likewise, in the $S=1$ sector, we find three triplets very close in energy. 
Two of these belong to the two-fold IR ``$(\pi,\pi).\{0,\pi\}$'' expected for the collinear tower, but we can also include the third state ``0.B1'' 
to complete the three states expected for the orthogonal tower. 
A similar situation occurs in the $S=2$ sector.  
Altogether, we find that the low-E excitations contain both towers of states which conforms with the previous picture from the GS correlations, that 
the collinear and orthogonal phases compete with each other but the 24-site cluster is too small to discriminate between the two.

\section{The intermediate non-magnetic phase.}\label{Sec:IntPhase} 
\subsection{Identifying the relevant effective term which drives the intermediate phase} 
We now turn to intermediate values of $x$. Looking back at the low-E spectra shown in the left panel of Fig.~\ref{fig:Pentag24_LESpectrumvsx} 
we see that there is a GS level crossing in the singlet sector around $x\sim 1.2$. Similar GS level crossings are also found in the other Cairo clusters that we studied. 
For the 24-site cluster the new singlet GS belongs to the ``0.B1'' sector which 
belongs to neither the collinear nor the orthogonal towers of states (Tables \ref{tab:ortho} and \ref{tab:col}), 
which is the first strong evidence that the system enters a new phase. 

\begin{figure}[!t]
\includegraphics[clip=true,width=0.94\linewidth]{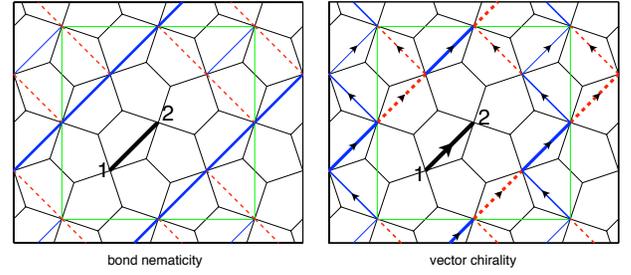}
\caption{Ground state bond nematic $\langle (\vec{S}_1\cdot \vec{S}_2) (\vec{S}_k\cdot \vec{S}_l)\rangle$ 
and vector-chiral $\langle (\vec{S}_1\times \vec{S}_2)_z ~(\vec{S}_k\times \vec{S}_l)_z\rangle$ 
correlations at $x=0.1$ for the 24-site Cairo pentagonal cluster. The thick bond $(1,2)$ is the reference bond.  
Solid (blue) lines denote positive correlations and dashed (red) lines denote negative correlations. 
The width of each line is proportional to corresponding expectation value.}
\label{fig:Pentag24_NematicityChirality}
\end{figure}

Another evidence is the reorganization of the low-E excitation spectrum.  
In particular as we see in left panel of Fig.~\ref{fig:Pentag24andEffModels}, the energies of the three lowest spin sectors form a concave envelope, 
which suggests a tower of states formed by even spin sectors only. 
Indeed the magnetization (not shown here) grows in $\delta S_z =2$ steps until we reach the 1/3-ferrimagnetic phase.    
Provided that these $\delta S=2$ excitations form a tower that collapses in the thermodynamic limit, 
the resulting state would not break time-reversal invariance, as it happens e.g. in a spin-nematic state.\cite{Shannon}

 \begin{figure*}[!t]
\includegraphics[clip=true,width=0.85\linewidth]{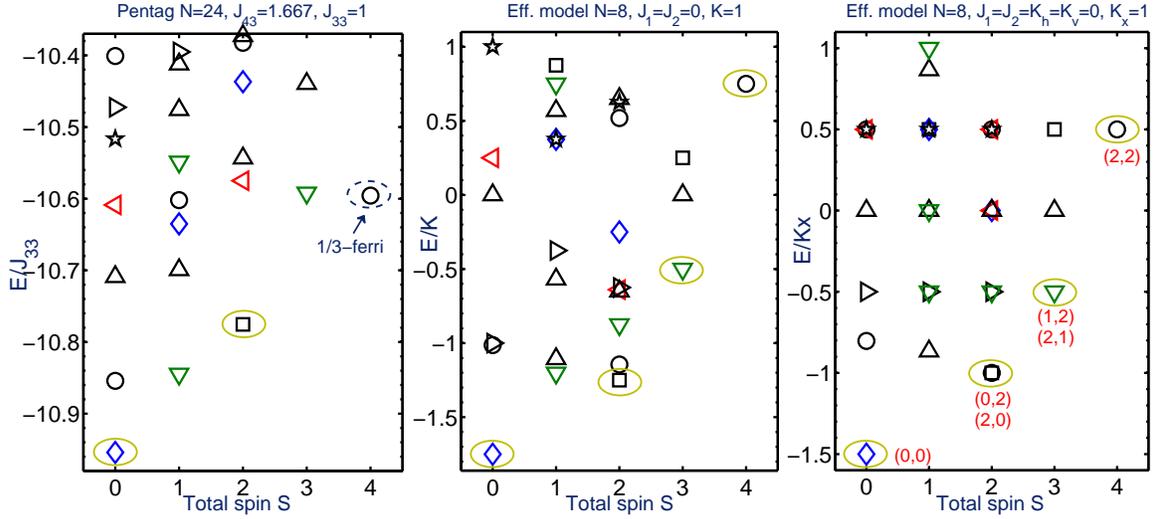}
\caption{Low-E spectra as a function of total spin $S$ for three different cases. The left is the spectra of the 24-site cluster for $x=1.667$, i.e. deep inside the intermediate phase. 
The middle panel shows the spectrum of the 8-site unconstrained $J_1$-$J_2$-$K$ model in the large $K/J_{1,2}$ regime. The third panel shows the 
spectrum of the 8-site effective model but this time we have kept only the $K_x$ term among the three plaquette terms of Eq.~(\ref{Kp1.Eq}). 
The pairs of numbers inside the parentheses indicate the total spins $S_A$ and $S_B$ of the two sublattices which are separately conserved in this SU(2)$\times$SU(2) model.
In all three cases, the ovals indicate the sequence of GS's visited by the system in a field, while the different symbols follow the same convention as in the right panel of 
Fig.~\ref{fig:Pentag24_LESpectrumvsx}.}
\label{fig:Pentag24andEffModels}
\end{figure*}

It turns out that we can actually learn more by tracing the intermediate ``0.B1'' state back to its original place in the $x=0$ spectrum.  
First of all, Fig.~\ref{fig:Pentag24_LESpectrumvsx} tells us that the ``0.B1'' state is a member of the $x=0$ GS manifold where $J_{33}$-dimers form singlets, 
in contrast to the 1/3-ferrimagnetic state which is clearly not. So the intermediate phase is more closely related to the physics of the $x=0$ limit.  
Secondly, we know explicitly the low spin states of the 24-site cluster at $x=0$ 
and so we may find out exactly which excitation at $x=0$ evolves into the intermediate ``0.B1'' phase.    
We find that it is the following combination among two of the 
singlets with energy $2 J$ above the GS (see right panel of Fig.~\ref{fig:plaquettespectrum}): 
\be\label{eqn:phi}
|\Phi\rangle = \frac{1}{\sqrt{2}} \left( |1,1,0\rangle_A \otimes |0,0,0\rangle_B+|0,0,0\rangle_A \otimes |1,1,0\rangle_B \right) ~.
\ee
This state has an important property (not shown explicitly here):  It minimizes simultaneously all plaquette terms of the type $\mc{K}_x$ of Eq.~(\ref{Kp2.Eq}), with eigenvalue -3/16. 
This provides evidence that the instability mechanism that triggers the transition to the intermediate phase is related to the four-spin exchange term $\mc{K}_x$. 
At the level of the effective model of Sec.~\ref{Sec:EffModel}, this suggests that the $\mc{K}_x$ terms dominate in much higher orders of perturbation theory in $x$, 
but we are not able to check this explicitly.  

We can still however make progress using only the three couplings that we know so far from the fourth-order theory. 
The idea is to compare the low-E spectrum of the intermediate phase of the Cairo model with 
that of the unconstrained $J_1$-$J_2$-$K$ model as we visit different regions in the $J_1/J_2$-$K/J_2$ plane.  
We have performed ED in the effective lattice model (which has 2 sites per unit cell) using clusters with 8, 10, 16, 18, 20, 26, and 32 sites. 
The fully symmetric clusters with 8, 16 and 32 sites give the most clear and systematic evidence so we shall only discuss these clusters here.  
We begin with the 8-site effective cluster since it can be directly compared to the 24-site Cairo cluster (the 4-fold sites of the latter form the 8-site effective cluster). 
By an inspection of the low-E spectra in various regions of the $J_1/J_2$-$K/J_2$ plane, we have located the region of large $K/J_{1,2}$ 
as the one with very similar low-E spectral features with that in the intermediate phase. This is demonstrated by the first two panels of Fig.~\ref{fig:Pentag24andEffModels}. 

Given now the special role of $\mc{K}_x$ regarding the above state $|\Phi\rangle$ (see Eq.~(\ref{eqn:phi})),  
we next check what happens if among the three terms included in $K$ (see Eq.~(\ref{Kp1.Eq})) we keep only the $\mc{K}_x$ term.  
The corresponding 8-site spectrum is shown in the third panel of Fig.~\ref{fig:Pentag24andEffModels} and demonstrates that 
the spectrum retains the same features at low energies whether we keep only $\mc{K}_x$ or not. 
This shows that $\mc{K}_x$ is indeed the most relevant plaquette term for the intermediate phase. 

\begin{figure*}[!t]
\includegraphics[clip=true,width=0.678\linewidth]{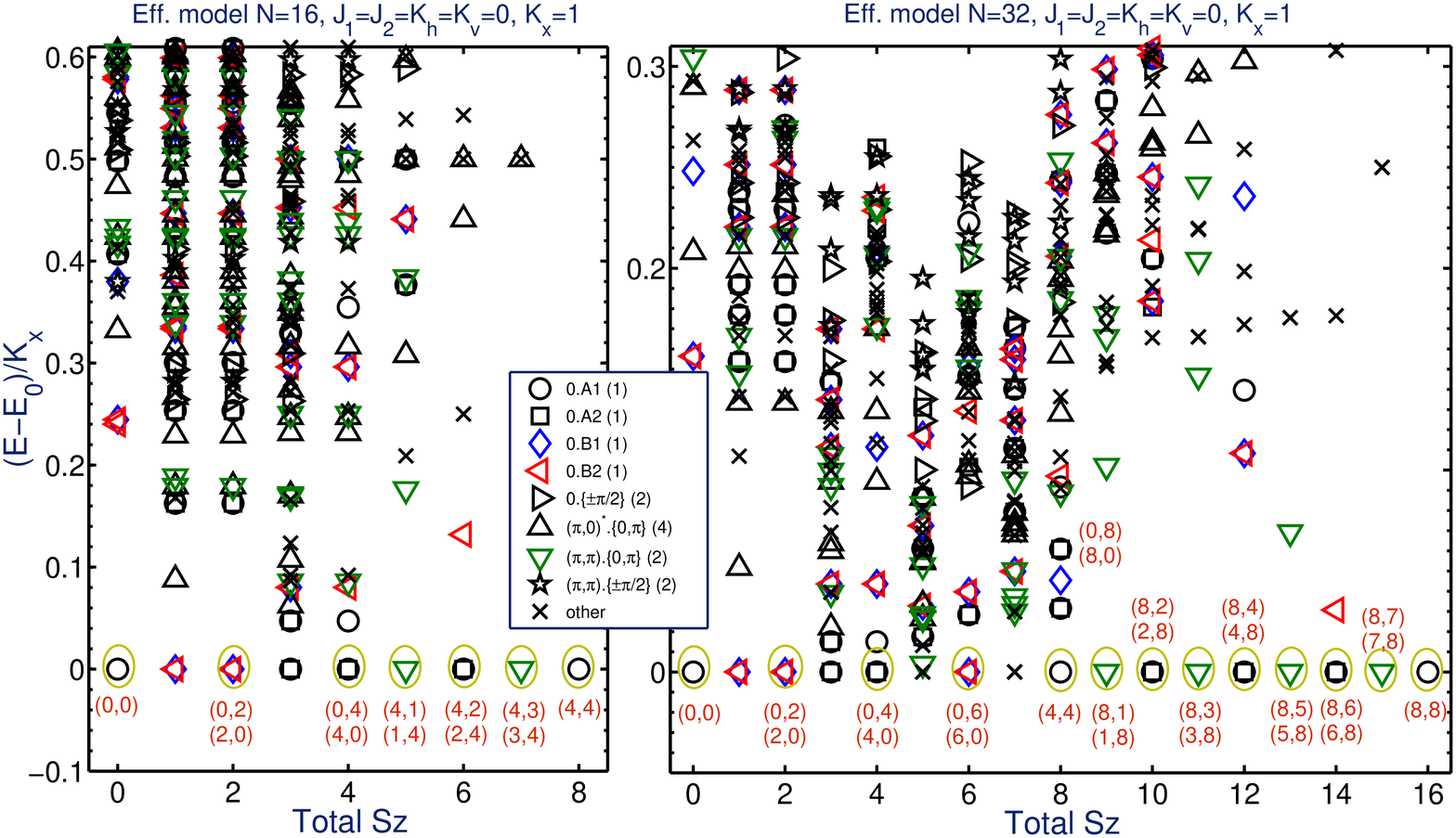}~~\includegraphics[clip=true,width=0.322\linewidth]{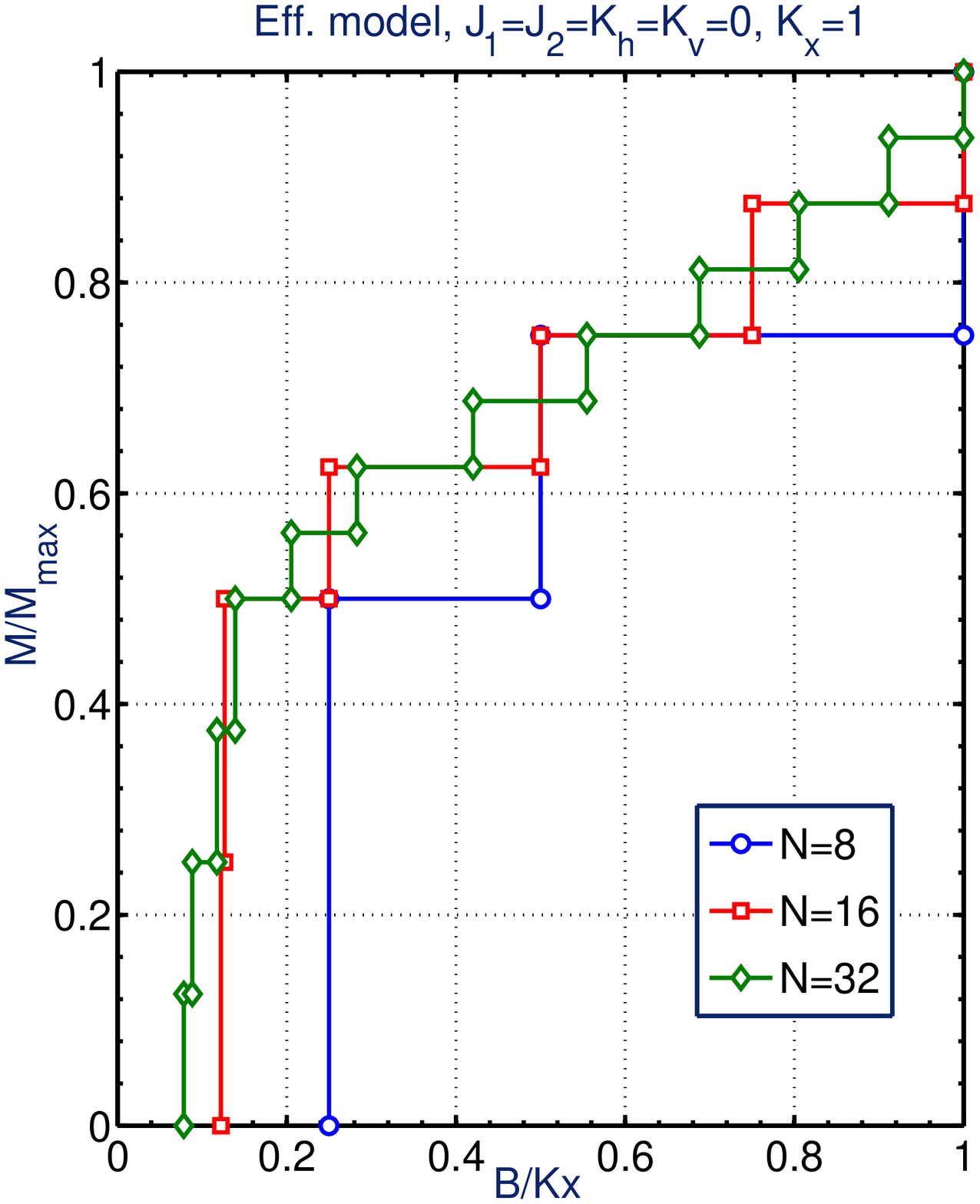}
\caption{(Color online) First two panels: Low-E spectra as a function of total $S_z$ for the 16- and 32-site clusters using only the effective $K_x$ plaquette term. 
Here all energies are measured for the GS energy $E_0(S_z)$. 
The ovals indicate the sequence of GS's visited by the system in a field while the pairs of numbers inside the parentheses indicate the total spins $S_A$ and $S_B$ of the two sublattices in these states. Note that for the 32-site case, the 1/2-magnetization states with $(S_A,S_B)=(0,8)$ and $(8,0)$ are not the GS's. 
Right panel: Magnetization process in a field for the clusters with 8, 16 and 32 sites.}\label{fig:Eff16and32spectra}
\end{figure*}

\subsection{The $\mc{K}_x$-model}\label{Sec:Kx}
For the above reasons we shall restrict ourselves to the $\mc{K}_x$-model in the following. The Hamiltonian 
reads
\be\label{eqn:Kx}
\mc{H}_{\mc{K}_x} = \sum_{\text{plaquettes}} (\vec{S}_\alpha \cdot \vec{S}_\gamma)~(\vec{S}_\beta \cdot \vec{S}_\delta)
\ee
where the sum is over all plaquettes of the square lattice, and $(\alpha, \beta, \gamma, \delta)$ label the spins around a plaquette clockwise.  
This model can be thought of as a 2D generalization of the well-studied\cite{SO1,SO2,SO3,SO4,SO5} 1D spin-orbital model.

We first discuss the classical limit of this model. It is easy to show (and we have also checked it by Classical Monte Carlo simulations)  
that the energy can be minimized by any collinear configuration with ``3up-1down'' or ``3down-1up'' spins 
in every plaquette. In particular, starting from one such GS we can generate others by flipping all the spins along any horizontal or vertical line of the square lattice.  
This leads to a sub-extensive number of GS's (with variant magnetizations) which are not related by global spin rotations.  
So the classical limit is highly frustrated and thus we anticipate rich physics from thermal (not studied here) or quantum fluctuations. 

It is natural that the above classical GS manifold appears also if we use Ising spins in Eq.~(\ref{eqn:Kx}). 
Actually, as found by Rojas {\it et al.},\cite{Rojas} the same GS's arise also in the Ising version of Eq.~(\ref{eqn:Ham}). 

Turning now to the quantum limit, let us denote by A and B the two interpenetrating square sublattices 
of the full square lattice (in Fig.~\ref{Fig:PlaqSqLattice} of App.~\ref{App:SPE}, the A and B sublattices consist of the $(\alpha,\gamma)$- and $(\beta,\delta)$-bonds respectively). 
This model has a $C_4$ rotation symmetry around the center of the plaquettes, contrary to the full $J_1$-$J_2$-$K$ model.  
More importantly, the model has an enlarged SU(2)$\times$SU(2) symmetry since 
we can make independent spin rotations in the two sublattices A and B without changing the energy. 
So the total spins $S_A$ and $S_B$ of the two sublattices are good quantum numbers.  
This also means that within a fixed $(S_A,S_B)$ manifold the energy is independent of the total spin $S$.  
In addition, the model has an Z$_2$-invariance under interchanging the two sublattices A and B, 
so the manifolds $(S_A, S_B)$ and ($S_B, S_A$), with $S_A\neq S_B$, must be degenerate.

Let us now try to establish some systematic spectral features in the $\mc{K}_x$-model by looking at the larger 16- and 32-site symmetric clusters. 
Their spectra are shown in the first two panels of Fig.~\ref{fig:Eff16and32spectra} as a function of total $S_z$. 
All energies are now measured from the GS energy $E_0(S_z)$ so that we better retrieve the details of the spectra. 
The states that are visited by the system in the presence of a magnetic field are highlighted by ovals (tower of states in the following), 
while the pairs of numbers inside the parentheses show the total spins $(S_A, S_B)$ in these states. 

All towers of states, including the one for 8-sites (third panel of Fig.~\ref{fig:Pentag24andEffModels}) exhibit common features.  
Apart from the singlet GS with $(S_A,S_B)=(0,0)$, and the fully polarized state $(N/4,N/4)$, 
all other states show a two-fold degeneracy which is related to the Z$_2$-symmetry mentioned above. 
More importantly, there is a very clear difference between the states below and above 1/2-magnetization.  
The tower below 1/2 comprises only states with even $S$, while above 1/2 we have $\delta S=1$ steps. 
This is also demonstrated in the third panel of Fig.~\ref{fig:Eff16and32spectra} which shows the magnetization process in a field.  Altogether these results reveal two different states above and below 1/2-magnetization. 

The physics above 1/2-magnetization can be easily understood by noting that the corresponding states have either $S_A$ or $S_B$ equal to $N/4$. 
Thus one of the two subsystems is fully polarized above the 1/2-magnetization. It is straightforward to show that in this case 
the $\mc{K}_x$-model reduces to the two-spin exchange model in the other sublattice with a nearest-neighbor coupling equal to $K_x/4$. 
We have checked that the tower of states above the 1/2-magnetization match exactly 
(both in the symmetries and the actual energies) the corresponding tower of states  
of the square lattice AFM with half the number of sites and a nearest-neighbor coupling equal to $K_x/4$.

The physics below 1/2-magnetization is much more interesting. 
We first note that the tower below 1/2 contains states where both $S_A$ and $S_B$ are even, and that one of the two is always zero. 
This suggests that the SU(2) symmetry is broken down to U(1) only in one of the two sublattices. 
Namely that we have some kind of spin nematic state in one sublattice and a spin liquid state in the other sublattice. 
In addition, the angular momentum alternates between $l=0$ (s-wave) and $l=\pi$ (d-wave) as we go down in $S_z$ starting from the 1/2-magnetization state which is s-wave.  Hence the zero-field GS is s-wave for 16 and 32 sites but it is d-wave for 8 sites, which conforms with the d-wave property of the intermediate ``0.B1'' state in the 24-site Cairo cluster.

The alternation between s-wave and d-wave GS symmetry at low-magnetizations is exactly what happens 
in the model studied by Shannon {\it et al.} in Ref.~[\onlinecite{Shannon}], 
and suggests that the intermediate spin-nematic phase has a d-wave symmetry. 
We have checked this numerically by a calculation of the GS bond-nematic correlations of the type $\langle (S_i^+S_j^+) (S_k^-S_l^-)\rangle$  
where $(i,j)$ and $(k,l)$ denote two different bonds of the lattice. The results are shown in Fig.~\ref{Fig:nematic} and confirm the presence of a d-wave signal in one sublattice only. The correlations between bonds of different sublattices vanish exactly because the GS belongs to the sector $(S_A,S_B)=(0,0)$.

\begin{figure}[!t]
\includegraphics[clip=true,width=0.8\linewidth]{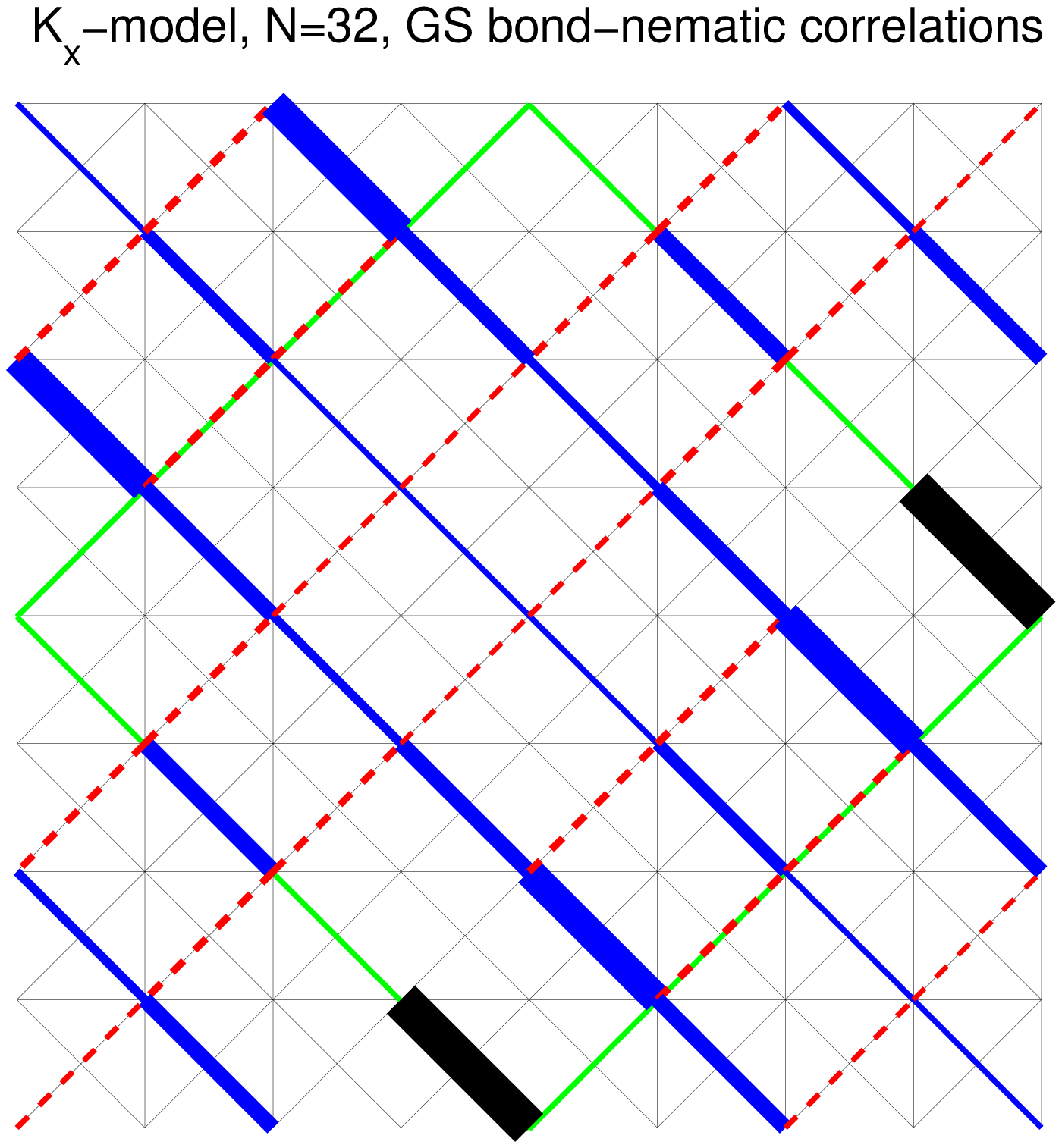}
\caption{(Color online) GS nematic correlations $\langle (S_i^+S_j^+) (S_k^-S_l^-) \rangle$ for the $\mc{K}_x$-model in the 32-site cluster. 
The cluster is denoted by the (green) rectangle. The think (black) line is the reference bond $(ij)$. The solid (blue) lines denote the bonds $(kl)$ with positive correlations 
while the dashed (red) lines denote negative correlations. The width of the bonds is proportional to the actual strength of the correlations, the largest of which is equal to $0.0228$. 
Note that there are no correlations between the reference bond and bonds belonging to a different sublattice.}\label{Fig:nematic}
\end{figure}

Let us now try to identify the low-E degrees of freedom that could give rise to such a nematic state.  
Since we are dealing with spins 1/2, we need some composite degrees of freedom that can sustain a finite quadrupolar order parameter. 
For the spin-nematic phase of Ref.~[\onlinecite{Shannon}], these might be the triplets formed on nearest-neighbor bonds,\cite{Shannon} or the plaquette $S=2$ states, 
as discussed by Ueda and Totsuka.\cite{Ueda} 
In the present case we can also identify some triplet degrees of freedom by considering the effect of $\mc{K}_x$ on a single plaquette $(\alpha, \beta, \gamma, \delta)$. 
It is easy to see that there are two possible ways to minimize the energy, either by forming a singlet on the $(\alpha,\gamma)$-bond and a triplet on the $(\beta,\delta)$-bond or vice versa. Both states have energy $-3/16$ and have total spin $S=1$, so they can indeed sustain a finite quadrupolar order parameter.  

Now, there are two factors that impose nontrivial constraints into the way these triplet degrees of freedom interact with each other in the full lattice. 
The first is that, by their nature, the two triplet GSs impose that a triplet in one diagonal bond must come with a singlet on the other diagonal bond. 
The second complication is that any given site participates into four plaquette interactions. 
One approach that deals with the first problem (but not with the second) is to perform a strong-coupling expansion 
around the limit where only 1/4 of the plaquettes have a finite $\mc{K}_x$ and are decoupled from each other (see Fig.~\ref{Fig:PlaqSqLattice}).\cite{Ueda,Lecheminant}  
The details of this approach are provided in App.~\ref{App:SPE} and follow closely in spirit the treatment by Lecheminant and Totsuka\cite{Lecheminant} 
of a very similar situation in a two-leg ladder system where a different pair of plaquette triplets emerges at low energies. 
This approach delivers an effective model which is similar to the well-known Kugel-Khomskii Hamiltonian\cite{KK} for orbital degenerate systems. 
Here the role of the orbitals is played by the two types of triplets.   

A natural variational treatment of this effective model delivers a spin-nematic GS of a novel kind where both types of triplets are entangled (see App.~\ref{App:SPE}). 
Despite this success, there are two problems with this wavefunction. 
The first is that both sublattices of the square lattice participate in this state (see App.~\ref{App:TOSvar} for the symmetry decomposition of this state), 
while the numerical tower of states suggest that the symmetry breaking occurs only in one sublattice.     
The second problem is that the variational state breaks translational invariance which is also in disagreement with the numerical tower of states.   
In particular, the d-wave character of the correlations is built-in from the outset since, by construction, the ``plaquettized'' lattice distinguishes between 
the two different directions in each sublattice (see App.~\ref{App:SPE}).\footnote{One may restore translational invariance by constructing a (zero-momentum) linear superposition 
of the four variational states corresponding to the four different plaquettized versions of the square lattice.}

\section{Summary}
In this article we have presented an extensive analytical and numerical study of the Heisenberg model on the Cairo pentagonal lattice.   
We have shown that by tuning the spin $S$ and the ratio $x$ of the two inequivalent exchange couplings of the lattice, we can drive the 
system through a number of competing phases, some of which have a strong quantum-mechanical origin. We have demonstrated that 
the rich physics of this model arises, to a large extent, from the presence of two inequivalent sites in the Cairo lattice. 
To highlight this central feature and to better understand the nature of the various phases we have followed a series of different approaches which build 
quantum fluctuations in a gradual way starting from the classical limit. 

The classical phase diagram contains three phases, the orthogonal phase for $x<\sqrt{2}$, 
the 1/3-ferrimagnetic phase for $x>2$, and a mixed non-coplanar phase for $\sqrt{2}< x <2$ which combines the orthogonal and the ferrimagnetic.  
The orthogonal state is the configuration found experimentally for the Bi$_2$Fe$_4$O$_9$ compound,\cite{Ressouche} 
which suggests that this phase is quite robust in a more extended parameter space compared to our symmetric version of the model.

In the quantum $S=1/2$ limit, we have found an underlying order-by-disorder mechanism 
which is active at small $x$ and which stabilizes a collinear magnetic configuration.   
The 1/3-ferrimagnet phase seems to survive quantum fluctuations down to $S=1/2$ in the large-$x$ regime. 
On the other hand, the mixed phase becomes unstable for low enough $S$ and disappears completely from the phase diagram for $S=1/2$. 

Our ED results provide strong evidence of a new intermediate phase with spin-nematic correlations. 
We argued that this phase is stabilized by an effective 4-spin plaquette interaction that first appears in fourth-order perturbation theory in $x$. 
This term has an enlarged SU(2)$\times$SU(2) symmetry structure similar to spin-orbital models, and favors two species of triplet GSs on a single plaquette.   
The symmetry structure of the low-E spectra in the corresponfing plaquette model 
suggests that these triplet degrees of freedom order in a non-trivial fashion, whereby  
one of the two square sublattices forms a translationaly invariant spin-nematic structure with d-wave symmetry (similar to the one found by Shannon {\it et al.}\cite{Shannon}),  
while no symmetry-breaking seems to occur in the second sublattice. 

One of the novel aspects of the intermediate phase is its response under a magnetic field.  
The sublattice where the spin-nematic order takes place responds by developing a dipolar moment until full saturation.   
At this point, which corresponds to 1/2 moment of the full system, the low-energy physics 
of the second sublattice effectively reduces to that of the square lattice N\'eel AFM.   

A simple physical picture for the intermediate phase is still lacking.  
We have discussed e.g. a strong-plaquette expansion that builds upon the low-energy triplet degrees of freedom  
of a single plaquette. A variational treatment of the resulting model does deliver a spin-nematic phase, but this seems to entangle both sublattices 
and in addition it has symmetry properties that are not compatible with our low-energy spectra from ED.   
It is our hope that further studies will shed more light in this direction. 

At a more general level, this study demonstrates that pentagonal lattice models can feature a wealth of strong correlation phenomena. 
Compared to the well-explored triangle-based models (triangular, kagome, etc.), here we have a qualitatively different degree of frustration and complexity 
since the low-energy physics is not related to the local single-pentagon physics in any obvious way. 
Hence pentagonal lattice models are interesting in their own right and provide a distinct platform for exploring and realizing novel phases 
of matter in frustrated magnetism.

\section{Acknowledgments}
We are grateful to K. Penc for fruitful comments and suggestions on the spin+bond-wave method. 
We would also like to acknowledge our stimulating discussions with T. Toth, O. Starykh, M. Zhitomirsky, G. Uhrig, N. B. Perkins,  S. Wenzel, and E. Bergholtz. 

\appendix

\section{Spin Wave Theory (LSWT)}\label{App:LSWT}
\subsection{General procedure}
Here we outline the general procedure for diagonalizing the quadratic bosonic Hamiltonians obtained from LSWT around the three classical ground states.  
The Cairo pentagonal lattice has 6 sites (two 4-fold and four 3-fold coordinated) per unit cell.   
So we introduce six bosonic operators denoted by $a_i, b_i, c_i, d_i, e_i, f_i$ to describe the harrmonic fluctuations on each site of the $i$-th unit cell. 
Quite generally, up to quadratic order the semiclassical expansion gives
\be\label{eqn:Hlswt}
\mc{H} = E_{cl} + \delta E_1 + \frac{S}{2} \sum_k \vec{A}_k^+ \cdot \vec{H}_k \cdot  \vec{A}_k 
\ee
where $\delta E_1$ is a constant (see below),  
\be
\vec{A}_k^+ = \left(
a_k^+,  b_k^+, c_k^+, d_k^+, e_k^+, f_k^+, a_{-k}, b_{-k}, c_{-k}, d_{-k}, e_{-k}, f_{-k} 
\right)
\ee
and 
$\vec{H}_k \!=\! \left(\begin{array}{cc} 
\vec{C}_k & \vec{D}_k \\
\vec{D}_k^+ & \vec{C}_{-k}^T
\end{array}\right)$, 
where both $\vec{C}_k$ and $\vec{D}_k$ are $6\times 6$ matrices.  
To diagonalize the Hamiltonian we search for a new set of bosonic operators $\tilde{\vec{A}}_k$ 
given by the generalized Bogoliubov transformation $\vec{A}_k = \vec{V}_k \!\cdot\! \tilde{\vec{A}}_k$, such that   
the matrix $\vec{V}_k^+  \vec{H}_k \vec{V}_k \equiv \vec{\Omega}_k$ becomes diagonal. 
The transformation must also preserve the bosonic commutation relations, which can be expressed compactly as 
$\vec{g}=\tilde{\vec{g}} = \vec{V}_k \cdot \vec{g}\cdot  \vec{V}_k^+ $, where $\vec{g}$ is the ``commutator'' matrix 
\be
\vec{g} = \vec{A}_k \cdot \vec{A}_k^+ - \left( \left(\vec{A}_k^+\right)^T \cdot \vec{A}_k^T \right)^T= 
\left(\begin{array}{c c}
{\bf 1}_6 & 0 \\
0 & -{\bf 1}_6 
\end{array}\right)
\ee
and ${\bf 1}_6$ stands for the $6\times 6$ identity matrix. The above two conditions give 
\be
(\vec{g} \vec{H}_k)\cdot \vec{V}_k = \vec{V}_k \cdot \left( \vec{g} \vec{\Omega}_k \right) \equiv \vec{V}_k \cdot \vec{\Omega}'_k 
\ee
which is an eigenvalue equation in matrix form (the columns of $\vec{V}_k$ contain the eigenvectors of $\vec{g} \vec{H}_k$). 

One can further show\cite{Blaizot} that if $\vec{H}_k$ is semi-definite positive, then 
$\vec{\Omega}_k \!=\! \left(\begin{array}{cc}
\boldsymbol{\omega}_k & 0 \\
0 & \boldsymbol{\omega}_k 
\end{array}\right)$,  
where $\boldsymbol{\omega}_k$ is a diagonal matrix with non-negative entries $\omega_{1k}$-$\omega_{6k}$. This in turn leads to 
\bea
\mc{H} &=& E_{cl} + \delta E_1 + \delta E_2 \nonumber\\
&+& S \sum_{k} \left( \omega_{1k} \tilde{a}_{k}^+ \tilde{a}_{k} +\! \ldots\! +\omega_{6k} \tilde{f}_{k}^+ \tilde{f}_{k} \right)~,
\eea 
with $\delta E_2 = \frac{S}{2} \sum_{k}\left( \omega_{1k}+\!\ldots\!+ \omega_{6k}\right)$, 
which represents the total zero-point energy from all harmonic oscillators in the theory. 
The total quadratic correction to the GS energy is then given by $\delta E=\delta E_1+\delta E_2$.

We now turn to the quadratic correction to the local spin lengths.  
Let us consider the spin $\vec{S}_a$ operator inside the unit cell $i=(0,0,0)$. We have
\be
\vec{S}_a^z= S-a_{i=0}^+ a_{i=0} = S-\frac{1}{N_{uc}} \sum_{k,q} a_k^+ a_q  ~,
\ee
where $N_{uc}=N/6$ is the number of unit cells. Using the transformation $\vec{A}_k = \vec{V}_k\cdot\tilde{\vec{A}}_k$ we get
\be
\vec{S}_a^z= S-\frac{1}{N_{uc}} \sum_{k,q} \sum_{nm} V_k(1,n)^\ast V_q(1,m) ~\tilde{A}_k^+(n) \tilde{A}_k(m) ~.
\ee
In the vacuum GS the only non-vanishing expectation values are of the type $\langle \tilde{a}_{k} \tilde{a}_{k}^+ \rangle=1$.
Thus from the above sum we may keep only terms with $n=m=7-12$ and $k=q$. Namely
\be
\langle \vec{S}_a^z \rangle = S-\frac{1}{N_{uc}} \sum_k \sum_{n=7}^{12} | V_k(1,n) |^2 ~,
\ee 
and similarly for the 3-fold sites: 
\be
\langle \vec{S}_b^z \rangle = S-\frac{1}{N_{uc}} \sum_k \sum_{n=7}^{12} | V_k(2,n) |^2 ~.
\ee

\subsection{LSWT in the Cairo pentagonal lattice}
In the following we provide the explicit expressions of the $6\times 6$ matrices $\vec{C}_k$ and $\vec{D}_k$ defined above. 
We shall make use of the following definitions: 
$k_{xy}=k_x+k_y$, $q_\pm = \frac{x}{2}(1\pm \frac{1}{\sqrt{2}})$, 
$z_\pm = \frac{1\pm i \sin\theta'(x)}{\sqrt{2}}$, and $x_\pm=\frac{x}{2}\pm 1$. 
 
In the orthogonal phase $E_{cl} = N \varepsilon_{\text{ortho}}$ , $\delta E_1 = N \varepsilon_{\text{ortho}} /S$, 
\begin{widetext}
{\small 
\be
\vec{C}_k \!=\! \left(
\begin{array}{cccccc}
2\sqrt{2}x&q_-&q_-&q_-&q_-&0\\ 
q_-&\sqrt{2}x+1&0&0&0&q_-e^{ik_{xy}}\\ 
q_-&0&\sqrt{2}x+1&0&0&q_-e^{ik_y}\\ 
q_-&0&0&\sqrt{2}x+1&0&q_-\\ 
q_-&0&0&0&\sqrt{2}x+1&q_-e^{ik_x}\\ 
0&q_-e^{-ik_{xy}}&q_- e^{-ik_y}&q_-&q_-e^{-ik_x}&2\sqrt{2}x
\end{array}
\!\right)\!,
~\vec{D}_k \!=\! -\left(
\begin{array}{cccccc}
0&q_+&q_+&q_+&q_+&0\\ 
q_+&0&0&e^{ik_y}&0&q_+e^{ik_{xy}}\\ 
q_+&0&0&0&e^{-ik_x}&q_+e^{ik_y}\\ 
q_+&e^{-ik_y}&0&0&0&q_+\\ 
q_+&0&-e^{ik_x}&0&0&q_+e^{ik_x}\\ 
0&q_+e^{-ik_{xy}}&q_+e^{-ik_y}&q_+&q_+e^{-ik_x}&0 
 \end{array}
 \right) . 
\ee}
For the 1/3-ferrimagnetic phase we find  
$E_{cl}=N\varepsilon_{\text{ferri}}$, $\delta E_1=N\varepsilon_{\text{ferri}}/S$, and 
{\small
\be
\vec{C}_k \!=\! \left(
\begin{array}{cccccc}
4x &0&0&0&0&0\\ 
0&2x-1&0&e^{i k_y}&0&0\\ 
0&0&2x-1&0&e^{-i k_x}&0\\ 
0&e^{-i k_y}&0&2x-1&0&0\\
0&0&e^{i k_x}&0&2x-1&0\\ 
0&0&0&0&0&4x 
\end{array}
\right)\!,
~\vec{D}_k \!=\! -\left(
\begin{array}{cccccc}
0&x&x&x&x&0\\
x&0&0&0&0&x e^{i k_{xy}}\\
x&0&0&0&0&x e^{i k_y}\\
x&0&0&0&0&x\\
x&0&0&0&0&x e^{i k_x}\\
0&x e^{-i k_{xy}}&x e^{-i k_y}&x&x e^{-i k_x}&0 
 \end{array}
 \right)~.  
\ee}
Finally, in the mixed phase $E_{cl} = N \varepsilon_{\text{mixed}}$,  $\delta E_1 = N \varepsilon_{\text{mixed}} /S$, and
{\small
\be
\vec{C}_k = \left(
\begin{array}{cccccc}
2x^2 & z_+x_-&z_- x_- &z_+ x_-&z_-x_-&0\\ 
z_-x_-&3&0&\frac{x^2-2}{2} e^{i k_y}&0&-z_+x_-e^{ik_{xy}}\\ 
z_+x_-&0&3&0&\frac{x^2-2}{2} e^{-i k_x}&z_-x_-e^{i k_y}\\ 
z_-x_-&\frac{x^2-2}{2} e^{-i k_y}&0&3&0&-z_+x_-\\ 
z_+x_-&0&\frac{x^2-2}{2} e^{i k_x} &0&3&z_-x_- e^{i k_x}\\ 
0 & - z_- x_- e^{-i k_{xy}} & z_+ x_- e^{-i k_y} & - z_- x_- & z_+ x_- e^{-i k_x} & 2 x^2 
\end{array}
\right), 
\ee}
{\small 
\be
\vec{D}_k = \left(
\begin{array}{cccccc}
0& z_-x_+ &z_+x_+&z_-x_+ &z_+x_+&0\\ 
z_-x_+ &0&0&\frac{4-x^2}{2}e^{i k_y}&0&-z_+x_+e^{i k_{xy}}\\ 
z_+x_+&0&0&0&\frac{4-x^2}{2}e^{-i k_x}&z_-x_+e^{i k_y}\\ 
z_-x_+ &\frac{4-x^2}{2}e^{-i k_y}&0&0&0&-z_+x_+\\ 
z_+x_+&0&\frac{4-x^2}{2}e^{i k_x}&0&0&z_-x_+e^{i k_x}\\ 
0&-z_+x_+ e^{-i k_{xy}} &z_-x_+e^{-i k_y}&-z_+x_+&z_-x_+e^{-i k_x}&0 
 \end{array}
 \right)~. 
\ee}
\end{widetext}

\subsection{LSW Theory around the orthogonal phase in the effective model}
The unit cell of the effective model has two sites whose corresponding bosonic operators are labeled by $a_i$ and $f_i$.  
So the dimension of the corresponding matrices $\vec{C}_k$ and $\vec{D}_k$ is equal to 2.
For the theory around the orthogonal phase these matrices are given by  
{\small 
\be
\vec{C}_k \!=\! \left(
\begin{array}{cc}
4 (J_2-\tilde{K}) & J_1 p(k) \\
J_1 p(-k)&4 (J_2-\tilde{K})
\end{array}
\!\right)~,  
\ee
}
and 
{\small 
\be
\vec{D}_k \!=\! -
\left(
\begin{array}{cc}
2 (J_2-\tilde{K}) r(k) & J_1 p(k) \\
J_1 p(-k)&2 (J_2-\tilde{K}) r(k)
\end{array}
\!\right),   
\ee
}
where $\tilde{K}=K S^2$, $r(k)=\cos k_x+\cos k_y$, and $p(k)=\left( 1+e^{i k_x}+e^{i k_y} + e^{i (k_x+k_y)} \right) /2$. 
Finally, the classical GS energy per site is $\varepsilon_{\text{ortho}}' /S^2= -2 J_2 + \tilde{K}$, while $\frac{\delta E_1}{N} =  -2 J_2+2 \tilde{K}$.

\subsection{Non-linear Spin Wave Theory around the collinear phase of the effective model}\label{App:NLSWT}
The explicit forms of the 2$\times$2 matrices $\vec{C}_k$ and $\vec{D}_k$ that enter the mean-field decoupled quadratic theory of Sec.~\ref{Sec:NLSWT} are the following: 
\begin{widetext}
{\small
\be 
\vec{C}_k = \left(
\begin{array}{c c}
4 J_2 +4\tilde{K} (\lambda-2) + \xi_1/S & \Big( J_1 + \tilde{K}  (1-\lambda) +\xi_2/S \Big) (1+e^{i k_{xy}})  \\ 
\Big( J_1 + \tilde{K}  (1-\lambda) +\xi_2/S \Big) ( 1+e^{-i k_{xy}} ) & 4 J_2 +4\tilde{K} (\lambda-2) +\xi_1/S 
\end{array}
\right)~,
\ee
\be 
\vec{D}_k = \left(
\begin{array}{c c}
2(\tilde{K}-J_2+\xi_5/S ) \cos k_x + 2(\tilde{K}-J_2+\xi_4/S) \cos k_y  &\Big( -J_1 +\tilde{K} (1-\lambda) +\xi_3/S \Big) (e^{i k_x} + e^{i k_y})\\
\Big( -J_1 +\tilde{K} (1-\lambda) + \xi_3/S\Big) (e^{-i k_x} + e^{-i k_y}) &  2(\tilde{K}-J_2+\xi_5/S) \cos k_x + 2(\tilde{K}-J_2+\xi_4/S) \cos k_y 
 \end{array}
 \right)~,
\ee
}
where we have introduced the parameters 
\bea
&&
\xi_1 = 4 N_0 \Big(   3(2-\lambda)\tilde{K} - J_2 \Big) - 2 N_1\Big( 3\tilde{K}(1-\lambda)+J_1\Big) -2 L_1 \Big( 3\tilde{K} (1-\lambda)-J_1 \Big) +  4 L_2 (J_2-3\tilde{K}) \nonumber\\
&&
\xi_2 = N_1 \Big( (9-5\lambda)\tilde{K}  + J_1 \Big) - N_0 \Big( 3\tilde{K}(1-\lambda)+J_1 \Big) -4 L_1 \tilde{K} -2 (1-\lambda) \tilde{K} L_2 \nonumber\\
&&
\xi_3 = L_1 \Big( (9-5\lambda)\tilde{K} -J_1 \Big) -N_0 \Big( 3\tilde{K} (1-\lambda)-J_1 \Big) - 4 N_1 \tilde{K}-2 (1-\lambda)\tilde{K} L_2    \\
&&
\xi_4 = L_2  \Big( \tilde{K}(5-2\lambda) - J_2 \Big) + N_0 \Big(J_2-3\tilde{K} \Big) + 2\lambda \tilde{K} L_1 -2 \tilde{K} N_1 \nonumber\\
&&
\xi_5 = L_2 \Big( \tilde{K} (5-2\lambda) -J_2\Big)+N_0 \Big(J_2-3\tilde{K} \Big) +2\lambda \tilde{K} N_1 -2\tilde{K} L_1 \nonumber ~.
\eea
\end{widetext}

\section{QM problem of a single AFM dimer in a staggered field} \label{App:Dimer}
The minimization of the variational ansatz described in Sec.~\ref{Sec:Variational} showed that the 
4-fold spins remain coplanar and orthogonal to each other up to $x=2$. 
In this configuration the two exchange fields that are exerted on the two sites of each $J_{33}$-bond are 
antiparallel and have magnitude $h_s=\sqrt{2} S x$ (where we have used a length $S$ for the classical 4-fold spins). 
One then realizes that this problem is equivalent to that of an AFM dimer in the presence of a staggered field.  
A similar situation appears for the experimental compound Cu$_2$Cd(BO$_3$)$_2$ in Ref.~[\onlinecite{Janson}].

The Hamiltonian of an AFM dimer in a staggered field $h_s$ is given by
\be
\mc{H} = J ~\vec{S}_1 \cdot \vec{S}_2 -h_s (S_1^z - S_2^z) ~.
\ee 
Importantly the staggered field does not commute with the exchange interaction, and therefore it can polarize the system immediately. 
This is in contrast to the case where we have a uniform field, where one must exceed the singlet-triplet gap $J$ to polarize the system. 
The triplets $|t_1\rangle= |\!\uparrow\uparrow\rangle$ and $|t_{-1}\rangle=|\!\uparrow\downarrow\rangle$ are eigenstates of $\mc{H}$ with energy $J/4$, while 
the singlet $|s\rangle=\frac{|\!\uparrow\downarrow\rangle-|\!\downarrow\uparrow\rangle}{\sqrt{2}}$,   
and the triplet $|t_0\rangle=\frac{|\!\uparrow\downarrow\rangle+|\!\downarrow\uparrow\rangle}{\sqrt{2}}$ are admixed as follows: 
$\mc{H} |t_0\rangle = \frac{J}{4} |t_0\rangle -h_s |s\rangle$, and $\mc{H} |s\rangle = -\frac{3J}{4} |s\rangle -h_s |t_0\rangle$. 
A straightforward diagonalization in this manifold gives the following eigenstates and eigenvalues 
\bea
&& |\psi_1\rangle= u |s\rangle + v |t_0\rangle, ~~ \epsilon_1 = -J/4-\sqrt{J^2/4+h_s^2}  \label{Eq:e1} \\ 
&& |\psi_2\rangle= v |s\rangle - u |t_0\rangle, ~~  \epsilon_2 = -J/4+\sqrt{J^2/4+h_s^2}  \label{Eq:e2}~,
\eea
with $u=\cos\theta$, $v=\sin\theta$, and $\tan (2\theta) = 2h_s/J$.  As expected,  $|\psi_1\rangle\!\to\!|s\rangle$ for $h_s\!\to\!0$, while for 
$h_s\!\gg\!J$, $|\psi_1\rangle\!\to\!|\!\uparrow\downarrow\rangle$. 
The GS expectation values of the local polarizations and the exchange energy are given by
\bea
\langle \psi_1| S_{1,2}^z | \psi_1\rangle &=& \pm u v \\ 
\langle \psi_1 | \vec{S}_1\!\cdot\!\vec{S}_2 | \psi_1 \rangle  &=& -\frac{3}{4} u^2 + \frac{1}{4} v^2 ~.
\eea

\section{Quadratic fluctuations around the variational GS}\label{App:SpinBondWaves}
Here we provide the explicit form of the 8$\times$8 matrices $\vec{C}_k$ and $\vec{D}_k$  that appear in Eq.~(\ref{eq:Hmixed}).  
The various constants that appear below are defined as follows: 
$q=x\sqrt{S M}/4$, $y_\pm = q (v\pm u)$, $\xi_\pm = 1\pm 1/\sqrt{2}$.
\begin{widetext}
{\small
\bea
\vec{C}_k =  \left(
\begin{array}{ccccc}
\frac{4 x v u M}{\sqrt{2}} &0& -r (e^{i k_x} +1) &-(\xi_+ y_- e^{i k_x} +\xi_-y_+)&\xi_- y_+e^{i k_x}+\xi_+y_-   \\  
0&\frac{4 x v u M}{\sqrt{2}}& r (e^{-i k_y}+1)&-(\xi_+y_- e^{-i k_y}+ \xi_-y_+)&\xi_- y_+e^{-i ky}   +\xi_+y_-   \\
-r (e^{-i k_x} +1) & r (e^{i k_y}+1)& \epsilon_2-\epsilon_1 &0&0 \\
 -(\xi_+ y_- e^{-i k_x} +\xi_-y_+) &-(\xi_+y_- e^{i k_y}+ \xi_-y_+)&0& 1/4-\epsilon_1 & 0\\
\xi_- y_+e^{-i k_x}+\xi_+y_- &\xi_- y_+e^{i ky}   +\xi_+y_-&0&0& 1/4-\epsilon_1  \\
 r(e^{-i k_y}+1) & - r(e^{i k_x}+1)&0&0&0  \\
-(\xi_+y_- e^{-i k_y} + \xi_-y_+ )&- (\xi_-y_+ e^{i k_x} + \xi_+y_-)&0&0&0  \\
\xi_-y_+ e^{-i k_y} +\xi_+y_- &\xi_+y_- e^{i k_x} + \xi_-y_+&0&0&0 
\end{array}
\right) \nonumber
\eea

\bea
\ldots~~~\left(
\begin{array}{ccc}
 r(e^{i k_y}+1)&-(\xi_+y_- e^{i k_y} + \xi_-y_+)&\xi_-y_+ e^{i k_y} +\xi_+y_-  \\
 - r(e^{-i k_x}+1)&- (\xi_-y_+ e^{-i k_x} + \xi_+y_-)&\xi_+y_- e^{-i k_x} + \xi_-y_+  \\ 
0&0&0  \\ 
0&0&0 \\ 
0&0&0 \\
\epsilon_2-\epsilon_1 &0&0  \\
0& 1/4-\epsilon_1 &  0\\ 
0&0& 1/4-\epsilon_1 
\end{array}
\right)~,
\eea
}
and
{\small
\bea
\vec{D}_k = 
\left(
\begin{array}{ccccc}
0&0& -r (e^{i k_x}+1)&\xi_-y_- e^{i k_x} + \xi_+y_+&-(\xi_+y_+ e^{i k_x} + \xi_-y_-) \\
0&0& r (e^{-i k_y}+1)&\xi_-y_- e^{-i k_y} + \xi_+y_+&-(\xi_+y_+ e^{-i k_y} + \xi_-y_-) \\
 -r (e^{-i k_x}+1)& r (e^{i k_y}+1)&0&0&0 \\
\xi_-y_- e^{-i k_x} + \xi_+y_+&\xi_-y_- e^{i k_y} + \xi_+y_+&0&0&0  \\
-(\xi_+y_+ e^{-i k_x} + \xi_-y_-)&-(\xi_+y_+ e^{i k_y} + \xi_-y_-)&0&0&0 \\
 r (e^{-i k_y}+1)& -r (e^{i k_x}+1)&0&0&0 \\
\xi_-y_- e^{-i k_y} + \xi_+y_+&\xi_+y_+ e^{i k_x} + \xi_-y_-&0&0&0  \\
-(\xi_+y_+ e^{-i k_y} + \xi_-y_-)&-(\xi_-y_- e^{i k_x} + \xi_+y_+)&0&0&0 
\end{array}
\right)\nonumber
\eea

\bea
\ldots~~~
\left(
\begin{array}{ccc}
 r (e^{i k_y}+1)&\xi_-y_- e^{i k_y} + \xi_+y_+&-(\xi_+y_+ e^{i k_y} + \xi_-y_-) \\
- r (e^{-i k_x}+1)&\xi_+y_+ e^{-i k_x} + \xi_-y_-&-(\xi_-y_- e^{-i k_x} + \xi_+y_+) \\
0&0&0 \\
0&0&0 \\
0&0&0 \\
0&0&0 \\
0&0&0 \\
\end{array}
\right)~. 
\eea
}
\end{widetext}


\begin{table*}[!t]\caption{Local effective terms that are generated up to fourth order in the strong coupling expansion in $x=J_{43}/J_{33}$. 
Only the nine clusters that give finite terms are shown (cf. first column). Dashed (solid) lines denote $J_{43}$ ($J_{33}$) couplings. The last three columns show the 
total contribution of each cluster to $J_1$, $J_2$ and the amplitude $K$ of the plaquette term of Eq.~(\ref{Kp2.Eq}). }
\begin{tabular}{ccccccc}
\toprule[0.7pt] 
\multicolumn{1}{|c|}{ Cluster }
&\multicolumn{1}{|c}{ 2nd }
&\multicolumn{1}{|c}{ 3rd } 
&\multicolumn{1}{|c|}{ 4th }
&\multicolumn{1}{|c}{ contr. to $J_1$ }
&\multicolumn{1}{|c}{ contr. to $J_2$ }
&\multicolumn{1}{|c|}{ contr. to $K$ }\\
\midrule[0.8pt] 
\multicolumn{1}{|c|}{ \parbox{1.8in}{\includegraphics[clip=true,scale=0.18]{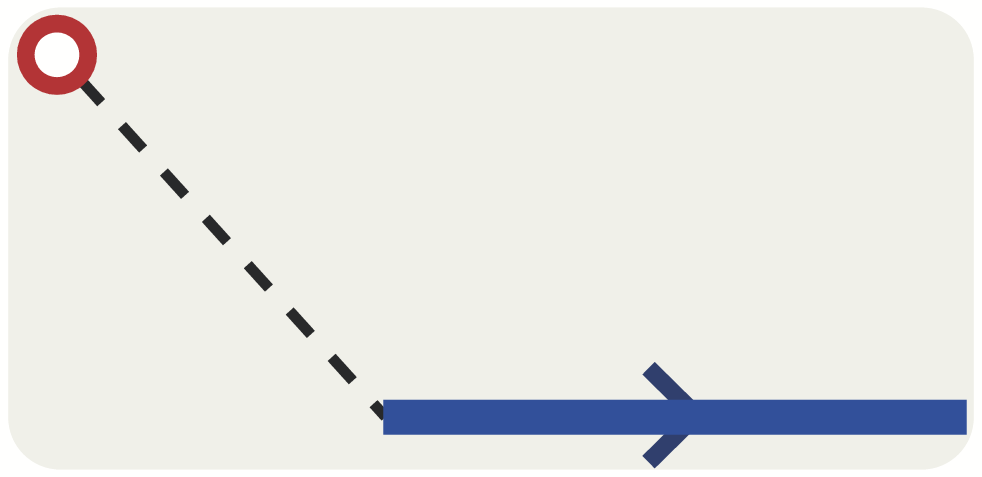}} }
&\multicolumn{1}{|c}{ $-\frac{3}{16}$ }
&\multicolumn{1}{|c}{ $-\frac{3}{32}$  }
&\multicolumn{1}{|c|}{ $-\frac{3}{256}$ } 
&\multicolumn{1}{|c}{}
&\multicolumn{1}{|c}{}
&\multicolumn{1}{|c|}{}\\
\hline
\multicolumn{1}{|c|}{\parbox{1.8in}{\includegraphics[clip=true,scale=0.18]{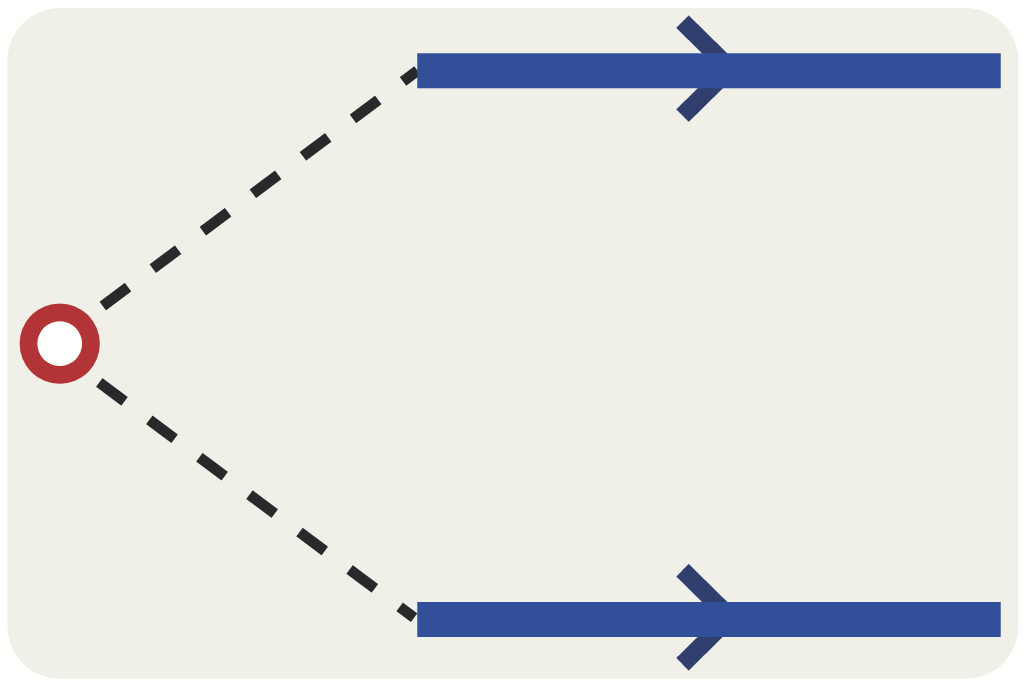}} }
&\multicolumn{1}{|c}{} 
&\multicolumn{1}{|c}{} 
&\multicolumn{1}{|c|}{ $\frac{3}{64}$  }
&\multicolumn{1}{|c}{}
&\multicolumn{1}{|c}{}
&\multicolumn{1}{|c|}{}\\
\hline
\multicolumn{1}{|c|}{\parbox{1.8in}{\includegraphics[clip=true,scale=0.18]{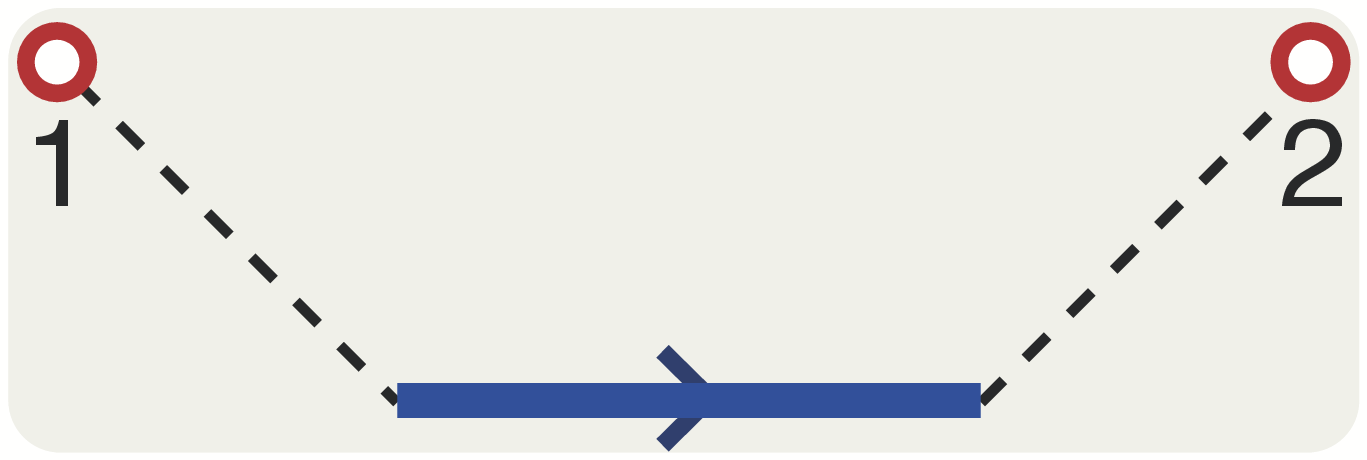}} }
&\multicolumn{1}{|c}{ $\frac{1}{2} \vec{s}_1 \!\cdot\! \vec{s}_2$  }
&\multicolumn{1}{|c}{ $\frac{3}{4} \vec{s}_1 \!\cdot\! \vec{s}_2$  }
&\multicolumn{1}{|c|}{ $\frac{1}{4}\vec{s}_1 \!\cdot\! \vec{s}_2 + \frac{3}{128} $  } 
&\multicolumn{1}{|c}{ $\frac{1}{2}x^2+\frac{3}{4}x^3+\frac{1}{4}x^4$ }
&\multicolumn{1}{|c}{ $\frac{1}{2}x^2+\frac{3}{4}x^3+\frac{1}{4}x^4$}
&\multicolumn{1}{|c|}{}\\
\hline 
\multicolumn{1}{|c|}{ \parbox{1.8in}{\includegraphics[clip=true,scale=0.18]{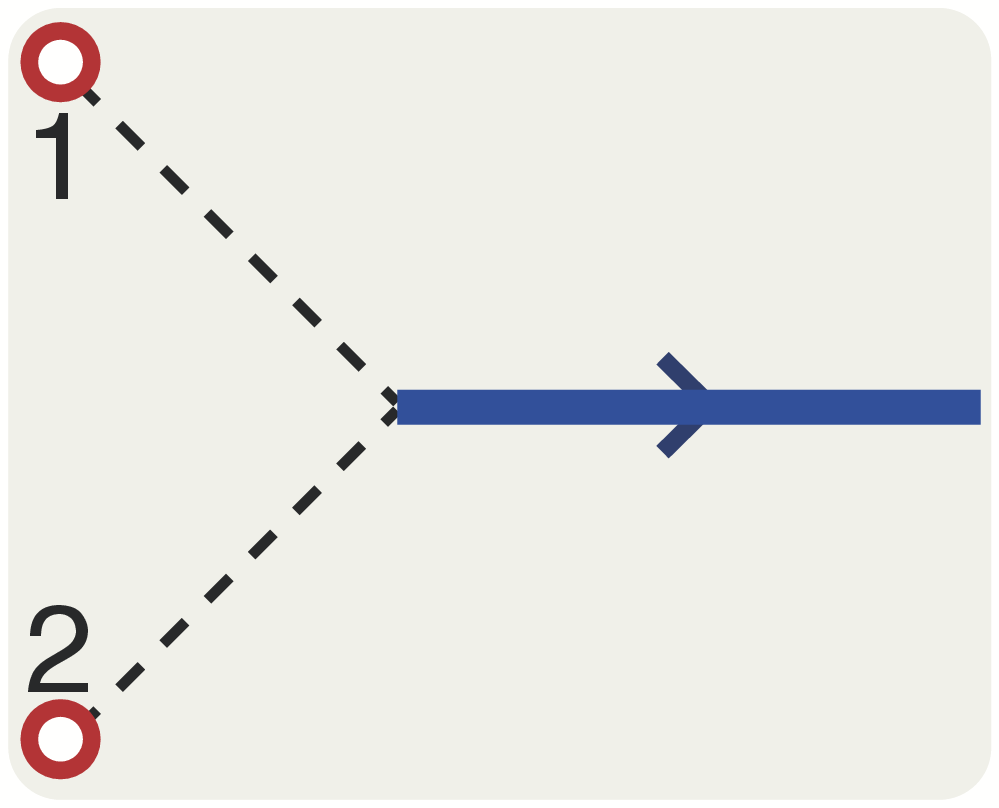}} } 
&\multicolumn{1}{|c}{ $-\frac{1}{2} \vec{s}_1 \!\cdot\! \vec{s}_2$ }
&\multicolumn{1}{|c}{ $-\frac{1}{4} \vec{s}_1 \!\cdot\! \vec{s}_2$ }
&\multicolumn{1}{|c|}{ $\frac{1}{8}\vec{s}_1 \!\cdot\! \vec{s}_2+\frac{15}{128}$ }  
&\multicolumn{1}{|c}{ $-\frac{1}{2}x^2-\frac{1}{4}x^3+\frac{1}{8}x^4$ }
&\multicolumn{1}{|c}{} 
&\multicolumn{1}{|c|}{} \\ 
\hline 
\multicolumn{1}{|c|}{ \parbox{1.8in}{\includegraphics[clip=true,scale=0.18]{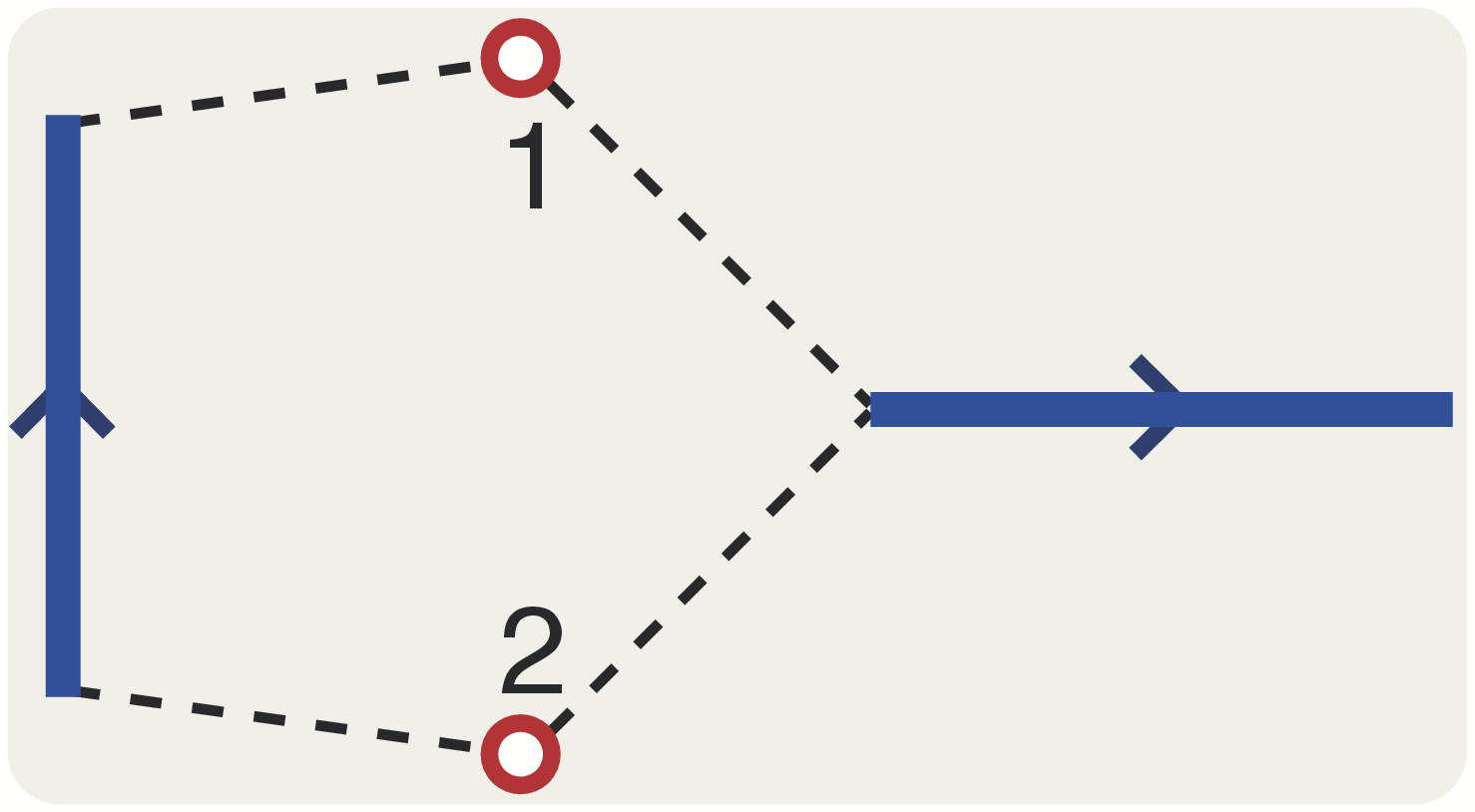}} }
&\multicolumn{1}{|c}{} 
&\multicolumn{1}{|c}{} 
&\multicolumn{1}{|c|}{ $\frac{3}{8}\vec{s}_1 \!\cdot\! \vec{s}_2$  }
&\multicolumn{1}{|c}{ $\frac{3}{8}x^4$ }
&\multicolumn{1}{|c}{}
&\multicolumn{1}{|c|}{}\\
\hline
\multicolumn{1}{|c|}{ \parbox{1.8in}{\includegraphics[clip=true,scale=0.18]{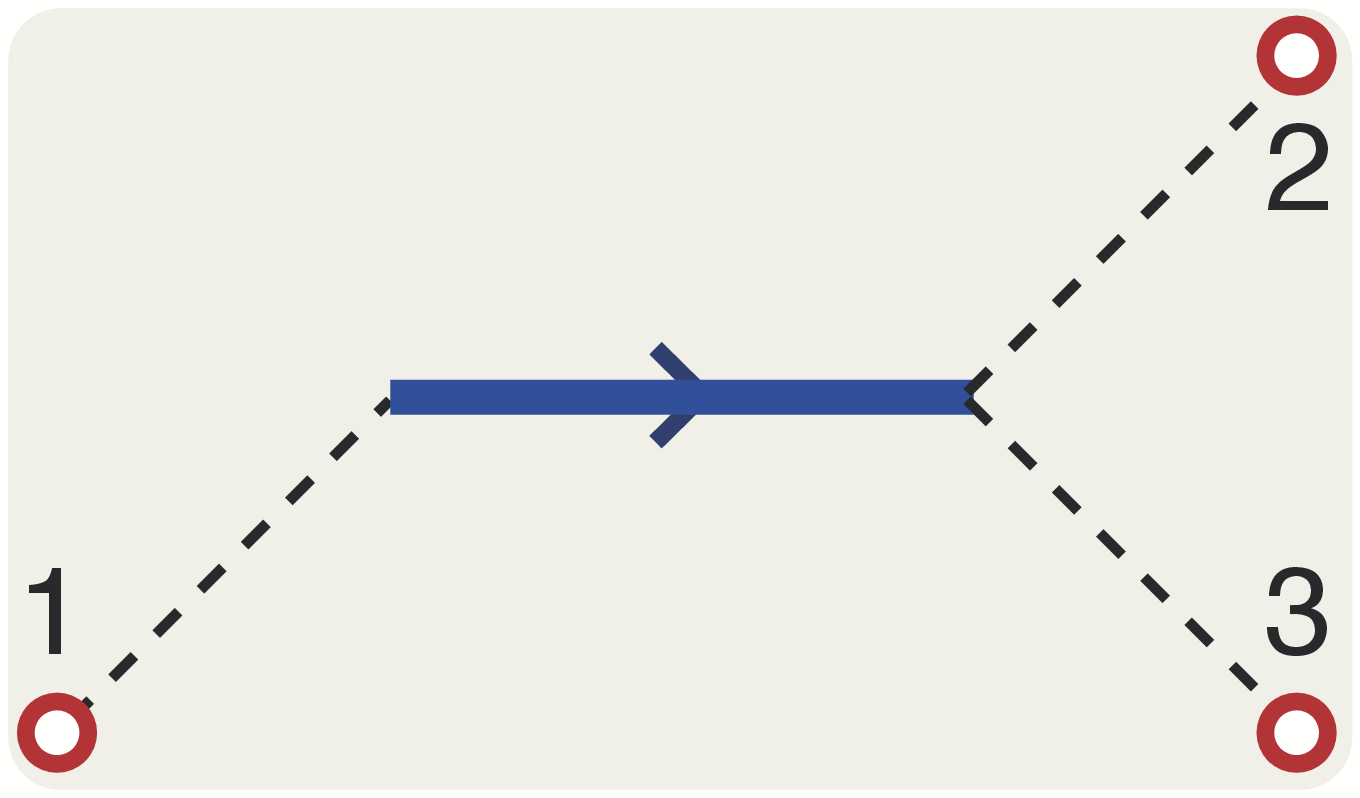}} }
&\multicolumn{1}{|c}{} 
&\multicolumn{1}{|c}{} 
&\multicolumn{1}{|c|}{
\parbox{1in}{\begin{eqnarray*}
+\!&\frac{1}{16}& \! \vec{s}_2 \!\cdot\! \vec{s}_3 \\
-\!&\frac{5}{16}& \! \vec{s}_1 \!\cdot\! \left( \vec{s}_2+ \vec{s}_3 \right) 
\end{eqnarray*}}  
}
&\multicolumn{1}{|c}{ $2 \times (\frac{1}{16}-\frac{5}{16} ) x^4$ }
&\multicolumn{1}{|c}{ $2 \times (-\frac{5}{16}) x^4$ }
&\multicolumn{1}{|c|}{}\\
\hline 
\multicolumn{1}{|c|}{ \parbox{1.8in}{\includegraphics[clip=true,scale=0.18]{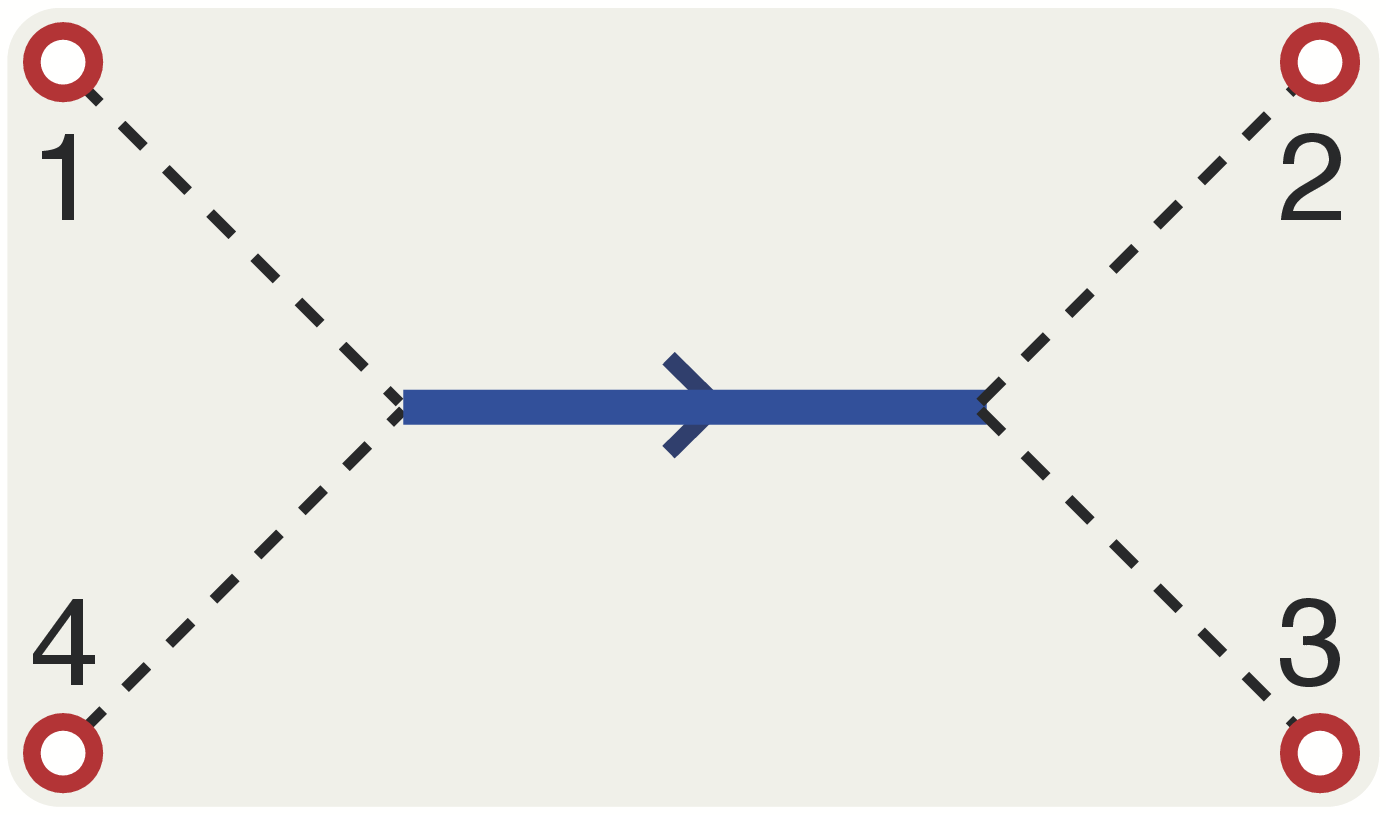}} }
&\multicolumn{1}{|c}{} 
&\multicolumn{1}{|c}{} 
&\multicolumn{1}{|c|}{ 
\parbox{1in}{\begin{eqnarray*} 
+( \vec{s}_1 \!\cdot\! \vec{s}_2 ) ( \vec{s}_3 \!\cdot\! \vec{s}_4 ) \\
+( \vec{s}_1 \!\cdot\! \vec{s}_3 ) ( \vec{s}_2 \!\cdot\! \vec{s}_4 ) \\
-\frac{1}{2}( \vec{s}_1 \!\cdot\! \vec{s}_4 ) ( \vec{s}_2 \!\cdot\! \vec{s}_3 ) 
\end{eqnarray*}}
}
&\multicolumn{1}{|c}{}
&\multicolumn{1}{|c}{}
&\multicolumn{1}{|c|}{ $x^4$ }\\
\hline
\multicolumn{1}{|c|}{\parbox{1.8in}{\includegraphics[clip=true,scale=0.18]{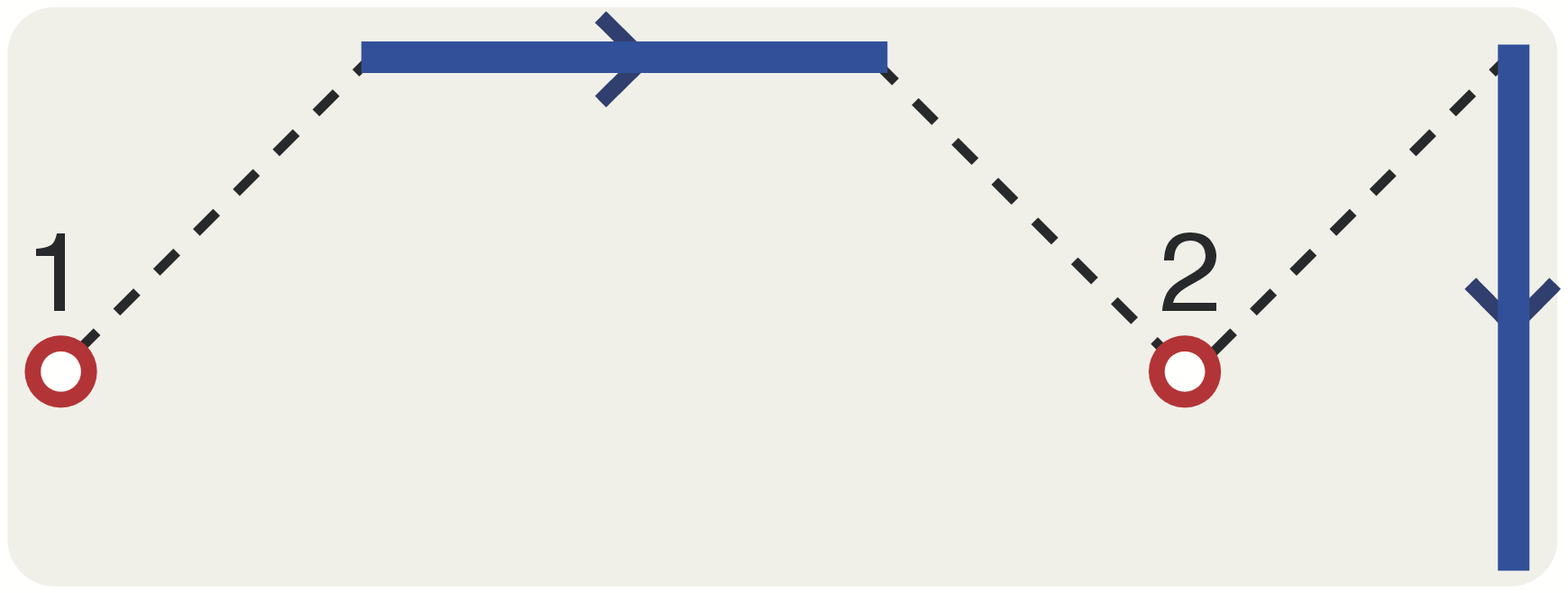}} }
&\multicolumn{1}{|c}{} 
&\multicolumn{1}{|c}{} 
&\multicolumn{1}{|c|}{ $-\frac{1}{8} \vec{s}_1 \!\cdot\! \vec{s}_2$ }
&\multicolumn{1}{|c}{ $6\times (-\frac{1}{8}) x^4$ }
&\multicolumn{1}{|c}{ $6\times (-\frac{1}{8}) x^4$ }
&\multicolumn{1}{|c|}{} \\ 
\hline
\multicolumn{1}{|c|}{ \parbox{1.8in}{\includegraphics[clip=true,scale=0.18]{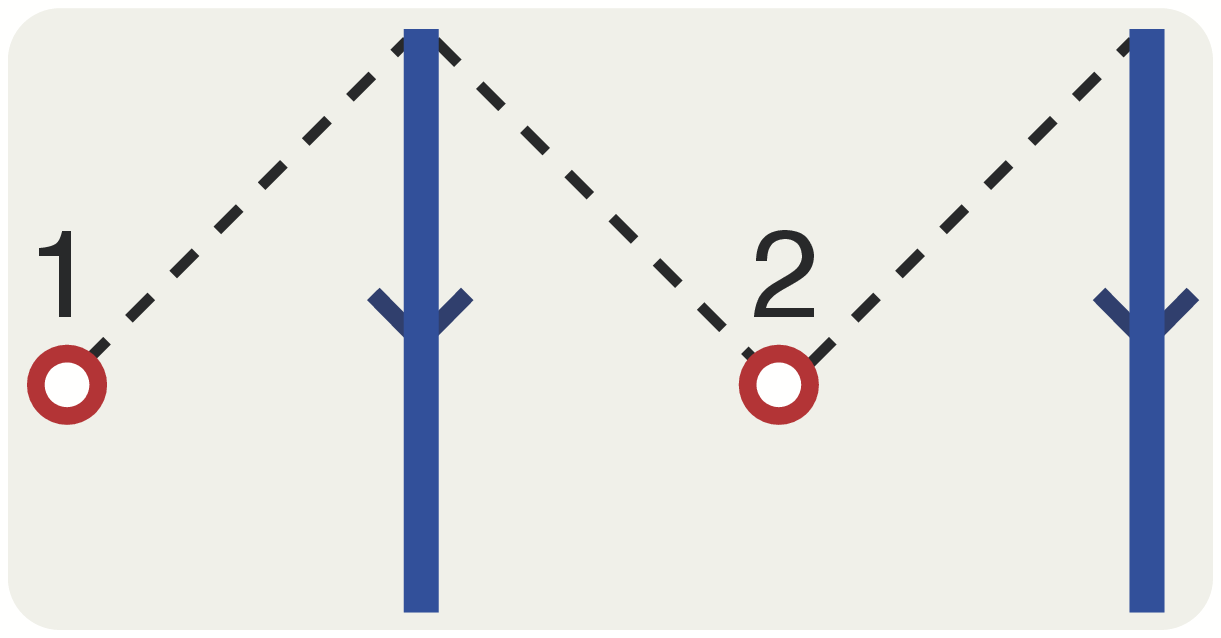}} }
&\multicolumn{1}{|c}{} 
&\multicolumn{1}{|c}{} 
&\multicolumn{1}{|c|}{ $\frac{1}{8} \vec{s}_1 \!\cdot\! \vec{s}_2$}
&\multicolumn{1}{|c}{ $6\times \frac{1}{8} x^4$ }
&\multicolumn{1}{|c}{} 
&\multicolumn{1}{|c|}{} \\ 
\bottomrule[0.7pt]
&&& 
&\multicolumn{1}{|c}{} 
&\multicolumn{1}{|c}{} 
&\multicolumn{1}{|c|}{} \\
&&& Total contributions: 
&\multicolumn{1}{|c}{ $\frac{1}{2}x^3+\frac{1}{4}x^4$} 
&\multicolumn{1}{|c}{ $\frac{1}{2}x^2+\frac{3}{4}x^3-\frac{9}{8}x^4$ }
&\multicolumn{1}{|c|}{ $x^4$} \\ 
&&&
&\multicolumn{1}{|c}{} 
&\multicolumn{1}{|c}{} 
&\multicolumn{1}{|c|}{} \\ \cline{5-7}
\end{tabular}
\label{T1}
\end{table*}

\section{Strong coupling expansion at small $x=J_{43}/J_{33}$}\label{App:PertTheory}
Here we provide some details of the degenerate perturbation theory around $x=0$ discussed in Sec.~\ref{Sec:EffModel}.
For $x=0$, the nearest-neighbor 3-fold sites form isolated singlets and the 4-fold sites are free to point up or down. For finite $x$, the 4-fold sites 
begin to interact with each other through the virtual fluctuations of the $J_{33}$-bonds out of the singlet manifold.  
The resulting effective Hamiltonian acts on the 4-fold sites only and can be obtained by degenerate perturbation theory around $x=0$.
This expansion can be organized in a linked cluster way as for non-degenerate perturbation theory.\cite{Oitmaa} 
The effective Hamiltonian is a sum of contributions from a sequence of finite clusters of sites. For each given order of perturbation theory only 
clusters up to a given size may contribute.

More explicitly, an $n$-th order process that lives on a given cluster $g$ can be obtained by 
\be
\mc{H}_{\text{eff}}^{(n)} \!=\! \sum_{r_1,\ldots,r_{n-1} \ge 0}^{r_1+\cdots+r_{n-1}=n-1} \!c_{\{r\}}~ \! \mc{P} V\mc{R}^{(r_1)} V\cdots V\mc{R}^{(r_{n-1})} V\mc{P} ~.
\ee 
Here $\mc{P}$ is the projection in the unperturbed manifold,  $\mc{Q}=1-\mc{P}$, 
$\mc{R}_0 = \left( E_0-\mc{Q}\mc{H}_0\mc{Q} \right)^{-1}$ is the resolvent operator, $E_0$ is the unperturbed energy, 
$\mc{R}^{(0)}=-\mc{P}$, while $\mc{R}^{(r\ge 1)}= \left(\mc{R}_0\right)^r$. 
Finally, the coefficients $c_{\{r\}}$ can be found tabulated (up to sixth order) in Ref.~[\onlinecite{Klein}].    

We have performed the above expansion up to fourth order in $x$ using all relevant clusters. 
In order to avoid having to subtract processes that live on the subclusters of $g$,\cite{Oitmaa} 
we enforce that we only keep processes that invoke all elements of the cluster $g$.     
The first column of Table \ref{T1} shows all clusters that generate a finite interaction between the 4-fold sites (written explicitly in columns 2-4).
By including all possible ways that we can embed each cluster on the lattice one obtains the corresponding contribution to the 
effective couplings $J_1$, $J_2$ and $K$ (columns 5-7).

{\it Polarization on the 3-fold sites}:---
The above effective description lives in a projected Hilbert space where pairs of n.n. 3-fold sites pair-up forming exact singlet wavefunctions.  
However the true GS of the problem has also a non-vanishing component on the orthogonal manifold. This component must be taken into account if we want to 
find e.g. the polarization on the 3-fold sites. Specifically, as soon as the 4-fold sites order magnetically they will exert a finite exchange field on the 3-fold sites. 
For example, in the orthogonal phase the total exchange field will be staggered in all dimers and thus it will immediately admix 
a triplet component into the singlet GS and give rise to a finite (staggered) polarization. 
This is in contrast to the uniform field case where one must exceed a critical value (the singlet-triplet gap $J$) in order to polarize an AFM dimer.  

The component of the GS wavefunction out of the singlet manifold $\mc{Q} |\Psi\rangle$ can be expressed as
\be
\mc{Q} |\Psi\rangle = \mc{R} V \mc{P} |\Psi\rangle,
\ee 
where  $\mc{R}= \left( E- \mc{Q} \mc{H} \mc{Q} \right)^{-1}$ is the full resolvent. Thus 
\be
|\Psi\rangle= \left( 1+\mc{R} V \right)  |\mc{P} \Psi\rangle= \sqrt{z_0} \left( 1+\mc{R} V \right)  |\Psi^0\rangle,
\ee
where $|\Psi^0\rangle$ is the normalized GS of the effective model (times the product of $J_{33}$-singlets), 
and $z_0=1/(1+\langle \Psi^0| V \mc{R}^2 V |\Psi^0\rangle)$ plays the role of a ``wavefunction renormalization factor" 
(specifically $1-z_0$ measures the degree of admixture inside the GS from states outside the unperturbed GS manifold). 

Now, the GS expectation value of the magnetization say at site $1$ of a given dimer is then given by
\be\label{eqn:s1a}
\langle S_1^\alpha \rangle = 2 ~z_0 ~\text{Re} ~ \langle \Psi^0 |  S_1^\alpha \mc{R} V |\Psi^0 \rangle .
\ee  
To lowest order in $V$ we may replace $z_0 \simeq 1$ and $\mc{R} \simeq \mc{R}_0$, which leads to the standard expression from linear response theory. 
In particular, the result coincides with the linear order contribution found for the problem of an AFM dimer in a staggered field studied in App.~\ref{App:Dimer}.
 
We should remark here that in the collinear phase the total exchange field vanishes in half of the $J_{33}$-dimers, see e.g. the site labeled by $c_i$ in Fig.~\ref{Fig:methods}(b).  
So the polarization in these dimers must grow quadratically and not linearly with $x$.

\section{Strong-coupling approach in the $\mc{K}_x$-model}\label{App:SPE} 
Here we give the details of the strong-coupling expansion around the limit where $1/4$ of the plaquettes are decoupled from each other.   
Following the notation of Fig.~\ref{Fig:PlaqSqLattice}, an isolated strong plaquette has the two triplet GS's with energy -3/16:  
\be
|\psi_1^m\rangle = |s\rangle_{\alpha\gamma} \otimes |t^m\rangle_{\beta\delta}~,~~~ |\psi_2^m\rangle =|t^m\rangle_{\alpha\gamma} \otimes |s\rangle_{\beta\delta}~,
\ee
where $|s\rangle$ is the singlet and $|t^m\rangle$ are the three S=1 triplets ($m=\pm 1, 0$). 
So we have a GS manifold with degeneracy $6^{N_p}$, where $N_p=N/4$ is the number of the strong plaquettes. 
The splitting of this manifold by the inter-plaquette $\mc{K}_x$ terms can be captured by first order degenerate perturbation theory. 
The resulting effective model can be written in a convenient form by introducing a pseudospin $\tau=1/2$ and a spin $T=1$ object in each strong plaquette. 
The direction of the pseudospin denotes which of the two types of triplets is taken, while the spin $T=1$ object carries the physical spin of the plaquette.   
This approach follows closely in spirit the treatment by Lecheminant and Totsuka\cite{Lecheminant} of a very similar situation where a different pair of plaquette triplets emerges at 
low energies in a two-leg ladder model. 
\begin{figure}[bp]
\includegraphics[clip=true,width=0.7\linewidth]{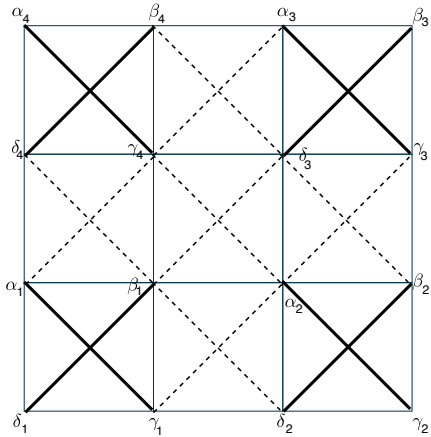}
\caption{\label{Fig:PlaqSqLattice} The plaquettized version of the square lattice. Thick diagonal bonds denote the strong $\mc{K}_x$-terms, 
while dashed ones are treated with first-order perturbation theory.}
\end{figure}
The inter-plaquette interactions are always of the type $\left( \vec{S}_\alpha \cdot \vec{S}_\gamma \right) \left( \vec{S}_\beta \cdot \vec{S}_\delta \right)$
where $\alpha, \beta, \gamma, \delta$ do not all belong to the same strong plaquette. 
In particular there are two-plaquette interactions (see fig.~\ref{Fig:PlaqSqLattice})
\bea
V_{1x} &=& \left( \vec{S}_{\alpha'} \cdot \vec{S}_{\gamma} \right) \left( \vec{S}_\beta \cdot \vec{S}_{\delta'} \right)  = \sum_{n,m} S_\beta^n S_\gamma^m ~S_{\alpha'}^m S_{\delta'}^n \\
V_{1y} &=& \left( \vec{S}_\alpha \cdot \vec{S}_{\gamma'} \right) \left( \vec{S}_\beta \cdot \vec{S}_{\delta'} \right)= \sum_{n,m} S_\beta^n S_\alpha^m ~S_{\gamma'}^m S_{\delta'}^n~,
\eea
and 4-plaquette interactions (see Fig.~\ref{Fig:PlaqSqLattice})
\bea
V_2 = \left( \vec{S}_\alpha \cdot \vec{S}_{\gamma'} \right) \left( \vec{S}_{\beta''} \cdot \vec{S}_{\delta'''} \right). 
\eea
So in the following we shall treat these interactions with first order perturbation theory, i.e. we shall write down an effective Hamiltonian 
which is formally given by
\be
\mc{H}_{\text{eff}} = P~ V ~P ~,
\ee 
where $P=\prod_{p} P_p$ is the projection operator that projects into the GS manifold, and
\be
P_p = \sum_{m=\pm 1, 0}  \left( |\psi_1^m\rangle\langle\psi_1^m| +  |\psi_2^m\rangle\langle\psi_2^m| \right)_p ~.
\ee
To continue we introduce a pseudospin $\tau=1/2$, whose direction specifies the type of the triplet, 
and a spin $T=1$ object which carries the total physical spin of the plaquette. This 
defines the following mapping 
\be
|\psi_1^m\rangle \rightarrow  |\!\upa\rangle \otimes |T^m\rangle~,~~~ |\psi_2^m\rangle \rightarrow  |\!\dna\rangle \otimes |T^m\rangle~.
\ee
It is straightforward to show that the pseudospin operators are the following scalar operators of the original spins of the plaquette: 
\bea
\tau_z &=& \frac{1}{2} \left(  -\vec{S}_\alpha \cdot \vec{S}_{\gamma} + \vec{S}_{\beta} \cdot \vec{S}_{\delta} \right) \\
\tau_x &=& \vec{S}_\alpha \cdot \vec{S}_{\beta} + \vec{S}_{\gamma} \cdot \vec{S}_{\delta} = \vec{S}_\alpha \cdot \vec{S}_{\delta} + \vec{S}_{\beta} \cdot \vec{S}_{\gamma} \\
\tau_y &=& \vec{N}_{\beta\delta} \cdot \vec{S}_\alpha \times \vec{S}_\gamma + \vec{N}_{\alpha\gamma} \cdot \vec{S}_\beta \times \vec{S}_\delta \\
{\bf 1}  &=& -2 \left( \vec{S}_\alpha \cdot \vec{S}_{\gamma} + \vec{S}_{\beta} \cdot \vec{S}_{\delta} \right)
\eea
where $\vec{N}_{\alpha\gamma}=\vec{S}_\alpha-\vec{S}_\gamma$ and $\vec{N}_{\beta\delta}=\vec{S}_\beta-\vec{S}_\delta$. 

We now turn to the operators that are needed for the derivation of the effective Hamiltonian. The following relations hold for single-site operators 
\begin{eqnarray*}
&& P~\vec{S}_\alpha~P = P~\vec{S}_\gamma~P = |\!\dna\rangle \langle\dna\!| \otimes \frac{1}{2} \vec{T} = (\frac{1}{2}-\tau_z) \otimes \frac{1}{2}\vec{T}\\
&& P~\vec{S}_\beta~P = P~\vec{S}_\delta~P = |\!\upa\rangle \langle\upa\!| \otimes \frac{1}{2} \vec{T} =(\frac{1}{2}+\tau_z) \otimes \frac{1}{2}\vec{T}~.
\end{eqnarray*}
For operators on the bond ($\beta,\gamma$) we find: 
\bea
P ( S_\beta^n S_\gamma^m  +S_\beta^m S_\gamma^n ) P &=& \frac{1}{2} \tau_x \otimes (Q^{nm}-\frac{2}{3} \delta^{nm}) \\
P ( \vec{S}_\beta \times \vec{S}_\gamma ) P &=& \frac{1}{2} \tau_y \otimes \vec{T} 
\eea
where we have introduced the quadrupolar tensor of the spin $T=1$ object:   
\be
Q^{nm} = T^n T^m + T^m T^n -\frac{2}{3}T(T+1)\delta^{nm} ~.
\ee 
Similarly for the bond $(\alpha,\delta)$: 
\bea
P ( S_\alpha^n S_\delta^m  +S_\alpha^m S_\delta^n ) P &=& \frac{1}{2} \tau_x \otimes (Q^{nm}-\frac{2}{3} \delta^{nm}) \\
P ( \vec{S}_\alpha \times \vec{S}_\delta)   P &=& -\frac{1}{2} \tau_y \otimes \vec{T} 
\eea
and for the bond ($\alpha,\beta$): 
\bea
P ( S_\alpha^n S_\beta^m  +S_\alpha^m S_\beta^n ) P &=& -\frac{1}{2} \tau_x \otimes (Q^{nm}-\frac{2}{3} \delta^{nm}) \\
P ( \vec{S}_\alpha \times \vec{S}_\beta)   P &=& \frac{1}{2} \tau_y \otimes \vec{T} ~.
\eea
Finally, the corresponding mappings for the bond ($\gamma,\delta$) are the same with that of ($\alpha,\beta$).

Using the above mappings we may write down the first order effect from the inter-plaquette interactions. We have
\be\label{form1}
P~V_{1x}~P = \frac{1}{8} \tau_x \tau_x' \otimes \left( \vec{Q} \cdot \vec{Q}' + \frac{2}{3}\right) + \frac{1}{8} \tau_y \tau_y' \otimes \vec{T}\cdot\vec{T}'~,
\ee
where the vector $\vec{Q}$ contains the following elements of the quadrupolar tensor\cite{Penc}
\be
\vec{Q}^T = \left(
\begin{array}{ccccc}
\frac{1}{2}(Q^{xx}-Q^{yy}),  
\frac{\sqrt{3}}{2} Q^{zz}, 
Q^{xy}, 
Q^{yz}, 
Q^{xz}
\end{array}
\right) ~.
\ee
An equivalent way to write the above interaction is  
\be\label{form2}
P  V_{1x} P = \frac{3}{8} ( \tau_x \tau_x' - \tau_y \tau_y' )  \otimes \Pi_{S=0} + \frac{1}{8} ( \tau_x \tau_x' + \tau_y \tau_y' ) \otimes P_{ij} ~,
\ee
where 
\be
\Pi_{S=0} = \frac{1}{6} ( \vec{Q} \cdot \vec{Q}' + \frac{2}{3} - \vec{T}\cdot\vec{T}' )~,
\ee 
is the projector into the total singlet state, and 
\be 
P_{ij} = \frac{1}{2} (\vec{Q} \cdot \vec{Q}' + \frac{2}{3} + \vec{T}\cdot\vec{T}' )=-2 ~\Pi_{S=1} +1~,
\ee
is the permutation operator which switches the states of two triplets.\cite{Penc}

To switch from $V_{1x}$ to $V_{1y}$, we need to map $\alpha\leftrightarrow\gamma$ in both plaquettes, which 
corresponds to $|\!\upa\rangle \rightarrow -|\!\upa\rangle$, and thus $\tau_{\pm}\rightarrow -\tau_{\pm}$ and $\tau_{x,y} \rightarrow -\tau_{x,y}$.  
Since we need to do this in both plaquettes the minus signs cancel each other and thus $P ~V_{1y}~P$ has exactly thesame form as $P ~V_{1x}~P$. 

The remaining 4-plaquette interaction term, denoted by $V_2$ above, gives  
\bea
P~V_{2}~P &=& (\frac{1}{2}-\tau_z)~(\frac{1}{2}-\tau_z') ~(\frac{1}{2}+\tau_z'')~(\frac{1}{2}+\tau_z''') \nonumber\\
&\otimes&  \frac{1}{2^4} ( \vec{T}\cdot \vec{T}' )~ (\vec{T}''\cdot \vec{T}''' )~.
\eea
It is interesting to note that the spin-1 portion of this interaction looks exactly the same with the original four-spin interaction $\mc{K}_x$,  
but now the four sites have spin 1 and not 1/2.

\subsection{Symmetries}
The original Hamiltonian has an SU(2)$\times$SU(2) symmetry, i.e. we can rotate all spins labeled by $\alpha$ and $\gamma$ independently from the spins 
$\beta$ and $\delta$.\footnote{We should note here that the unperturbed part $\mc{H}_0$ has a much higher symmetry since we can do independent SU(2)$\times$SU(2) rotations in every strong plaquette.} The corresponding generators are the total spins of the two sublattices 
\be
\vec{S}_A = \sum_{p=1}^{N_p} \vec{S}_{\alpha\gamma}(p), ~~~\vec{S}_B = \sum_{p=1}^{N_p} \vec{S}_{\beta\delta}(p)
\ee
and the group elements are parametrized by two vectors $\vec{\Omega}_A$ and $\vec{\Omega}_B$: 
\be
R(\vec{\Omega}_A,\vec{\Omega}_B) = R_a (\vec{\Omega}_A)~R_b (\vec{\Omega}_B)
= e^{-i (\vec{S}_A \cdot \vec{\Omega}_A + \vec{S}_B \cdot \vec{\Omega}_B)}~, 
\ee
or, more explicitly $R (\vec{\Omega}_A,\vec{\Omega}_B) = \prod_{p} R_p(\vec{\Omega}_A,\vec{\Omega}_B)$, with
\be
R_p(\vec{\Omega}_A,\vec{\Omega}_B) = |\!\upa\rangle_p\langle\upa\!|\otimes e^{-i \vec{T}_p\cdot\vec{\Omega}_B} 
+ |\!\dna\rangle_p\langle\dna\!| \otimes e^{-i \vec{T}_p \cdot \vec{\Omega}_A} ~. 
\ee
As expected, all the effective terms written above retain the SU(2)$\times$SU(2) invariance of the original $\mc{K}_x$-model.   
To see this e.g. for the term $P~V_{1x}~P$, it is convenient to make use of Eq.~(\ref{form2}) rather than Eq.~(\ref{form1}).

\subsection{Quantum-mechanical solution for two neighboring plaquettes}
For two neighboring plaquettes $p_1=(1234)$ and $p_2=(5678)$ and for both open and periodic boundary conditions the QM GS is the following: 
\be
|\psi\rangle = \frac{1}{\sqrt{2}} ( |\upa_{1}\upa_{2}\rangle - |\dna_{1}\dna_{2}\rangle ) \otimes | T_{1}=1, T_{2}=1, T=0\rangle~.
\ee 
The GS energy is -3/16, i.e. $|\psi\rangle$ minimizes fully the interplaquette interaction.  
We should note here that in this wavefunction, the spin-1 objects make a singlet with $\langle\vec{T}_{1}\cdot\vec{T}_{2}\rangle=-2$,  
and $\langle\vec{Q}_{1}\cdot\vec{Q}_{2}+2/3\rangle =4$. 
The latter value is twice the one we would get by using a product state of two parallel director wavefunctions.

\subsection{Classical Variational solution to $V_1^{\text{eff}} = P(V_{1x}+V_{1y})P$ }\label{App:SP}  
In the following we shall present a variational treatment of the terms $V_{1x}$ and $V_{1y}$, disregarding $P~V_2~P$. 
The variational wavefunction is a product of plaquette wave-functions with a pseudospin portion and a spin-1 portion. 
The pseudospin-1/2 portion is treated classically, i.e. it is parametrized by two direction angles $\theta, \phi$. 
On the other hand, the spin-1 portion is parametrized by a complex vector $\vec{d}=\vec{u}+i \vec{v}$, 
with the constraints $u^2+v^2=1$, and $\vec{u}\cdot\vec{v}=0$.\cite{Penc} 

A numerical minimization of this variational state delivers a GS in which the $\tau=1/2$ pseudospins order AFM 
in a N\'eel state with their moments along the x-axis, while the spin-1 objects develop a ferro-quadrupolar ordering without any dipolar moment.  
Choosing the common director along the $z$-axis, we can write this state explicitly as
\be\label{Eqn:zz}
|\vec{z},\vec{z}\rangle \equiv \prod_{p=1}^{N_p} \left( |s\rangle_{\alpha\gamma}\otimes |\vec{z}\rangle_{\beta\delta} + (-1)^{\vec{Q}\cdot \vec{R}_p} |\vec{z}\rangle_{\alpha\gamma}\otimes |s\rangle_{\beta\delta} \right) ~,
\ee
where $\vec{Q}=(\pi,\pi)$ is the N\'eel ordering wavevector of the pseudospins.  
We should note here that one can actually generate a continuous family of equivalent variational wavefunctions 
by SU(2)$\times$SU(2) rotations of $|\vec{z},\vec{z}\rangle$. These states are of the general form 
\be\label{Eqn:nanb}
|\vec{d}_a,\vec{d}_b\rangle \equiv \prod_{p=1}^{N_p} \left( |s\rangle_{\alpha\gamma}\otimes |\vec{d}_b\rangle_{\beta\delta} + (-1)^{\vec{Q}\cdot\vec{R}_p} |\vec{d}_a\rangle_{\alpha\gamma}\otimes |s\rangle_{\beta\delta} \right) ~,
\ee 
where now we have two (in general) different directors pointing along $\vec{d}_a$ and $\vec{d}_b$ in the two sublattices. 
By symmetry all these wavefunctions give the same variational energy and so we can choose any one of them to work with.

In the following we take the state $|\vec{z},\vec{z}\rangle$ and look at the nematic order parameters for the initial spin-1/2 degrees of freedom. 
Using the relations 
\bea
&&P~S_\alpha^{\pm} S_\gamma^{\pm}~P = |\upa\rangle \langle\upa| \otimes \frac{1}{2} (T^{\pm})^2 \\
&& P~S_\beta^{\pm} S_\delta^{\pm}~P = |\dna\rangle \langle\dna| \otimes \frac{1}{2} (T^{\pm})^2 ~,
\eea
and
\bea
&&\langle \tau_x = \pm \frac{1}{2} | \left( \frac{1}{2}+\tau_z \right) | \tau_x = \pm \frac{1}{2}\rangle  = \frac{1}{2} \\
&&\langle \tau_x = \pm \frac{1}{2} | \left( \frac{1}{2}-\tau_z \right)  | \tau_x = \pm \frac{1}{2}\rangle  = \frac{1}{2} ~,
\eea
and looking back at Fig.~\ref{Fig:PlaqSqLattice} we find: 
\be\label{eqn:n1}
\langle S_{\alpha_i}^+S_{\gamma_i}^+ \rangle = \langle S_{\beta_i}^+S_{\delta_i}^+\rangle \neq 0 ~,
\ee
but e.g. $\langle S_{\alpha_3}^+S_{\gamma_4}^+\rangle = 0$.  There are two important points to note here. 
First, according to Eq.~(\ref{eqn:n1}), the above variational GS has a finite spin-nematic order parameter in both sublattices.  
This is however not in line with our ED results in the full $\mc{K}_x$-model, since  
the low-E tower of states suggests that the spin-nematic ordering occurs only in one of the two sublattices (see Sec.~\ref{Sec:Kx}).    
Secondly, the fact that $\langle S_{\alpha_1}^+S_{\gamma_1}^+\rangle$ is not equal to $\langle S_{\alpha_3}^+S_{\gamma_4}^+\rangle$ 
is in line with the d-wave symmetry shown in Fig.~\ref{Fig:nematic}. However, as we comment in Sec.~\ref{Sec:Kx}, this result is to some extent 
artificial since the bonds ($\alpha_1,\gamma_1$) and ($\alpha_3,\gamma_4$) become inequivalent once we plaquetize our lattice.  

Finally, we discuss the influence of an external magnetic field. 
Here the field reduces the symmetry of the Hamiltonian down to U(1)$\times$U(1). 
Following Refs.~[\onlinecite{Ivanov,LauchliPenc,Penc}], we expect that an infinitesimal field will pin the two directors on the $xy$-plane
and will also induce a finite longitudinal dipolar moment in both sublattices. 
This picture is confirmed by a numerical minimization of the corresponding variational state in a field.  
In particular, we find that the spin-1 objects become fully polarized at a critical field $H_c=K_x/32$.

\section{Tower of states spectroscopy}\label{App:Towers}
\subsection{Method}
Here we provide the group-theory derivation of the low-energy tower of states corresponding to some of the states that we encountered in this work.
The method is quite general and is based on the following recipe from group theory. 
By applying all elements of a symmetry group $G$ on a given state $|c\rangle$ we generate a family (or orbit) of symmetry-equivalent states. 
By construction, this family of states provides a representation ${\bf O}$ of $G$ and can be decomposed into IR's ${\bf D}^\alpha$ of $G$ as follows 
\be
{\bf O}(g) = \bigoplus _{\alpha}~ m_\alpha ~{\bf D}^\alpha(g)
\ee  
where the number of times a given IR labeled by $\alpha$ appears in this decomposition is given by the well known formula\cite{Tinkham,Batanouny,Inui}  
\be\label{eqn:ma1}
m_\alpha = \frac{1}{|G|} \sum_{g \in G} \chi^\alpha (g)~\text{Tr} [ {\bf O}(g) ]^\ast ~.
\ee 
Let us now assume that the starting state $|c\rangle$ is a classical magnetic state 
(i.e. a site-factorized product of coherent states where each spin points to some fixed direction), and let us 
choose $G=G_r \times$ SO(3) as the product of a real space group $G_r$ times the global SO(3) rotations in spin space.  
The generated orbit ${\bf O}$ is then a continuous family of states and its decomposition into symmetry sectors 
can be obtained from Eq.~(\ref{eqn:ma1}), if we can find a way to calculate the trace Tr$[{\bf O}(g)]$ for each $g \in G$.  
For finite systems this is a quite involved task given that one first needs to find an orthonormal basis that spans the orbit. 
One alternative method is to use projection operators (which project onto specific rows of irreducible representations),   
but this relies on performing a number of numerical integrations over the group manifold for each symmetry sector and finite cluster separately. 

A much better alternative is to keep Eq.~(\ref{eqn:ma1}), but make use of the thermodynamic limit.   
Here the generated orbit of the classical states themselves give an orthonormal basis because, in contrast to the finite-size case, 
two classical states that are related even by an infinitesimal global rotation are strictly orthogonal to each other.  
One immediate consequence of this feature is that out of the continuous sum over $g$ in Eq.~(\ref{eqn:ma1}) 
only a much smaller (and usually discrete) subset of operations give a non-vanishing trace. 
Specifically, these are the operations $h$ that leave the state $|c\rangle$ invariant up to a global phase, namely $h |c\rangle= e^{i \phi_h} |c\rangle$.  
These elements form a subgroup $H_c$ of $G$, called the stabilizer of $|c\rangle$.  
The above orthonormal basis is actually in one-to-one correspondence with the elements of the factor group $G/H_c$,  
and Eq.~(\ref{eqn:ma1}) reduces to\footnote{In somewhat more technical terms, 
the classical state must belong to a given one-dimensional representation 
$D^\lambda$ of $H_c$ ($\lambda$ is not always the identity representation of $H_c$ because of the non-trivial phase factors $e^{i \phi_h}$), 
and ${\bf O}$ is the so-called ``induced representation'' $D^E\!\uparrow\!G$. 
So what we are actually doing here amounts to a decomposition of $D^E\!\uparrow\!G$ into symmetry sectors of $G$.}\footnote{The derivation of Eq.~(\ref{eqn:ma2}) also appears in App.~B of Ref.~[\onlinecite{polytopes}] 
but without the phase factors $e^{i \phi_h}$ which were overlooked. This however does not affect the results of Ref.~[\onlinecite{polytopes}] since $e^{i \phi_h}=1$ in all cases studied there.} 
\be\label{eqn:ma2}
m_{\alpha} = \frac{1}{|H_c|} \sum_{h \in H_c} e^{-i \phi_h} \chi^{\alpha}(h)~.
\ee
This is our central formula for the derivation of the tower of states for a magnetic state. 

A few important comments are in order here. 
First, the above derivation remains valid whether we take for $G$ the complete group of the Hamiltonian or any subgroup of it.  
Second, one should always check whether the orbit generated by the elements of $G$ covers the full family of classical states that we want to consider. 
For example, the family of the collinear states that are stabilized for large $J_2/J_1$ 
comprises two sub-families of states. In the first (resp. second), the spins are FM (resp. AFM) along the line $y'$ of Fig.~\ref{fig:ColOrtho} and AFM (resp. FM) 
along the line $x'$. These two sets are not related to each other by a global rotation in spin space nor by any primitive translation of the lattice. 
So if we take $G=$SO(3) alone or $G=\mc{T}\otimes$ SO(3) (where $\mc{T}$ is the translation group) we generate two distinct invariant orbits and each one must be decomposed separately. 
By contrast, the above two sets of collinear states can be mapped to each other by a $C_4$ spatial rotation. 
So if we include $C_4$ in $G$, then the generated orbit will comprise both sets of states and so we have to do the decomposition just once. 

Our final comment on Eq.~(\ref{eqn:ma2}) is about the phase factors $e^{i \phi_h}$. 
We may understand the origin and significance of these phase factors by taking as an example the 2-sublattice N\'eel AFM on the square lattice where half of the spins point 
along $+\vec{z}$ and the other half along $-\vec{z}$. 
Then one element of the stabilizer is a combination of a $C_2$ spin rotation around e.g. the y-axis with a spatial operation that maps one sublattice to the other 
(e.g. a translation by one lattice constant). 
In this case we pick up a phase factor of $e^{i\pi/2}$ from the action of the spin rotation on each spin site. 
With an even number $N$ of spins, this gives $e^{i\phi_h} = (-1)^{N/2}$. 
The same phase factor arises for a $C_2$ spin rotation around any axis on the xy-plane.  
Overall we get a different decomposition for the square lattice N\'eel AFM depending on whether $N=4p$ or $N=4p+2$ (where $p$ is an integer).\footnote{This signifies that there might 
exist different types of clusters that approach the thermodynamic limit in a different way.}

In the following we shall make use use of the central formula (\ref{eqn:ma2}) to decompose the orthogonal and the collinear families of states into 
IR's of the group $G=\mc{T}~C_4 \otimes$ SO(3). Then we shall derive some further symmetry properties by including the glide operations $(\sigma_i|\boldsymbol{\tau})$ as well.

\begin{figure}[!t]
\includegraphics[clip=true,width=0.9\linewidth]{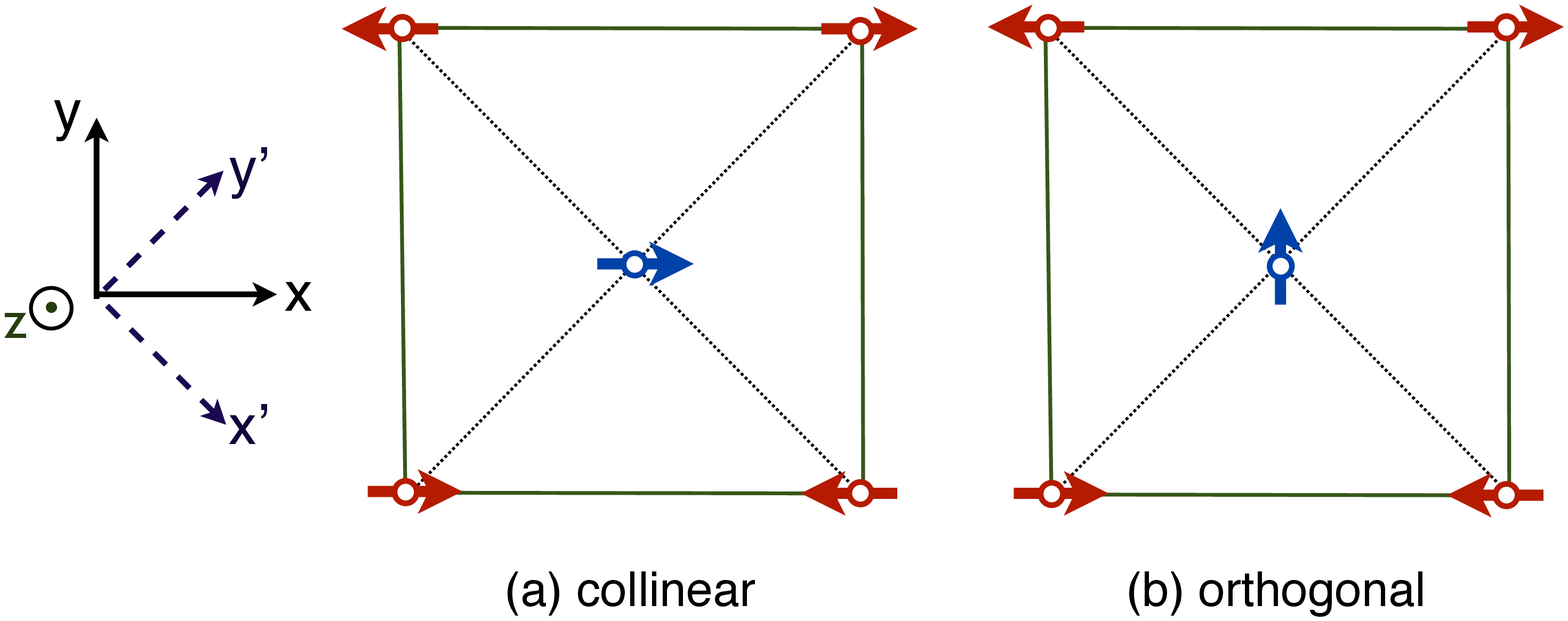}
\caption{Classical collinear and orthogonal states.}
\label{fig:ColOrtho}
\end{figure}

\subsection{Orthogonal phase}
For the orthogonal state we take the global spin orientations shown in Fig.~\ref{fig:ColOrtho}(b). 
The stabilizer of this state consists of the following operations $h$
\bea
&&\{ t_x^{2n}~ t_y^{2m},~t_x^{2n+1}~ t_y^{2m+1} \}~\otimes \{E, C_2 \}\nonumber\\
&&\{ t_x^{2n+1}~ t_y^{2m},~t_x^{2n}~ t_y^{2m+1} \}~\otimes \{E, C_2 \}~ \otimes R_z(\pi)\nonumber\\
&&\{ t_x^{2n}~ t_y^{2m},~t_x^{2n+1}~ t_y^{2m+1} \}~\otimes \{C_4, C_4^{-1} \}~ \otimes R_y(\pi) \nonumber \\
&&\{ t_x^{2n+1}~ t_y^{2m},~t_x^{2n}~ t_y^{2m+1} \}~\otimes \{C_4, C_4^{-1} \}~ \otimes R_x(\pi) \nonumber
\eea
where $n,m=1,\cdots L/2$,  $R_{x,y,z}(\pi)$ are spin rotations by $\pi$ around the axes $x$, $y$, and $z$ shown in Fig.~\ref{fig:ColOrtho}. 
The phase factors $e^{i \phi_h}$ are provided in Table \ref{tab:phase}. 
In the following we shall consider systems where the number of sites $N$ is such that all phase factors reduce to 1. 
 
Adding the contributions from the above operations $h$ and using $\chi^S(R(0))=2S+1$, and $\chi^S(\pi)=(-1)^S$, 
we find that the only non-vanishing integers $m_{\vec{k},l,S}$ are those for $\vec{k}=0$ and $(\pi,\pi)$, with
\bea
&&
m_{\vec{k}=0,l,S} =\frac{1}{4} \delta_{l,\{0,\pi\}} \Big[ 2S+1 + (-1)^S \Big( 1 + 2 e^{i l} \Big) \Big] \nonumber \\
&&
m_{\vec{k}=(\pi,\pi),l,S} =\frac{1}{4} \delta_{l,\{0,\pi\}} \Big[  2S+1 - (-1)^S \Big] ~.\nonumber
\eea 
We note in particular that 
\be
m'_{S}   \equiv \sum_{\vec{k},l} m_{\vec{k},l,S} = 2S+1 ~,
\ee
so we have 2S+1 different total S-states for each given S.   
This important result can be also arise by a decomposition of the classical states into IR's of SO(3) alone. 
Since the orthogonal state breaks all global spin symmetries, and so the stabilizer consists only of the identity $R(0)$, whose character gives the $(2S+1)$-multiplicity.  

We should remark here that the above symmetry structure has also been reported for the orthogonal phase found in Ref.~[\onlinecite{Lauchli}]. 

\begin{table}[!t]
\caption{Phase factors $e^{i\phi_h}$ that appear in the symmetry decomposition of the orthogonal and the collinear phase (see Fig.~\ref{fig:ColOrtho}). 
Here $h_s$ stands for the spin-space portion of the given symmetry operation $h$ of the stabilizer $H_c$.}\label{tab:phase}
\begin{ruledtabular}
\begin{tabular}{ll || ll}
orthogonal  &  & collinear &\\
\hline
$h_s$ & $e^{i \phi_h}$ & $h_s$ & $e^{i \phi_h}$ \\
\hline
$R(0)$ & 1 & R(0)& 1\\
$R_z(\pi)$ & $(-i)^N$ & $R_x(\phi)$& 1\\
$R_{x,y}(\pi)$,$R_z(\pm \pi/2)$ & $(-1)^{N/4}$ & $R_{\vec{a}}(\pi)$& $(-1)^{N/2}$
\end{tabular}
\end{ruledtabular}
\end{table}

\textit{Inclusion of glides $(\sigma_i|\boldsymbol{\tau})$}.--
So far we have not exploited the four non-symorphic operations $(\sigma_i|\boldsymbol{\tau})$ of the lattice.   
The states with $\vec{k}=(\pi,\pi)$ belong necessarily to the 2-dimensional IR ``$(\pi,\pi).\{0,\pi\}$''. 
So it is sufficient to discuss the states with $\vec{k}=0$ only. 
In addition to the elements of the stabilizer listed above, now we may add the following symmetries that involve a glide operation
\bea
&&
\{ t_x^{2n}~ t_y^{2m},~t_x^{2n+1}~ t_y^{2m+1} \}~\otimes (\sigma_{x, y}|\boldsymbol{\tau}) \otimes R_z(-\pi/2) \nonumber \\
&&
\{ t_x^{2n}~ t_y^{2m},~t_x^{2n+1}~ t_y^{2m+1} \}~\otimes (\sigma_{d, d'}|\boldsymbol{\tau}) \otimes R_{y'}(\pi) \nonumber\\
&&
\{ t_x^{2n+1}~ t_y^{2m},~t_x^{2n}~ t_y^{2m+1} \}~\otimes  (\sigma_{x, y}|\boldsymbol{\tau}) \otimes R_z(\pi/2)\nonumber \\ 
&&
\{ t_x^{2n+1}~ t_y^{2m},~t_x^{2n}~ t_y^{2m+1} \}~\otimes  (\sigma_{d, d'}|\boldsymbol{\tau}) \otimes R_{x'}(\pi) \nonumber 
\eea
where $R_{x',y'}(\pi)$ are spin rotations by $\pi$ around the $x', y'$ directions shown in Fig. \ref{fig:ColOrtho}. 
By including these operations in the stabilizer group we get the following multiplicities at zero momentum:  
\bea
m_{\vec{k}=0,\Gamma,S} = \frac{1}{16} &\Big[& \!(2S+1) \Big( \chi^\Gamma(E)+\chi^\Gamma(C_2)\Big) \nonumber \\
&+& (-1)^S \Big( \chi^\Gamma(E)+\chi^\Gamma(C_2)+4 \chi^\Gamma(C_4) \Big) \nonumber \\
&+& 2 (-1)^S \Big( \chi^\Gamma(\sigma_d|\boldsymbol{\tau})+ \chi^\Gamma(\sigma_{d'}|\boldsymbol{\tau})\Big)  \nonumber \\
&+& 2  \chi^S(\pi/2) \Big( \chi^\Gamma(\sigma_x|\boldsymbol{\tau})+\chi^\Gamma(\sigma_y|\boldsymbol{\tau}) \Big) \nonumber
 \Big]~,
\eea
where $\Gamma$ labels the IR's of $C_{4v}$ shown in Table \ref{tab:C4v}.  
In Table \ref{tab:col} we show the specific multiplicities for the lowest four total spin sectors $S$.

\begin{table}[!t]
\caption{Group theory predictions for the content of the low-lying tower of states in the orthogonal phase. }\label{tab:ortho}
\begin{ruledtabular}
\begin{tabular}{c|ccccc}
S & 0 & 1 & 2 & 3 & 4 \\
\hline
\hline
$0.A1 ~(l=0) $        & 1 &    & 1 &    & 2 \\
$0.A2 ~(l=0) $        &    &    & 1 & 1 & 1 \\
$0.B1 ~(l=\pi) $      &    & 1 &    & 1 & 1 \\
$0.B2 ~(l=\pi) $      &    &    &1  & 1 &1  \\
\midrule[0.8pt]
$(\pi,\pi).\{l=0,\pi\}$           &    &  1  &  1 &  2 & 2\\
\midrule[0.8pt]
Total & 1 & 3 & 5 & 7 & 9 
\end{tabular}
\end{ruledtabular}
\end{table}

\subsection{Collinear phase}
We turn now to the collinear phase with the global spin orientations shown in Fig.~\ref{fig:ColOrtho}(a). 
As we mentioned above, by including the spatial $C_4$ rotation in the group $G$ we generate an orbit which includes both sets of collinear states.
The stabilizer of any member of this orbit consists of the following operations
\bea
&&
\{ t_x^{2n}~ t_y^{2m}, ~t_x^{2n+1}~ t_y^{2m+1}\} ~\otimes \{E, C_2 \}~\otimes \{ R_x(\phi) \} \nonumber\\
&&
\{ t_x^{2n+1}~ t_y^{2m},~t_x^{2n}~ t_y^{2m+1}\}~\otimes \{E, C_2 \}~~ \otimes  \{ R_{\vec{a}}(\pi) \}  \nonumber
\eea
where $\vec{a}$ is any axis perpendicular to the direction $\vec{x}$ of the spins (see Fig.~\ref{fig:ColOrtho}), and $\phi \in [0,2\pi)$. 
Again in the following we are considering system sizes $N$ such that all phase factors $e^{i\phi_h}$ corresponding to the collinear phase (see Table \ref{tab:phase}) reduce to 1. 
We have $|\{R_z(\phi)\}|=|\{R_{\vec{a}}(\pi)\}|=\int_0^{2\pi} d\phi = 2\pi$, and 
\be\label{e1}
\frac{1}{2\pi} \int_0^{2\pi} d\phi ~\chi^S (R_z(\phi)) = \frac{1}{2\pi} \int_0^{2\pi} d\phi ~\frac{\sin (S+\frac{1}{2})\phi}{\sin \frac{\phi}{2}} =  1 
\ee
for integer $S$ (this is our case since we consider clusters with an even number of sites), and similarly 
\be\label{e2}
\frac{1}{2\pi} \int_0^{2\pi} d\phi ~\chi^S (R_\vec{a}(\pi)) = \frac{1}{2\pi} \int_0^{2\pi} d\phi ~ (-1)^S =  (-1)^S ~.
\ee
So we find that the only non-vanishing integers $m_{\vec{k},l,S}$ are those with $\vec{k}=0$ and $(\pi,\pi)$ with 
\bea
&&m_{\vec{k}=0,l,S} = \delta_{l,\{0,\pi\}} \frac{1+(-1)^S}{2}, \nonumber \\
&& m_{\vec{k}=(\pi,\pi),l,S} = \delta_{l,\{0,\pi\}} \frac{1-(-1)^S}{2}~, \nonumber 
\eea 
and in particular 
\be
m'_{S}   \equiv \sum_{\vec{k},l} m_{\vec{k},l,S} = 2~,
\ee
so we have 2 different total S-states for each given S, one with $l=0$ and another with $l=\pi$. 
This result also follows from the decomposition of the classical states into IR's of SO(3) alone. Here the stabilizer 
consists of the U(1) rotations $\{ R_x(\phi) \} $, whose weighted integral over the character gives $1$. An extra factor of 2 comes from the
fact that we have to consider each family of collinear states as a separate orbit (representation) since, as we mentioned above, 
they are not related to each other by SO(3) alone.  

\begin{table}[!t]
\caption{Group theory predictions for the content of the low-lying tower of states in the collinear phase. }\label{tab:col}
\begin{ruledtabular}
\begin{tabular}{c|ccccc}
S & 0 & 1 & 2 & 3 & 4 \\
\hline
\hline
$0.A1 ~(l=0)$                   & 1 &  & 1 &  & 1 \\
$0.B2 ~(l=\pi)$                  & 1&  & 1 &  & 1 \\
\midrule[0.8pt]
$(\pi,\pi).\{l=0,\pi\}$  &    &  1  &  & 1 & \\
\midrule[0.8pt]
Total & 2 & 2 & 2 & 2 & 2 
\end{tabular}
\end{ruledtabular}
\end{table}

\textit{Inclusion of glides $(\sigma_i|\boldsymbol{\tau})$}.--
Let us now exploit the four non-symorphic operations $(\sigma_i|\boldsymbol{\tau})$ of the lattice in order to make further predictions 
for the symmetry properties of the collinear tower of states under these operations.  
As we did above, it is sufficient to discuss the states with $\vec{k}=0$ which appear for even values of $S$ here.  
Since the collinear tower has only two states per spin sector $S$ it is sufficient to look at the decomposition 
of the classical family of collinear states 
into IR's of $C_{4v}$ alone, i.e. disregarding the operations in spin space and the translations.    
To this end one needs the corresponding (stabilizer) subgroup of $C_{4v}$ with the operations that leave a given collinear state invariant. 
It is easy to check that this is the $C_{2v}$ subgroup which comprises in addition to $E$ and $C_2$ the glides $(\sigma_d|\boldsymbol{\tau})$ and $(\sigma_{d'}|\boldsymbol{\tau})$. 
Using the characters shown in Table \ref{tab:C4v} one then finds that only the IR's $A_1$ and $B_2$ should appear in the tower of states when $\vec{k}=0$. 
This is because only $A_1$ and $B_2$ remain invariant under the action of $(\sigma_d|\boldsymbol{\tau})$ and $(\sigma_{d'}|\boldsymbol{\tau})$. 
So for each even value of $S$, we have one $A_1$ and one $B_2$ state. The first is symmetric with respect to all non-symmorphic operations, 
while $B_2$ is even with respect to $(\sigma_d|\boldsymbol{\tau})$ and $(\sigma_{d'}|\boldsymbol{\tau})$ but odd with respect to $(\sigma_x|\boldsymbol{\tau})$ and $(\sigma_y|\boldsymbol{\tau})$.

\subsection{The spin-nematic variational state of App.~\ref{App:SP}}\label{App:TOSvar}
Here we derive the spin symmetry decomposition of the spin-nematic variational state $|\vec{z},\vec{z}\rangle$ of Eq.~(\ref{Eqn:zz}).  
To this end, we first need to find the subgroup of SU(2)$\times$SU(2) rotations which leave the state $|\vec{z},\vec{z}\rangle$ 
invariant up to a global phase. One set of operations that belong to the stabilizer is the U(1)$\times$U(1) rotations around the z-axis 
\be
R(\phi_a,\phi_b) = R_a(\phi_a \vec{z}) R_b(\phi_b \vec{z})~.
\ee 
Another set of operations is that of simultaneous $\pi$-rotations $C_{\infty \text{v}}\times C_{\infty \text{v}}$ in the two sublattices around any pair  $\vec{n}_a, \vec{n}_b$ of axes in the xy-plane 
\be
R(\pi \vec{n}_a, \pi \vec{n}_b) = R_a(\pi \vec{n}_a) R_b(\pi  \vec{n}_b)~.
\ee 
This can be shown by first noting that a $\pi$-rotation around an axis in the xy-plane reverses the sign of a triplet $|t_0\rangle$, e.g. 
$e^{-i \pi S_x} |t_0\rangle = - |t_0\rangle$. Hence, to restore a plaquette wavefunction of the type 
\be
|\psi_p\rangle = \frac{1}{\sqrt{2}} \left( |s\rangle_{\alpha\gamma}|t_0\rangle_{\beta\delta} \pm |t_0\rangle_{\alpha\gamma}|s\rangle_{\beta\delta} \right)
\ee
(see Eq.~(\ref{Eqn:zz})), we must perform a $\pi$-rotation in both sublattices A and B. 
Since this gives a minus sign for a single plaquette, we get an overall phase of $(-1)^{N_p}$, where $N_p=N/4$ is the number of strong plaquettes. 

Having established the stabilizer of the state $|\vec{z},\vec{z}\rangle$, we may now make use of  Eq.~(\ref{eqn:ma2}) in order to decompose it 
into symmetry sectors of SU(2)$\times$SU(2), which are labeled by the total spins $S_A$ and $S_B$ of the two sublatices. 
For clusters with even $N/4$, we get  
\be
m_{S_A,S_B} = \frac{1}{2} \left( 1 + (-1)^{S_A+S_B} \right)~. 
\ee
So, the tower of states corresponding to the variational state $|\vec{z},\vec{z}\rangle$ must consist only of sectors with even $S_A+S_B$. 
The numerical spectra for the $\mc{K}_x$-model (see Fig.~\ref{fig:Eff16and32spectra}) show a different symmetry pattern in the tower of states,   
so we believe that the variational state of Sec.~\ref{App:SP} does not describe the GS of the $\mc{K}_x$-model.

\subsection{U(1) spin-nematic phase}
The numerical tower of states shown of Fig.~\ref{fig:Eff16and32spectra} consist of sectors $(S_A,S_B)$ where one of the two spins is zero and the other is an even integer. 
The first suggests that the symmetry breaking occurs in one sublattice only, while the second suggests a spin-nematic state with U(1) symmetry.   
To demonstrate this we provide a simple derivation of the tower of states for a U(1) spin-nematic phase corresponding to a product of an even number of directors, all pointing along the $z$-axis.  

The stabilizer consists of global rotations around the $z$-axis as well as global $\pi$-rotations $C_{\infty \text{v}}$ around any axis on the $xy$-plane. 
Given Eqs.~(\ref{e1}) and (\ref{e2}) above for the corresponding characters, we arrive at 
\be
m_S = \frac{1+(-1)^S}{2}~,
\ee
namely, only even sectors $S$ participate in the tower. 
Furthermore, the fact that we get a single state per even S in the tower of states is related to the U(1) symmetry of the spin-nematic state described here.  
If the directors were not collinear, as it happens e.g. in the antiferro-quadrupolar phase of the $S=1$ billinear-biquadratic model 
in the triangular lattice,\cite{LauchliPenc} then a more complicated multiplicity appears in the tower of states. 



\begin{thebibliography}{99}
\bibitem{HFM} {\it Introduction to Frustrated Magnetism}, edited by C. Lacroix, P. Mendels, and F. Mila, Springer Series in Solid-State Sciences, Vol. 164 (Springer, Berlin, 2011). 
\bibitem{Diep} {\it Frustrated spin systems}, edited by H. T. Diep (World Scientific, Singapore, 2004). 
\bibitem{Richter} J. Richter, J. Schulenburg, and A. Honecker, in \textit{Quantum Magnetism}, 
edited by  U. Schollw\"ock, J. Richter,  D. J. J. Farnell, and  R. F. Bishop, 
Lect. Notes Phys. 645 (Springer, Berlin, 2004), Chap. 2. 

\bibitem{SchmidtLuban} H-J Schmidt and M. Luban, J. Phys. A: Math. Gen. {\bf 36}, 6351 (2003). 

\bibitem{Kepler} J. Kepler, {\it Harmonices Mundi} (J. Planck for G. Tampach, Linz, Austria, 1619). 
\bibitem{Tilings1} B. Gr\"unbaum and  G. C. Shephard, {\it Tilings and Patterns} (Freeman, New York, 1987). 
\bibitem{Tilings2} M. Gardner, {\it Tiling with Convex Polygons,} in {\it Time travel and other mathematical bewilderments} (Freeman, New York, 1988).

\bibitem{Moessner} R. Moessner and S. L. Sondhi, \prb {\bf 63}, 224401 (2001). 
\bibitem{Urumov} V. Urumov,  J. Phys. A: Math. Gen. 35 7317 (2002). 
\bibitem{Rojas} M. Rojas, O. Rojas, S. M. de Souza, e-print arXiv:1105.5130v2.
\bibitem{Ralko} A. Ralko, \prb {\bf 84}, 184434 (2011). 
\bibitem{Raman} K. S. Raman, R. Moessner, and S. L. Sondhi, Phys. Rev. B {\bf 72}, 064413 (2005). 

\bibitem{Ressouche} E. Ressouche, V. Simonet, B. Canals, M. Gospodinov, and V. Skumryev, Phys. Rev. Lett. {\bf 103}, 267204 (2009). 
\bibitem{Iliev} M. N. Iliev, A. P. Litvinchuk, V. G. Hadjiev, M. M. Gospodinov, V. Skumryev, and E. Ressouche, Phys. Rev. B {\bf 81}, 024302 (2010). 

\bibitem{QDM} R. Moessner and K. S. Raman, in {\it Introduction to Frustrated Magnetism}, edited by C. Lacroix, P. Mendels, and F. Mila, Springer Series in Solid-State Sciences, Vol. 164 (Springer, Berlin, 2011), Chap. 17.


\bibitem{Shamir} Shamir {\it et  al.}, Acta Crystallogr. Sect. A {\bf 34}, 662 (1978). 
\bibitem{Singh} A. K. Singh {\it et al.}, Appl. Phys. Lett. {\bf 92}, 132910 (2008). 



\bibitem{Bhatt} S. Sachdev and R. N. Bhatt, Phys. Rev. B {\bf 41}, 9323 (1990).
\bibitem{Judit} J. Romh\'anyi, K. Totsuka, and K. Penc, Phys. Rev. B {\bf 83}, 024413 (2011). 


\bibitem{Chandra} P. Chandra, and B. Doucot, \prb {\bf 38}, 9335 (1988).  
\bibitem{Schultz} H. J. Schultz, T. A. L. Ziman, and D. Poilblanc, J. Phys. I {\bf 6}, 675 (1996).  
\bibitem{Bishop} R. F. Bishop, D. J. J. Farnell, and J. B. Parkinson, \prb {\bf 58}, 6394 (1998).  
\bibitem{Caprioti} L. Caprioti, F. Becca, A. Parola, and S. Sorella, \prl {\bf 87}, 097201 (2001). 
\bibitem{Sirker} J. Sirker, Z. Weihong, O. P. Sushkov, and J. Oitmaa, \prb {\bf 73}, 184420 (2006). 
\bibitem{Darradi} R. Darradi, O. Derzhko, R. Zinke, J. Schulenburg, S. E. Kr\"uger, and J. Richter, \prb {\bf 78}, 214415 (2008).

\bibitem{Roger} M. Roger, J. H. Hetherington, and J. M. Delrieu, \rmp {\bf 55}, 1 (1983).



\bibitem{CCL} P. Chandra, P. Coleman and A. I. Larkin, Phys. Rev. Lett. {\bf 64}, 88 (1990). 
\bibitem{Weber} C. Weber, L. Capriotti, G. Misguich, F. Becca, M. Elhajal, and F. Mila, \prl {\bf 91}, 177202 (2003).  

\bibitem{Shannon} N. Shannon, T. Momoi, and P. Sindzingre, \prl {\bf 96}, 027213 (2006).


\bibitem{Penc} K. Penc and A. L\"auchli, in {\it Introduction to Frustrated Magnetism}, edited by C. Lacroix, P. Mendels, and F. Mila, Springer Series in Solid-State Sciences, Vol. 164 (Springer, Berlin, 2011), Chap. 13.


\bibitem{Chubukov} A. Chubukov, E. Gagliano, and C. Balseiro, \prb {\bf 45}, 7889 (1992). 
\bibitem{Lauchli} A. L\"auchli, J. C. Domenge, C. Lhuillier, P. Sindzingre, and M. Troyer, \prl {\bf 95}, 137206 (2005).


\bibitem{White} R. M. White, {\it Quantum Theory of magnetism}, 3rd ed. (Springer-Verlag, Berlin 2007), Chap. 8.
\bibitem{Blaizot} J.-P. Blaizot, G. Ripka, {\it Quantum Theory of Finite Systems} (MIT Press, Cambridge, MA, 1985), Chap. 3.

\bibitem{Anu} Jagannathan, R. Moessner, and S. Wessel, Phys. Rev. B {\bf 74}, 184410 (2006).

\bibitem{Uhrig} G. S. Uhrig, M. Holt, J. Oitmaa, O. P. Sushkov, and R. R. P. Singh, \prb {\bf 79}, 092416 (2009). 

\bibitem{Alicea} J. Alicea, A. V. Chubukov, and O. A. Starykh, \prl {\bf 102}, 137201 (2009).


\bibitem{Anderson} P. W. Anderson, {\it Basic Notions of Condensed Matter Physics}, Frontiers in Physics, Vol. 55 (Benjamin Cummings, London, 1984). 
\bibitem{Bernu1} B. Bernu, C. Lhuillier, and L. Pierre, Phys. Rev. Lett. {\bf 69}, 2590 (1992).
\bibitem{Bernu2} B. Bernu, P. Lecheminant, C. Lhuillier and L. Pierre, Phys. Rev. B {\bf 50}, 10048 (1994).
\bibitem{Lecheminant_J1J2} P. Lecheminant, B. Bernu, C. Lhuillier, and L. Pierre, Phys. Rev. B {\bf 52}, 6647 (1995).
\bibitem{Lecheminant_kagome} P. Lecheminant, B. Bernu, C. Lhuillier, L. Pierre, and P. Sindzingre, Phys. Rev. B {\bf 56}, 2521 (1997).
\bibitem{polytopes} I. Rousochatzais, A. L\"auchli, and F. Mila, Phys. Rev. B {\bf 77}, 094420 (2008). 
\bibitem{TOSMisguich} G. Misguich, P. Sindzingre, J. Phys. Condens. Matter {\bf 19}, 145202 (2007).


\bibitem{SO1} S. K. Pati, R. R. P. Singh, and D. I. Khomskii, \prl {\bf 81}, 5406 (1998). 
\bibitem{SO2} P. Azaria, A. O. Gogolin, P. Lecheminant, and A. A. Nersesyan, \prl {\bf 83}, 624 (1999).
\bibitem{SO3} C. Itoi, S. Qin, and I. Affleck, \prb {\bf 61}, 6747 (2000).
\bibitem{SO4} P. Azaria, E. Boulat, and P. Lecheminant, \prb {\bf 61}, 12112 (2000).  
\bibitem{SO5} Y. Yamashita, N. Shibata, and K. Ueda, J. Phys. Soc. Jpn. {69}, 242 (2000).


\bibitem{Ueda} H. T. Ueda and K. Totsuka, \prb {\bf 76}, 214428 (2007). 
\bibitem{Lecheminant} P. Lecheminant and K. Totsuka, \prb {\bf 74}, 224426 (2006).

\bibitem{KK} K. I. Kugel' and D. I. Khomskii, Sov. Phys. Usp. {\bf 25}, 231 (1982).

\bibitem{Janson} O. Janson, I. Rousochatzakis, A. A. Tsirlin, J. Richter, Yu. Skourski, H. Rosner, 
Phys. Rev. B {\bf 85}, 064404 (2012). 

\bibitem{Oitmaa} J. Oitmaa, C. Hamer, and W. Zheng, {\it Series Expansion Methods for Strongly Interacting Lattice Models} (Cambridge University Press, Cambridge, 2006).
\bibitem{Klein}  D. J. Klein, J. Chem. Phys. {\bf 61}, 786 (1974), Appendix A.

\bibitem{LauchliPenc} A. L\"auchli, F. Mila, and K. Penc, \prl {\bf 97}, 087205 (2006).
\bibitem{Ivanov} B. A. Ivanov and A. K. Kolezhuk, \prb {\bf 68}, 052401 (2003).  

\bibitem{Tinkham} M. Tinkham, {\it Group Theory and Quantum Mechanics} (Dover, New York, 2003). 
\bibitem{Batanouny}  M. El-Batanouny and F. Wooten, {\it Symmetry and Condensed Matter Physics, A Ccomputational Approach} (Cambridge University Press, Cambridge, 2008).
\bibitem{Inui} T. Inui, Y. Tanabe, and Y. Onodera, {\it Group Theory and Its Applications in Physics} (Springer-Verlag, Berlin, 1996).
 
\end{thebibliography}
\end{document}